\documentclass[a4paper,11pt]{article}
\usepackage{bm,latexsym,amsmath,amssymb,amsfonts,mathrsfs}
\usepackage[pdftex]{graphicx}
\usepackage[colorlinks=true, pdfstartview=FitV, linkcolor=black, citecolor=blue, urlcolor=black, pdftitle={Effective field theory for spacetime symmetry breaking},pdfauthor={Yoshimasa Hidaka, Toshifumi Noumi and Gary Shiu},pdfsubject={}, pdfkeywords={}]{hyperref}
\usepackage[nosort]{cite}
\usepackage{color}
\input{colordvi.tex}

\allowdisplaybreaks

\newcommand{\Mc}[1]{{\bar{#1}}}

\newcommand{\blue}[1]{{\color{blue} #1}}

\newcommand{\tem}[1]{#1_{\parallel}}
\newcommand{\spa}[1]{#1_{\bot}}
\begin{document}

\baselineskip=17pt

\begin{titlepage}
\rightline{RIKEN-MP-96, RIKEN-QHP-171, MAD-TH-14-10}
\begin{center}
\vskip 1.5cm
{\Large \bf {
Effective field theory for spacetime symmetry breaking}}\\[5mm]
\vskip 1cm

{\large {Yoshimasa Hidaka$^1$, Toshifumi Noumi$^1$ and Gary Shiu$^{2,3}$}}
\vskip 1.0cm
$^1${\it Theoretical Research Devision, RIKEN Nishina Center, Japan}\\[3mm]
$^2${\it Department of Physics, University of Wisconsin, Madison, WI 53706, USA}\\[3mm]
$^3${\it Center for Fundamental Physics and Institute for Advanced Study,\\ Hong Kong University of Science and Technology, Hong Kong}\\[5mm]
hidaka@riken.jp, toshifumi.noumi@riken.jp, shiu@physics.wisc.edu

\vskip 1.5cm
{\large {\bf Abstract}}
\end{center}
\vskip 0.5cm
\baselineskip 16pt

We discuss the effective field theory
for spacetime symmetry breaking
from the local symmetry point of view.
By gauging spacetime symmetries,
the identification of Nambu-Goldstone (NG) fields and the construction of the effective action are performed
based on the breaking pattern of
diffeomorphism, local Lorentz, and (an)isotropic Weyl symmetries as well as the internal symmetries including possible central extensions in nonrelativistic systems.
Such a local picture
distinguishes, e.g., whether the symmetry breaking condensations have spins
and provides a correct identification of the physical NG fields,
while the standard coset construction based on global symmetry breaking does not.
We illustrate
that the local picture becomes important
in particular
when we take into account massive modes associated with symmetry breaking,
whose masses are not necessarily high.
We also revisit the coset construction for spacetime symmetry breaking.
Based on the relation between the Maurer-Cartan one form and connections for spacetime symmetries,
we classify the physical meanings of the inverse Higgs constraints by the coordinate dimension of broken symmetries.
Inverse Higgs constraints for spacetime symmetries with a higher dimension remove the redundant NG fields,
whereas those for dimensionless symmetries can be further classified by the local symmetry breaking pattern.

\end{titlepage}

\setcounter{tocdepth}{2}

\tableofcontents

\newpage
\section{Introduction}
\setcounter{equation}{0}

Symmetry and its spontaneous breaking play an important role in various areas of physics.
In particular,
the low-energy effective field theory (EFT)
based on the underlying symmetry structures
provides a powerful framework for understanding the low-energy dynamics in the symmetry broken phase~\cite{Manohar:1996cq}.

\medskip
For internal symmetry breaking
in Lorentz invariant systems, the EFT based on coset construction had been established  in 1960's~\cite{Callan:1969sn,Coleman:1969sm}. 
When a global symmetry group $G$
is broken to a residual symmetry group $H$,
the corresponding Nambu-Goldstone (NG) fields $\pi(x)$
are introduced as the coordinates of the coset space $G/H$
and
the general effective action can be constructed
from the Maurer-Cartan one form,
\begin{align}
\label{MC_intro}
J_\mu dx^\mu=\Omega^{-1}\partial_\mu\Omega dx^\mu
\quad
{\rm with}
\quad
\Omega(x)=e^{\pi(x)} \in G/H
\,.
\end{align}
Such a coset construction was also extended
to spacetime symmetry breaking~\cite{Volkov:1973vd, Ogievetsky} accompanied by the inverse Higgs constraints~\cite{Ivanov:1975zq}
and
has been applied to various systems (see, e.g.,~\cite{McArthur:2010zm,Fedoruk:2011ua,Hinterbichler:2012mv,Gomis:2012ki,Nicolis:2013sga,Creminelli:2013fxa,Endlich:2013vfa,Brauner:2014aha, Goon:2014ika,Creminelli:2014zxa,Delacretaz:2014jka,Delacretaz:2014oxa,Brauner:2014jaa} for recent discussions).
Although the coset construction captures certain aspects of spacetime symmetry breaking,
its understanding seems incomplete compared to the internal symmetry case
and
as a result generated a lot of recent research activities~\cite{McArthur:2010zm,Fedoruk:2011ua,Hinterbichler:2012mv,Gomis:2012ki,Nicolis:2013sga,Creminelli:2013fxa,Endlich:2013vfa,Brauner:2014aha, Goon:2014ika,Creminelli:2014zxa,Delacretaz:2014jka,Delacretaz:2014oxa,Brauner:2014jaa,
Low:2001bw,Bluhm:2004ep,Bluhm:2007bd,Nicolis:2011pv,Watanabe:2013iia,Hayata:2013vfa,Kobayashi:2014xua,Takahashi:2014vua,Castillo:2014dta,Nitta:2014jta}.
It would then be helpful to
revisit the issue of spacetime symmetry breaking
based on an alternative approach,
providing a complementary perspective to the coset construction. 

\medskip
For this purpose,
let us first revisit the identification of NG fields
for spacetime symmetry breaking.
As in standard textbooks,
symmetry breaking structures are classified by
the type of order parameters
and their local transformations
generate the corresponding NG fields
(we refer to this as the {\it local picture}).
In Lorentz invariant systems,
since only the condensation of scalar fields is allowed,
we need not pay much attention to the type of order parameters. However, when  Lorentz symmetry is broken or does not exist, the type of order parameters becomes more important.
For example,
when the order parameter is a non-abelian charge density, there appear 
NG modes with a quadratic dispersion different from that in Lorentz invariant systems~\cite{Leutwyler:1993gf,Schafer:2001bq,Nambu:2004,Watanabe:2011ec,Watanabe:2012hr,Hidaka:2012ym,Amado:2013xya}.
In addition, if the charge density and the other order parameter that break the same symmetry coexist, some massive modes associated with the symmetry breaking appear~\cite{Nicolis:2013sga,Kapustin:2012cr,Gongyo:2014sra,Hayata:2014yga}.\footnote{
In this paper
we use the word ``NG fields"
to denote fields which transform nonlinearly under broken symmetries.
In general,
the NG fields can contain massive modes as well as massless modes.
We refer to the massless modes in NG fields
as the NG modes in particular. }

\medskip
For spacetime symmetry breaking, the standard coset construction based on global symmetry (refer to as the {\it  global picture}) 
does not distinguish the types of order parameters.
As is well-known in the case of conformal symmetry breaking~\cite{Ogievetsky,Ivanov:1975zq},
a naive counting of broken spacetime symmetries based on the global picture contains redundant fields and
causes a wrong counting of NG modes (see, e.g., Refs.~\cite{Low:2001bw,Watanabe:2013iia,Hayata:2013vfa}).
The inverse Higgs constraints are introduced to compensate such a mismatch of NG mode counting. 
As discussed in Refs.~\cite{Nicolis:2013sga,Endlich:2013vfa,Brauner:2014aha}, the inverse Higgs constraints eliminate not only the redundant fields but also the massive modes. Thus,  to identify the physical NG fields, 
we should take into account the
massive modes associated with the symmetry breaking in addition to the massless modes.
Such massive modes often play an important role, e.g., the smectic-A phase of liquid crystals near the smectic-nematic phase transition, in which the rotation modes are massive~\cite{deGennesText}.
In this paper, we would like to construct the effective action including these modes based on the local picture.

\medskip
To proceed in this direction,
it is convenient to recall
the relation between the coset construction
and gauge symmetry breaking
for internal symmetry.
When a gauge symmetry is broken,
the NG fields are eaten by the gauge fields, and the dynamics
is  captured
by the unitary gauge action for the massive gauge boson $A_\mu$.
Since the gauge boson mass is given by $m\sim gv$
with the gauge coupling $g$
and the order parameter~$v$,
the unitary gauge is not adequate
to discuss the global symmetry limit $g\to0$,
which corresponds to the singular massless limit.
Rather,
it is convenient to introduce NG fields
by the St\"{u}ckelberg method as
\begin{align}
\label{gauge_intro}
A_\mu\to A_\mu'=\Omega^{-1}A_\mu\Omega+\Omega^{-1}\partial_\mu\Omega
\quad
{\rm with}
\quad
\Omega(x)\in G/H
\,,
\end{align}
where $G$ and $H$ are the original and residual symmetry groups, respectively,
and $\Omega(x)$ describes the NG fields.
In this picture,
we can take the global symmetry limit smoothly
to obtain the same effective action constructed from the Maurer-Cartan one form \eqref{MC_intro}.
As this discussion suggests,
the unitary gauge is convenient
for constructing the general effective action.
Indeed,
it is standard to begin with the unitary gauge
in the construction of
the dilaton effective action
and the effective action for inflation~\cite{Cheung:2007st}.
Based on this observation,
we apply the following recipe of effective action construction
to spacetime symmetry breaking
in this paper:
\begin{enumerate}
\item gauge the (broken) global symmetry,
\item write down the unitary gauge effective action,
\item introduce NG fields by the St\"{u}ckelberg method
and decouple the gauge sector.
\end{enumerate}

\begin{table}[t]
\begin{center}
\scalebox{0.9}{
{\renewcommand\arraystretch{1.5}
\newcommand{\bhline}[1]{\noalign{\hrule height #1}}  
\begin{tabular}{|c||c|c|c|}
\hline
 relativistic symmetry & diffeomorphism & local Lorentz & isotropic Weyl \\
\bhline{1.5pt}
translation & \checkmark&&\\
\hline
isometry & \checkmark& \checkmark& \\
\hline
conformal &  \checkmark& \checkmark& \checkmark\\
\hline
\end{tabular}}}
\vspace{1mm}
\caption{
Embedding of relativistic spacetime symmetries.
In relativistic systems,
spacetime symmetries
can be classified into isometric
and conformal transformations,
by requiring spacetime isotropy.
They can then be embedded into diffeomorphism,
local Lorentz,
and isotropic Weyl transformations.}
\label{table:rela}
\vspace{7mm}
\scalebox{0.9}{
{\renewcommand\arraystretch{1.5}
\newcommand{\bhline}[1]{\noalign{\hrule height #1}}  
\begin{tabular}{|c||c|c|c|c|c|}
\hline
nonrelativistic symmetry&foliation preserving& local rotation & (an)isotropic Weyl & internal $U(1)$ \\
\bhline{1.5pt}
translation & \checkmark&&& \\
\hline
Galilean &  \checkmark& \checkmark&&\checkmark\\
\hline
Schr\"odinger &\checkmark& \checkmark& \checkmark&\checkmark\\
\hline
Galilean conformal&\checkmark& \checkmark& \checkmark&\\
\hline
\end{tabular}}}
\vspace{1mm}
\caption{
Embedding of nonrelativistic spacetime symmetries.
One difference from the relativistic case
is that nonrelativistic spacetime symmetries
should preserve the spatial slicing.
The corresponding coordinate transformations
are then foliation preserving diffeomorphisms.
Another difference is that nonrelativistic systems
admit central extensions of spacetime symmetry algebras.
Correspondingly,
we included the internal $U(1)$ gauge symmetry
in the above table.
See Appendix.~\ref{Sec:nonrela} for details.}
\label{table:nonrela}
\end{center}
\vspace{-3mm}
\end{table}

\medskip
Our starting point is that
any spacetime symmetry can be locally
generated by Poincar\'e transformations
and (an)isotropic rescalings.
Correspondingly,
we can embed any spacetime symmetry transformation
into diffeomorphisms (diffs), local Lorentz transformations, and (an)isotropic Weyl transformations (see Tables~\ref{table:rela} and \ref{table:nonrela}
for concrete embedding of global spacetime symmetry).
We then would like to gauge the original global symmetry
to local ones.
First, diffeomorphism invariance and local Lorentz invariance
can be realized by introducing the curved spacetime action
with the metric $g_{\mu\nu}$ and the vierbein $e_\mu^m$.\footnote{
We use Greek letters for the curved spacetime indices
and Latin letters for the (local) Minkowski indices.}
On the other hand,
there are two typical ways to realize isotropic Weyl invariance:
Weyl gauging and Ricci gauging.
In general,
we can gauge the Weyl symmetry by introducing a gauge field $W_\mu$
and defining the covariant derivatives appropriately (Weyl gauging),
whereas we can introduce a Weyl invariant curved space action
if the original system is conformal (Ricci gauging).
The anisotropic Weyl symmetry can be also gauged in a similar way.
These procedures for spacetime symmetry
allow us to gauge all the global symmetries
together with the internal
symmetries.

\medskip
Once gauging global symmetry,
we identify the broken local symmetry
from the condensation pattern
\begin{align}
\left\langle\Phi^A(x)\right\rangle=\bar{\Phi}^A(x)
\end{align}
and
construct the effective action
based on symmetry breaking structures.
Here and in what follows,
we use capital Latin letters for internal symmetry indices
and the spin indices
are implicit unless otherwise stated.
When the condensation is spacetime dependent,
diffeomorphism invariance is broken.
On the other hand,
local Lorentz invariance,
(an)isotropic Weyl invariance,
and internal gauge invariance
are broken
when the condensation has
the Lorentz charge (spin), scaling dimension, and internal charge,
respectively.
If the symmetry breaking pattern is given,
it is straightforward to take the unitary gauge
and construct the effective action
following the recipe.
We will first apply
our approach
to some concrete examples
to illustrate importance of 
the local viewpoint of
spacetime symmetry breaking.
We will then revisit the coset construction
from such a local perspective.
One important difference from the EFT for the internal symmetry breaking case in Lorentz invariant systems
is that the EFT constructed from the unitary gauge contains not only massless modes but also massive modes associated with spacetime symmetry breaking.
These massive modes transform nonlinearly  under the broken symmetries, i.e., they are NG fields.

\medskip
The organization of this paper is as follows.
In Sec.~\ref{sec:strategy},
we explain our basic strategy in more detail.
After reviewing the EFT for internal symmetry breaking,
we discuss how global spacetime symmetry can be gauged.
We then summarize how to construct the effective action based on the local symmetry breaking pattern.
In Sec.~\ref{sec:single},
we apply our approach to codimension one branes
to illustrate the difference between the global and the local picture of spacetime symmetry breaking.
In the global picture, one may characterize the branes by the spontaneous breaking of  translation and Lorentz invariance.
In the local picture, on the other hand, such a symmetry breaking pattern can be further classified by the spin of the condensation forming the branes.
We see that the
spectra of massive modes associated with symmetry breaking
depend on the spin of the condensation
and the mass of massive modes is not necessarily high.
If the masses are small compared with the typical energies of the system, the modes play a role
as low energy degrees of freedom.
Therefore, the local picture becomes important
in following
the dynamics of such massive modes appropriately.
In Sec.~\ref{section_multi}
we discuss a system with one-dimensional periodic modulation,
i.e., a system in which the condensation is periodic in one direction,
by applying the effective action constructed in Sec.~\ref{sec:single}.
We find that
the dispersion relations of NG modes for the broken diffeomorphism
are constrained by the minimum energy condition,
in contrast to the codimension one brane case.
In Sec.~\ref{Sec:mix}
such a discussion is extended to
the breaking of a mixture of spacetime and internal symmetries.
In Sec.~\ref{sec:CosetConstructionRevisited}
we revisit the coset construction
from the local symmetry picture.
We first show that the parameterization of NG fields in the coset construction is closely related to the local symmetry picture.
We then discuss the relation between the Maurer-Cartan one form and the connections for spacetime symmetries.
We finally classify the physical meanings of the inverse Higgs constraints based on the coordinate dimension of the broken symmetries.
In Sec.~\ref{EFT_gravity}
we make a brief comment on applications to gravitational systems.
The final section is devoted to a summary.
Details on the nonrelativistic case
are summarized in Appendix~\ref{Sec:nonrela}.
A derivation of the Nambu-Goto action is given in~Appendix~\ref{App:Nambu-Goto}.

\section{Basic strategy}
\setcounter{equation}{0}
\label{sec:strategy}

In this section,
we outline our basic strategy to construct the effective action
for symmetry breaking 
including spacetime ones.
In Sec.~\ref{subsec:internal},
we first review the relation between coset construction
and gauge symmetry breaking for internal symmetry,
and explain how the local picture can be used
to construct the effective action for global symmetry breaking.
To extend this discussion to spacetime symmetry,
in Secs.~\ref{subsec:local} and ~\ref{subsec:gaugespacetime},
we discuss how global spacetime symmetries can be embedded into local ones.
We then present our recipe for the effective action construction
in Sec.~\ref{subsection:recipe}.

\subsection{EFT for internal symmetry breaking}
\label{subsec:internal}
Let us first review
how the coset construction for internal symmetry breaking~\cite{Callan:1969sn,Coleman:1969sm}
can be reproduced
from the effective theory
for gauge symmetry breaking.
Suppose that a global symmetry group $G$
is spontaneously broken to a subgroup $H$
and
the coset space $G/H$ satisfying
\begin{align} \label{Lie_algebra_internal}
\mathfrak{g}=\mathfrak{h}\oplus\mathfrak{m}
\quad
{\rm with}
\quad
[\mathfrak{h},\mathfrak{m}]=\mathfrak{m}\,,
\end{align}
where $\mathfrak{g}$ and $\mathfrak{h}$ are
the Lie algebras of $G$ and $H$, respectively.
$\mathfrak{m}$ represent the broken generators.
In the coset construction, we introduce representatives of the coset space $G/H$ as $\Omega= e^{\pi}$ with $\pi\in \mathfrak{m}$,
whose left $G$ transformation is given by
\begin{equation}
\label{leftG}
\Omega(\pi)\to \Omega(\pi') = g \Omega(\pi) h^{-1}(\pi,g),
\end{equation}
where $g\in G$ and $h(\pi,g)\in H$. In general, $h^{-1}(\pi,g)$ depends on both $g$ and $\pi$.
The transformation from $\pi$ to $\pi'$ is nonlinear, so that it is called the nonlinear realization.
If $g$ is the element of the unbroken symmetry $H$, $g=h\in H$, $\pi$ linearly transforms: $\pi'(x)=h\pi(x)h^{-1}$.
Here, it is useful to introduce the Maurer-Cartan one form to construct the effective Lagrangian,  which is defined as
\begin{equation}
J_\mu \,dx^\mu\equiv \Omega^{-1}\partial_\mu \Omega \,dx^\mu\,.
\end{equation}
If we decompose $J_{\mu}$
into the broken component $J^{\mathfrak{m}}_{\mu}\in \mathfrak{m}$
and the unbroken component $J^{\mathfrak{h}}_{\mu}\in \mathfrak{h}$
as $J_{\mu}=J^{\mathfrak{m}}_{\mu}+J^{\mathfrak{h}}_{\mu}$,
each component transforms
as
\begin{align}
J^{\mathfrak{m}}_{\mu}\to h J^{\mathfrak{m}}_{\mu} h^{-1}\,,
\quad
J^{\mathfrak{h}}_{\mu}\to h J^{\mathfrak{h}}_{\mu}h^{-1} +h\partial_{\mu} h^{-1}\,,
\end{align}
under $G$ transformation.
Here note that
the broken component $J^{\mathfrak{m}}_\mu$
transforms covariantly.
In general, 
$\mathfrak{m}$ is reducible under $H$ transformation, and it can be decomposed into direct sums,
$\mathfrak{m}=\mathfrak{m}_1\oplus \mathfrak{m}_2\oplus\cdots\oplus\mathfrak{m}_N$.
At the leading order in the derivative expansion, the effective Lagrangian for Lorentz invariant systems is given by\footnote{
For simplicity, we do not consider matter fields in this paper.}
\begin{equation}
\mathcal{L}  = \sum_{a=1}^{N} F_a^2\mathrm{tr}\left[ J^{\mathfrak{m_{a}}}_{\mu} J^{\mathfrak{m_{a}}\,\mu}\right],
\end{equation}
where the trace is defined in a $G$-invariant way. 
$J^{\mathfrak{m_{a}}}_{\mu}$ and $F_a^2$ are  the component of  the Maurer-Cartan one form and the decay constant for each irreducible sector, respectively. By construction, this Lagrangian is invariant under $G$ transformation. 

\medskip
We next move on to the effective action construction
for gauge symmetry breaking.
In this case,
it is convenient to take the unitary gauge,
where the NG fields are eaten by the gauge field $A_\mu$
and do not fluctuate.
The general effective action
can then be constructed
only from the massive gauge field $A_\mu$ in an $H$ gauge invariant way.
In relativistic systems,
the effective Lagrangian takes the form
\begin{align}
\label{unitary_action_internal}
\mathcal{L}=\text{tr}\left[\frac{1}{2g^2} F^{\mu\nu} F_{\mu\nu}+v_a^2 A^{\mathfrak{m_{a}}}_{\mu}\,A^{\mathfrak{m_{a}}\,\mu}+\ldots\right]\,,
\end{align}
where $g$ and $v_a$ are the gauge coupling and the order parameters
for symmetry breaking, respectively,
and $A^{\mathfrak{m}}_{\mu}\in \mathfrak{m}$
is the gauge field in the broken symmetry sector.
Note that since we are considering the unitary gauge, Eq.~\eqref{unitary_action_internal} is not invariant under $G$ gauge transformation.
Because the gauge boson mass is given by $m\sim gv_a$,
the global symmetry limit $g\to 0$ for fixed $v_a$ corresponds
to the singular massless limit,
so that the unitary gauge is not appropriate to discuss
the global symmetry limit.
In order to take the global symmetry limit,
it is convenient to introduce the NG fields $\pi(x)\in \mathfrak{m}$
by performing a field-dependent gauge transformation
(the St\"{u}ckelberg method):
\begin{align}
\label{AOmega}
A_\mu\to \Omega^{-1}A_\mu \Omega +\Omega^{-1}\partial_\mu \Omega
\quad
{\rm with}
\quad
\Omega=e^{\pi(x)}\,.
\end{align}
The $G$ gauge invariance can be recovered
by assigning a nonlinear transformation rule on the NG fields $\pi(x)$
and
we can take the global symmetry limit $g\to0$ smoothly in this picture.
Since the gauge sector decouples from the NG fields in the global symmetry limit,
the effective action for the NG fields can be obtained
by the replacements:
\begin{align}
  v_a^2&\to F_a^2\,, \\
A_\mu&\to J_\mu=\Omega^{-1}\partial_\mu\Omega\,.
\end{align}
The latter is nothing but the Maurar-Cartan one form.
As this discussion suggests,
the unitary gauge is useful
to construct the general ingredients needed to obtain the effective action
for global symmetry breaking.
Note that Wess-Zumino terms in the coset construction
are reproduced by Chern-Simons terms in the unitary gauge action.

\subsection{Local properties of spacetime symmetries}
\label{subsec:local}

In the previous subsection,
we saw that the unitary gauge action for gauge symmetry breaking
can be used to construct the general effective action
for global symmetry breaking.
We now would like to extend such a discussion
to spacetime symmetry breaking.
For this purpose,
let us recall the
local properties of
(infinitesimal) spacetime symmetry transformations in this subsection.\footnote{
Though we concentrate on infinitesimal transformations
for simplicity,
extension to the finite case is straightforward.}
Any spacetime symmetry transformation
has an associated coordinate transformation
\begin{align}
\label{general_coord}
x^\mu\to x'^\mu=x^\mu-\epsilon^\mu(x)
\end{align}
and its local properties around a point $x^\mu=x_*^\mu$
can be read off by expanding the parameter $\epsilon^\mu(x)$ covariantly as
\begin{align}
\label{covariantly_expand}
\epsilon^\mu(x)=\epsilon^\mu(x_*)+(x^\nu-x_*^\nu)\nabla_\nu\epsilon^\mu(x_*)+\mathcal{O}\Big((x-x_*)^2\Big)\,.
\end{align}
The first term is the zeroth order in $x-x_*$
and
describes translations of the coordinate system.
On the other hand,
deformations of the coordinate system are encoded in the second term
(the linear order in $x-x_*$),
which can be decomposed~as
\begin{align}
\label{general_decomposition}
\nabla_\mu\epsilon^\nu
=\delta_\mu^\nu\,\lambda+s_\mu{}^\nu+\omega_\mu{}^\nu
\quad
{\rm with}
\quad
s_{\mu}{}^\mu=0\,,
\quad
s_{\mu\nu}=s_{\nu\mu}\,,
\quad
\omega_{\mu\nu}=-\omega_{\nu\mu}\,.
\end{align}
The trace part $\lambda$ and the symmetric traceless part $s_{\mu\nu}$
are local isotropic rescalings (dilatations)
and local anisotropic rescalings,
respectively.
The antisymmetric part $\omega_{\mu\nu}$ corresponds to
local Lorentz transformations.
Any spacetime symmetry can therefore be
locally decomposed into Poincar\'e transformations and isotropic/anisotropic rescalings.
Correspondingly, the symmetry transformations of local fields
are specified by their Lorentz charges and isotropic/anisotropic scaling dimensions.

\medskip
As is suggested by Eqs.~\eqref{covariantly_expand} and \eqref{general_decomposition},
we can embed global spacetime symmetry transformations
into
diffeomorphisms, local Lorentz transformations, and local isotropic/anisotropic Weyl transformations.
For simplicity,
let us consider the case of relativistic systems in this section
(see Appendix~\ref{Sec:nonrela} for extension to the nonrelativistic case).
In relativistic systems,
any spacetime symmetry transformation
can be locally decomposed into the Poincar\'e part and the dilatation part
because anisotropic rescalings are incompatible with the Lorentz symmetry:
\begin{align}
\nabla_\mu\epsilon^\nu
=\delta_\mu^\nu\,\lambda+\omega_\mu{}^\nu
\quad
{\rm with}
\quad
\omega_{\mu\nu}=-\omega_{\nu\mu}\,.
\end{align}
Note that we have conformal transformations for general functions $\lambda(x)$
and isometric transformations for $\lambda=0$
because the metric field transforms as
\begin{align}
\delta g_{\mu\nu}=\nabla_\mu\epsilon_\nu+\nabla_\nu\epsilon_\mu
=2g_{\mu\nu}\lambda\,.
\end{align}
The transformation rules of local fields
are then determined by their spin and scaling dimension.
When a field $\Phi(x)$ follows a representation $\Sigma_{mn}$
of the Lorentz algebra
and has a scaling dimension $\Delta_\Phi$,
its symmetry transformation is given by
\begin{align}
\label{decomposition1}
\Phi(x)\to\Phi'(x)&=\Phi(x)+\Delta_\Phi\lambda(x)\Phi(x)+\frac{1}{2}\omega^{mn}(x)\Sigma_{mn}\Phi(x)+\epsilon^\mu(x)\nabla_\mu\Phi(x)\,.
\end{align}
Here, the curved spacetime indices
(Greek letters)
and the local Lorentz indices
(Latin letters)
are converted
by the vierbein $e_\mu^m$ as
$\omega^{mn}=e^m_\mu e_\nu^n\omega^{\mu\nu}$.
The covariant derivative $\nabla_\mu\Phi$ is defined
in terms of the spin connection $S_\mu^{mn}$ as
\begin{align}
\label{covariant_derivative}
\nabla_\mu\Phi=\partial_\mu\Phi+\frac{1}{2}S_\mu^{mn}\Sigma_{mn}\Phi
\quad
\text{with}
\quad
S_\mu^{mn}=e_\nu^m\partial_\mu e^{\nu n}
+e_\lambda^m\,\Gamma^\lambda_{\mu\nu}\,e^{\nu n}\,,
\end{align}
with the Christoffel symbols $\Gamma^\lambda_{\mu\nu}$ defined by
\begin{equation}
\begin{split}
\Gamma^\lambda_{\mu\nu} \equiv \frac{g^{\lambda\rho}}{2}(\partial_\mu g_{\rho\nu}+\partial_\nu g_{\mu\rho}-\partial_\rho g_{\mu\nu}) \,.
\end{split}
\end{equation}
To identify the transformation \eqref{decomposition1}
as local symmetries,
it is convenient to rewrite it in the form,
\begin{align}
\label{decomposition2}
\Phi'(x)&=\Phi(x)+\Delta_\Phi\lambda(x)\Phi(x)
+\frac{1}{2}\Big(\omega^{mn}(x)+\epsilon^\mu(x) S_{\mu}^{mn}(x)\Big)\Sigma_{mn}\Phi(x)+\epsilon^\mu(x)\partial_\mu\Phi(x)\,.
\end{align}
We then notice that
the latter three terms
can be thought of as local Weyl transformations, local Lorentz transformations, and diffeomorphisms, respectively.
Since the transformation rule of $\Phi$, $g_{\mu\nu}$, and $e_\mu^m$
under each local transformation is given by
\begin{align}
\label{local_Weyl}
\text{local Weyl}:&\quad\delta \Phi=\Delta_\Phi\tilde{\lambda}\Phi\,,
\quad
\delta g_{\mu\nu}=-2\tilde{\lambda} g_{\mu\nu}\,,
\quad
\delta e^m_\mu=-\tilde{\lambda} e^m_\mu\,,
\\*
\label{local_Lorentz}
\text{local Lorentz}:&\quad\delta \Phi=\frac{1}{2}\tilde{\omega}^{mn}\Sigma_{mn}\Phi\,,
\quad
\delta g_{\mu\nu}=0\,,
\quad
\delta e^m_\mu=\tilde{\omega}^m{}_n e^n_\mu\,,
\\*
\label{diffs}
\text{diffeomorphism}:&\quad\delta \Phi=\tilde{\epsilon}^\mu\partial_\mu\Phi\,,
\quad
\delta g_{\mu\nu}=\nabla_\mu\tilde{\epsilon}_\nu+\nabla_\nu\tilde{\epsilon}_\mu\,,
\quad
\delta e^m_\mu=\nabla_\mu\tilde{\epsilon}^m-\tilde{\epsilon}^\nu S_{\nu }^m{}_n\, e^n_\mu\,,
\end{align}
we can reproduce the transformation \eqref{decomposition2}
by the parameter choice
\begin{align}
\label{embed_relativistic}
\tilde{\lambda}=\lambda\,,
\quad
\tilde{\omega}^{mn}=\omega^{mn}+\epsilon^\mu S_\mu^{mn}\,,
\quad
\tilde{\epsilon}^\mu=\epsilon^\mu\,.
\end{align}
Note that
the metric $g_{\mu\nu}$ and the vierbein $e_\mu^m$
are invariant under the original global transformation \eqref{embed_relativistic},  although they are not invariant under general local ones.
Any global spacetime symmetry in relativistic systems can
therefore be embedded into local Weyl transformations,
local Lorentz transformations, and diffeomorphisms.

\subsection{Gauging spacetime symmetry}
\label{subsec:gaugespacetime}
In the previous subsection
we discussed that
any spacetime symmetry transformation in relativistic systems is locally generated
by Poincar\'e and Weyl transformations.
Isometric transformations can be embedded in
diffeomorphisms and local Lorentz transformations,
whereas conformal transformations require local Weyl transformations as well.
Since NG fields correspond to local transformations
of the order parameters for broken symmetries,
we would like to construct the effective action
from this local symmetry point of view.
For this purpose,
let us summarize how we can gauge global spacetime symmetry
to those local symmetries.

\medskip
When the system is isometry invariant
before symmetry breaking,
we gauge the Poincar\'e symmetry
by introducing
the curved spacetime action
with the metric $g_{\mu\nu}(x)$
and the vierbein $e_\mu^m(x)$.
For example,
an action in a non-gravitational system on Minkowski space,
\begin{align}
\label{original_CFT_action}
S[\Phi]=\int d^4x\,\mathcal{L}[\Phi,\partial_m\Phi],
\end{align}
can be reformulated as
\begin{align}
\label{diffeomorphic_action}
S[\Phi]\to S[\Phi,g_{\mu\nu},e^m_\mu]
=\int d^4x\sqrt{-g}\,\mathcal{L}[\Phi,e_m^\mu\nabla_\mu\Phi]\,,
\end{align}
where the covariant derivative $\nabla_\mu$ is given by Eq.~\eqref{covariant_derivative}.
From the viewpoint of the curved space action~\eqref{diffeomorphic_action},
the original non-gravitational system can be reproduced
by taking the metric $g_{\mu\nu}$ and the vierbein $e^m_\mu$ as the Minkowski ones
with the gauge choice,
\begin{align}
\label{flat_gauge}
g_{\mu\nu}=\eta_{\mu\nu}\,,
\quad
e^m_\mu=\delta_\mu^m\,.
\end{align}
The original global Poincar\'e symmetry
can be also understood as the residual symmetries
under the gauge conditions \eqref{flat_gauge}.
The same story holds for non-gravitational systems
on curved spacetimes.

\medskip
On the other hand,
there are two typical ways to gauge the Weyl symmetry:
Weyl gauging and Ricci gauging (see, e.g., Ref.~\cite{Iorio:1996ad}).
When the system is conformal,
we can introduce a local Weyl invariant curved spacetime action,
essentially because the local Weyl invariance
is equivalent to the traceless condition of the energy-momentum tensor.
Such a procedure is called the Ricci gauging and
we need not introduce additional fields in this case.
When the Ricci gauging is not applicable,
we need to introduce a gauge field $W_\mu$ for Weyl symmetry
and the covariant derivative $D_\mu$ defined by
\begin{align}
\label{Weyl_covariant_der}
D_\mu\Phi
=\nabla_\mu\Phi+
\left(\Delta_\Phi \,\delta_\mu^\nu-\Sigma_\mu{}^\nu\right)W_\nu\Phi\,,
\end{align}
where $\Sigma_\mu{}^\nu=e_\mu^m\, \Sigma_m{}^n\,e_n^\nu$
and the local Weyl transformation rule is given by\footnote{
Note that the gauge field $W^\mu$ with an upper spacetime index transforms as
\begin{align}
W^\mu=g^{\mu\nu}W_\nu
\to W'^\mu=e^{-2\lambda}g^{\mu\nu} (W_\nu-\partial_\mu\lambda)
=e^{-2\lambda}(W^\mu-\partial^\mu\lambda)\,.
\end{align}}
\begin{align}
\Phi\to\Phi'=e^{\Delta_\Phi\lambda} \Phi\,,
\quad
g_{\mu\nu}\to g_{\mu\nu}'=e^{2\lambda}g_{\mu\nu}\,,
\quad
e_\mu^m\to e_\mu'^m=e^{\lambda}\,e_\mu^m\,,
\quad
W_\mu\to W_\mu'=W_\mu-\partial_\mu\lambda\,.
\end{align}
If the curved spacetime action is global Weyl invariant,
a local Weyl invariant action can be obtained
by replacing the covariant derivative $\nabla_\mu$
with the Weyl covariant derivative $D_\mu$.
For example,
Eq.~\eqref{diffeomorphic_action} is reformulated as
\begin{align}
\label{Weyl_gauge_action}
S[\Phi,g_{\mu\nu},e^m_\mu,W_\mu]
=\int d^4x\sqrt{-g}\,\mathcal{L}[\Phi,e_m^\mu\, D_\mu\Phi]\,.
\end{align}
Note that the original action \eqref{original_CFT_action}
can be reproduced by imposing the gauge condition
\begin{align}
\label{W_gauge}
g_{\mu\nu}=\eta_{\mu\nu}\,,
\quad
e^m_\mu=\delta^m_\mu\,,
\quad
W_\mu=0\,,
\end{align}
and symmetries of the action
are reduced to the original global ones.
Also,
while the first two conditions in Eq.~\eqref{W_gauge}
are always invariant under the original global symmetries,
the condition $W_\mu=0$
is not necessarily invariant
when the original system is conformal.
Indeed,
it is not invariant under the special conformal transformation.
Correspondingly,
the Weyl gauge field $W_\mu$ appears in a particular combination
in the action.
For example,
the action of a massless free scalar $\phi$
can be gauged as
\begin{align}
-\frac{1}{2}\int d^4x\sqrt{-g}g^{\mu\nu}(\partial_\mu+W_\mu)\phi
\,(\partial_\nu+W_\nu)\phi
&=
-\frac{1}{2}\int d^4x\sqrt{-g}\left[(\partial_\mu\phi)^2
-(\nabla_\mu W^\mu -W^2)\phi^2
\right]\,,
\end{align}
where $W_\mu$ appears in a special conformal invariant combination
$\nabla_\mu W^\mu -W^2$.

\subsection{EFT recipe}
\label{subsection:recipe}
\begin{table}[t]
\begin{center}
\scalebox{0.9}{
{\renewcommand\arraystretch{1.5}
\newcommand{\bhline}[1]{\noalign{\hrule height #1}}  
\begin{tabular}{|c||c||c||c|}
\hline
diffeomorphism & local Lorentz & local Weyl & internal gauge\\
\bhline{1.5pt}
\, spacetime dependence \,& spin & scaling dimension &\, internal charge \,\\
\hline
metric $g_{\mu\nu}$ & vierbein $e_\mu^m$ &\, Weyl gauge field $W_\mu$ \,& gauge field $A_\mu$\\
\hline
\end{tabular}}}
\end{center}
\caption{Broken symmetries, condensation patterns, and gauge fields.}
\label{table:breaking}
\end{table}

As we have discussed,
all the global symmetries
in relativistic systems
can be embedded into diffeomorphisms,
local Lorentz symmetries,
local Weyl symmetry,
and internal gauge symmetries.
We can also gauge the global symmetry
by the use of the procedures in the previous subsection
and the standard internal gauging.
Similar discussions hold for nonrelativistic systems
accompanied by local anisotropic Weyl symmetries
and internal symmetry associated with the possible central extension,
as we illustrate in Appendix~\ref{Sec:nonrela}.
We now extend the discussion in~Sec.~\ref{subsec:internal}
for internal symmetry
to spacetime symmetry.
First,
the symmetry breaking patterns
are classified by the condensation patterns:
\begin{align}
\left\langle\Phi^A(x)\right\rangle=\bar{\Phi}^A(x)\,.
\end{align}
When the condensation is spacetime dependent,
diffeomorphism invariance is broken.
On the other hand,
local Lorentz invariance,
local isotropic/anisotropic Weyl invariance,
and internal gauge invariance
are broken
when the condensation has
a Lorentz charge (spin), scaling dimension, and internal charge,
respectively.
Once we identify the symmetry breaking pattern,
we can construct the effective action
based on the following recipe
just as in
the case of internal symmetry breaking:
\begin{enumerate}
\item gauge the (broken) global symmetry,
\item write down the unitary gauge effective action,
\item introduce NG fields by the St\"{u}ckelberg method
and decouple the gauge sector.
\end{enumerate}
The first step can be performed
by introducing gauge fields
based on the procedure in Sec.~\ref{subsec:gaugespacetime}
(see also table~\ref{table:breaking}).
We then take the unitary gauge,
where the NG fields do not fluctuate.
Using the dynamical degrees of freedom
in the unitary gauge,
we construct the general unitary gauge effective action
invariant under the residual symmetries.
Finally,
we perform the St\"{u}ckelberg method
to introduce the NG fields
and restore the full gauge symmetry.
By decoupling the gauge sector,
we obtain the effective action for the NG fields.
In the following sections,
we apply this recipe
to concrete examples for spacetime symmetry breaking.

\medskip
We emphasize that the condensation pattern rather than the breaking pattern of global symmetries plays an important role to identify the NG fields, unlike the case for internal symmetry breaking in Lorentz invariant systems. 
The breaking pattern of global symmetries itself cannot distinguish the breaking of diffeomorphism, local Lorentz, and Weyl symmetries.
As will be seen in the next section, this difference becomes important when we discuss the massive modes originating from the symmetry breaking,
although the existence of massless modes does not depend on the condensation pattern.

\section{Codimension one brane}
\setcounter{equation}{0}
\label{sec:single}

\begin{figure}[t]
\begin{center}
\includegraphics[width=100mm, bb=0 0 580 262]{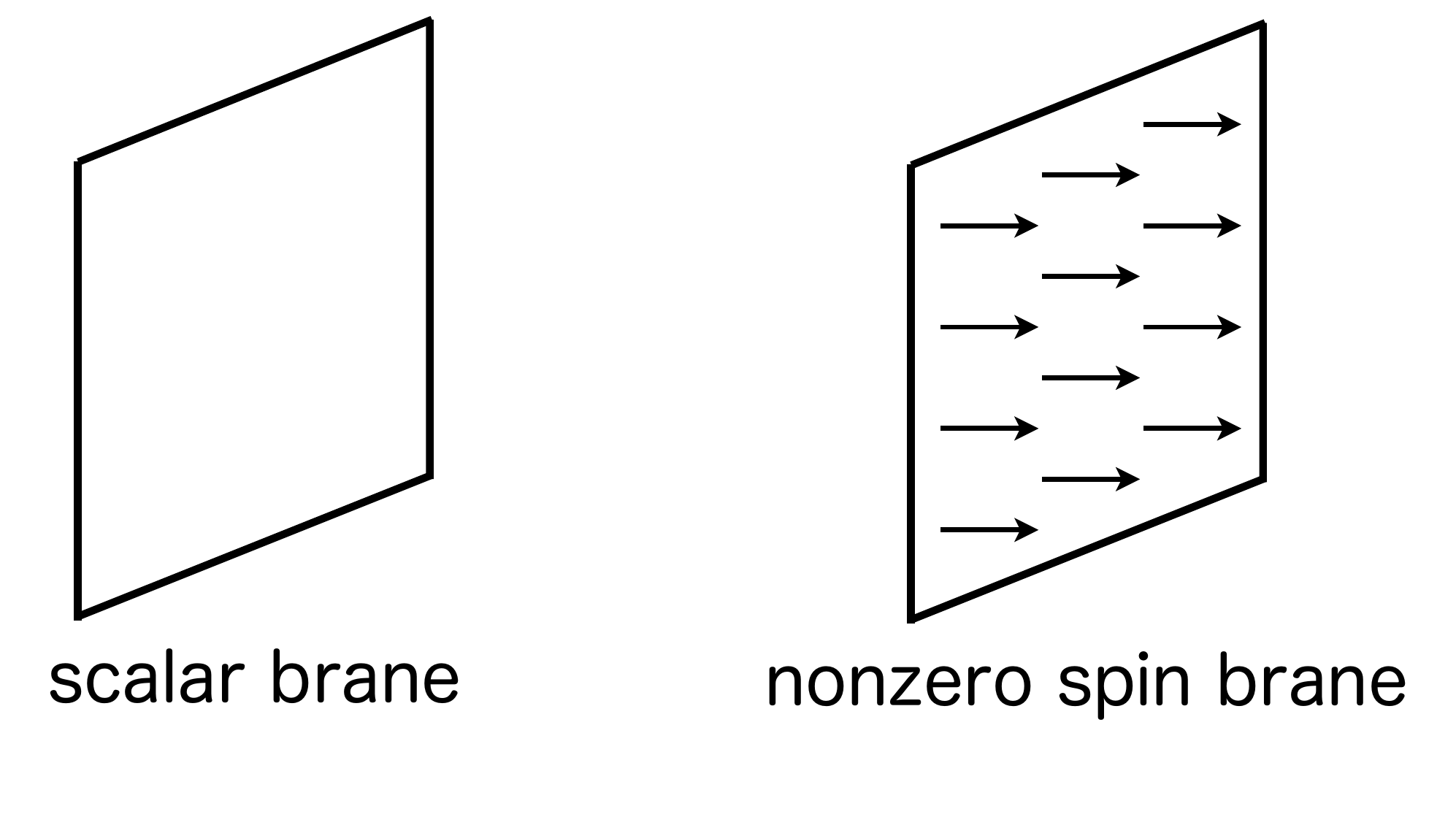}
\end{center}
\vspace{-8mm}
\caption{Scalar vs nonzero spin branes.
While the diffeomorphism symmetry is broken for both types of branes,
the local Lorentz symmetry is broken only
for the nonzero spin case.}
\label{fig:scalar_vs_nonzero}
\end{figure}

In this section,
we apply our approach to
codimension one branes on the Minkowski space
to illustrate the difference between
the global and the local picture of spacetime symmetry breaking.
In the global picture,
one may characterize the branes
by the spontaneous breaking of translation and Lorentz invariance.
In the local picture,
on the other hand,
such a symmetry breaking pattern can be further classified
by the spin of the condensation forming the brane
(see also Fig.~\ref{fig:scalar_vs_nonzero}):
\begin{enumerate}
\item Scalar brane

When a scalar field forms a codimension one brane,
the only broken symmetry is the diffeomorphism invariance
in the $z$-direction orthogonal to the brane.
In particular,
the local Lorentz symmetry is not broken
although the global one is.

\item Nonzero spin branes

When a nonzero spin field forms a codimension one brane
and
the condensation is aligned to the $z$-direction,
the local Lorentz invariance associated with the $z$-direction
is broken as well as the $z$-diffeomorphism invariance.

\end{enumerate}
Since those two cases are classified into the same category in the global picture,
the local picture is necessary to distinguish them.
In the rest of this section,
we discuss in which situation the difference becomes important, if we take into account the massive modes associated with symmetry breaking.

\medskip
In Sec.~\ref{subsec:real_scalar}
we first perform the tree-level analysis of NG fields around
scalar brane backgrounds,
to illustrate our strategy in the previous section concretely.
In Secs.~\ref{subsec:scalar_domain} and~\ref{subsec:spectra_scalar},
we construct the general effective action for the diffeomorphism symmetry breaking
and apply it to single scalar brane backgrounds.
In Secs.~\ref{subsec:inclusion}
and~\ref{subsec:qualitative},
we include local Lorentz symmetry breaking
in the effective action construction
and apply it to single nonzero spin brane backgrounds.
For single brane backgrounds,
it tuns out that
the dynamics in the low-energy limit
results in the same 
action regardless of the spin of the field that condenses.
However,
we find that
the degeneracy is resolved beyond the low-energy limit
and the resolving scale is not necessarily high.
We see that the effective action based on the local picture
can be used to investigate such an intermediate scale.

\medskip
Though discussion in this section
is only for single brane backgrounds,
the effective action constructed in Secs.~\ref{subsec:scalar_domain} and~\ref{subsec:inclusion}
is applicable to more general setups.
In Sec.~\ref{section_multi},
we discuss periodic modulation and
clarify the difference from the single brane case.

\subsection{Real scalar field model for scalar brane}
\label{subsec:real_scalar}

In order to illustrate our strategy,
let us begin with a real scalar field model,
\begin{align}
\label{real_scalar_model}
S=\int d^4x\left[
-\frac{1}{2}\partial_m\phi\partial^m\phi-V(\phi)\right]
\quad
{\rm with}
\quad
V=\frac{g^2}{2}(\phi^2-v^2)^2\,,
\end{align}
and perform the tree-level analysis of NG fields
around domain-wall configurations.
Here $g$ and $v$ are constant parameters
and
the potential $V(\phi)$ has two minima at $\phi=\pm v$.
The equation of motion
\begin{align}
\Box\phi-V'(\phi)=0
\end{align}
has the following domain-wall solution with the boundary conditions $\phi(z=\pm\infty)=\pm v$:
\begin{align}
\label{domain_config}
\phi(x)=\bar{\phi}(z)=v\tanh \beta z\,,
\end{align}
where $\beta=gv$ characterizes the thickness of the brane.
Note that
there exists a one-parameter family of domain-wall solutions 
$\phi(x)=\bar{\phi}(z-z_0)$
parameterized by the
brane
position~$z_0$
because of the translation invariance.
The domain-wall configuration Eq.~\eqref{domain_config}
breaks the translation invariance
and the corresponding NG field $\pi(x)$
can be obtained by promoting $z_0$
to a field as\footnote{
Our parameterization of the NG field $\pi(x)$
corresponds to 
the transformation parameter $\epsilon$ in Eq.~\eqref{decomposition1}.}
\begin{align}
\phi(x)=\bar{\phi}\left(z+\pi(x)\right)\,.
\end{align}
The action for the NG field is then given by
\begin{align}
\nonumber
S&=\int d^4x\left[
-\frac{1}{2}\partial_m\bar{\phi}(z+\pi)\partial^m\bar{\phi}(z+\pi)-V\left(\bar{\phi}(z+\pi)\right)\right]
\\
\label{scalar_domain_pre}
&=\int d^4x\left[
-\frac{\bar{\phi}'(z+\pi)^2}{2}\partial_m(z+\pi)\partial^m(z+\pi)-V\left(\bar{\phi}(z+\pi)\right)\right]\,.
\end{align}
Using the integrated version of the equation of motion,\footnote{
In general,
the integrated equation of motion takes the form
$\frac{1}{2}\bar{\phi}'^2-V(\bar{\phi})=\text{constant}$.
However,
the constant term vanishes
for the potential~\eqref{real_scalar_model} and the solution~\eqref{domain_config}
to obtain the second equation in Eq.~\eqref{integrated_eom}.}
\begin{align}
\label{integrated_eom}
\bar{\phi}''-V'(\bar{\phi})=0
\quad
\leftrightarrow
\quad
\frac{1}{2}\bar{\phi}'^2-V(\bar{\phi})=0\,,
\end{align}
we can further reduce the action~\eqref{scalar_domain_pre}
to the form
\begin{align}
\label{toy_NG}
S=\int d^4x\left[
-\frac{\bar{\phi}'(z+\pi)^2}{2}\Big(\partial_m(z+\pi)\partial^m(z+\pi)+1\Big)\right]
=
-\frac{1}{2}\int d^4x\,
\bar{\phi}'(z+\pi)^2\partial_m\pi\partial^m\pi\,,
\end{align}
where we dropped total derivative terms at the second equality.

\medskip
Let us then reproduce the action~\eqref{toy_NG}
along the line of our strategy.
Following the EFT recipe in the previous section,
we first gauge the translation symmetry
to the diffeomorphism symmetry
by introducing the curved coordinate action
\begin{align}
S=\int d^4x
\sqrt{-g}\left[
-\frac{1}{2}\partial_\mu\phi\partial^\mu\phi-V(\phi)\right]\,.
\end{align}
We next consider fluctuations around the domain-wall background~\eqref{domain_config}.
Since $z$-coordinate transformations of $\bar{\phi}(z)$ generate fluctuations of $\phi(x)$,
we can take the unitary gauge $\phi(x)=\bar{\phi}(z)$
at least as long as fluctuations are small.
In other words,
we can choose a coordinate frame
such that the constant-$\phi$ slices
coincide with the constant-$z$ slices.
In this coordinate frame,
the action is given by
\begin{align}
\label{toy_unitary}
S=\int d^4x
\sqrt{-g}\left[
-\frac{1}{2}g^{zz}\bar{\phi}'(z)^2-V(\bar{\phi})\right]
=-\frac{1}{2}
\int d^4x
\sqrt{-g}\,
\bar{\phi}'(z)^2\left(1+g^{zz}\right)\,,
\end{align}
where we used Eq.~\eqref{integrated_eom} in the second equality.
Note that
the action~\eqref{toy_unitary} enjoys
only the $(2+1)$-dimensional diffeomorphism symmetry
along the $t,x,y$-directions
\begin{align}
\label{2+1_diff}
x^\mu\to x'^\mu=x^\mu-\epsilon^\mu(x)
\quad
{\rm with}
\quad
\epsilon^z=0
\end{align}
and the NG field is eaten by the metric $g_{\mu\nu}$ in this gauge.
We then restore the $z$-diffeomorphism invariance by the field-dependent coordinate transformation
(the St\"uckelberg method)
\begin{align}
\label{stuckelberg_diff}
z\to \tilde{z}
\quad
{\rm with}
\quad
\tilde{z}+\tilde{\pi}(\tilde{x})=z\,.
\end{align}
After the transformation,
the action~\eqref{toy_unitary} takes the form
\begin{align}
\label{toy_stuckelberg}
S=-\frac{1}{2}
\int d^4x
\sqrt{-g}\,
\bar{\phi}'(z+\pi)^2\Big(1+g^{\mu\nu}\partial_\mu(z+\pi)\partial_\nu(z+\pi)\Big)\,,
\end{align}
where we dropped the tilde for simplicity.
The action (\ref{toy_stuckelberg})
is now invariant under the full diffeomorphism symmetry
by assigning the following nonlinear transformation rule
on the NG field $\pi$:
\begin{align}
\label{nonlinear_rule}
\pi(x)\to\pi'(x')=\pi(x)+\epsilon^z(x)
\quad
{\rm with}
\quad
x'^\mu=x^\mu-\epsilon^\mu(x)\,.
\end{align}
Finally, we remove gauge degrees of freedom
by taking the Minkowski coordinate.
Since we are working on the Minkowski space,
the full diffeomorphism invariance,
nonlinearly realized by the NG fields,
allow us to set the metric field as $g_{\mu\nu}=\eta_{\mu\nu}$.
The action~\eqref{toy_stuckelberg} is then reduced to Eq.~\eqref{toy_NG}.

\medskip
In this subsection
we illustrated our approach
by performing the tree-level analysis of NG fields
around domain-wall backgrounds
in the model~\eqref{real_scalar_model}.
As we have seen,
the introduction of the curved coordinate action~\eqref{toy_unitary}
allows us to impose the unitary gauge condition $\phi(x)=\bar{\phi}(z)$,
which breaks the $z$-diffeomorphism invariance.
The scalar $\phi(x)$ is then eaten by the metric field $g_{\mu\nu}$.
By performing the St\"uckelberg method
and removing the gauge degrees of freedom,
we obtained the action for NG fields.
More generally,
the action~\eqref{real_scalar_model} can be
modified with higher derivative terms
due to quantum corrections for example.
In the next subsection
we construct the general effective action for NG fields
by introducing the general unitary gauge action
consistent with the symmetry.

\subsection{Effective action for $z$-diffeomorphism symmetry breaking}
\label{subsec:scalar_domain}

We then construct the general effective action
for the $z$-diffeomorphism symmetry breaking,
by introducing the general unitary gauge action
consistent with the symmetry.
Just as in the previous real scalar model,
let us introduce the metric field and work in the general coordinate system.
We can then impose the unitary gauge condition,
which prohibits fluctuations of the NG field
and breaks the $z$-diffeomorphism symmetry.
In such a unitary gauge,
the dynamical degrees of freedom
are the metric field $g_{\mu\nu}$ only
(if there are no additional matter degrees of freedom)
and there remains the $(2+1)$-dimensional diffeomorphism symmetry.
Schematically,
we write this unitary gauge setup as
\begin{align}
g_{\mu\nu}(x)
\quad
+\quad
\text{(2+1)-dim diffs.}
\end{align}
The general effective action
is then constructed from the metric field
in a $(2+1)$-dimensional diffeomorphism invariant way.
This setup is essentially the same as the one in single-field inflation~\cite{Cheung:2007st}.
Following the results there,
ingredients of the unitary gauge effective action are given by
\begin{align}
\label{ing_domain}
\text{scalar functions of $z$},\,\,
g_{\mu\nu}\,,
\,\,
R_{\mu\nu\rho\sigma}\,,
\,\,
\text{and their covariant derivatives.}
\end{align}
The lowest few terms
of the
expansion
in fluctuations
around the Minkowski metric,
$g_{\mu\nu}-\eta_{\mu\nu}$,
and derivatives
are given by
\begin{align}
\label{scalar_simple}
S=-\frac{1}{2}\int d^4x\sqrt{-g}
\left[\alpha_1(z)+\alpha_2(z)g^{zz}+\alpha_3(z)\left(g^{zz}-1\right)^2\right]\,.
\end{align}
Here, $\alpha_i(z)$'s are scalar functions of $z$, which depend on the
details of the microscopic theory.
Note that
$g^{zz}$ arises from $\partial_\mu\phi\partial^\mu\phi=\bar{\phi}'^2g^{zz}$
in the previous real scalar model.
One may then identify 
$\alpha_3$ with higher derivative interactions in the real scalar model.

\medskip
We next introduce the NG field $\pi$ for the $z$-diffeomorphism
by the St\"{u}ckelberg method.
Just as we did in the previous subsection,
we perform a field-dependent coordinate transformation~\eqref{stuckelberg_diff}.
Practically,
this transformation can be realized by replacing
a function $f(z)$ with $f(z+\pi)$~\cite{Cheung:2007st},
where we dropped the tilde for simplicity.
Correspondingly,
the unitary gauge action \eqref{scalar_simple} is transformed~as
\begin{align}
\label{scalar_simple_stuckelberg}
S=-\frac{1}{2}\int d^4x\sqrt{-g}
\left[\alpha_1(z+\pi)+\alpha_2(z+\pi)\left(g^{zz}+2\partial^z\pi+\partial_\mu\pi\partial^\mu\pi\right)+\alpha_3(z+\pi)\left(2\partial^z\pi+\partial_\mu\pi\partial^\mu\pi\right)^2\right]\,,
\end{align}
where we used
\begin{align}
g^{zz}=g^{\mu\nu}\delta_\mu^z\delta_\nu^z\to g^{\mu\nu}\partial_\mu(z+\pi)\partial_\nu(z+\pi)=g^{zz}+2\partial^z\pi+\partial_\mu\pi\partial^\mu\pi\,.
\end{align}
The full diffeomorphism symmetry is now restored
accompanied
by the nonlinearly transformation rule~\eqref{nonlinear_rule}.
We then would like to remove the
gauge degrees of freedom
and construct the effective action for the NG field $\pi$ only.
Since we are working on the Minkowski space,
the full diffeomorphism invariance,
nonlinearly realized by the NG fields,
allow us to set the metric field as $g_{\mu\nu}=\eta_{\mu\nu}$.
In this Minkowski coordinate system,
the ingredients \eqref{ing_domain}
are given by
\begin{align}
\text{scalar functions of $z+\pi$},\,\,
g_{\mu\nu}=\eta_{\mu\nu}\,,
\,\,
R_{\mu\nu\rho\sigma}=0\,,\,\,
\text{and their derivatives,}
\end{align}
and the effective action \eqref{scalar_simple_stuckelberg}
can be expressed as
\begin{align}
\label{scalar_simple_pi}
S&=-\frac{1}{2}\int d^4x
\left[\alpha_1(z+\pi)+\alpha_2(z+\pi)\Big(1+2\partial_z\pi+\partial_m\pi\partial^m\pi\Big)
+\alpha_3(z+\pi)\Big(2\partial_z\pi+\partial_m\pi\partial^m\pi\Big)^2\right]\,,
\end{align}
where Latin indices
indicate that we use the Minkowski coordinate.

\medskip
So far,
we have not taken into account the background equation of motion. 
Indeed
Eq.~\eqref{scalar_simple_pi}
contains linear order terms in $\pi$:
\begin{align}
S&=-\frac{1}{2}\int d^4x
\left[
\alpha_1(z)+\alpha_2(z)
+\left(\alpha_1'(z)+\alpha_2'(z)\right)\pi+2\alpha_2(z)\partial_z\pi
+\mathcal{O}(\pi^2)\right]\,.
\end{align}
To remove such tadpole terms,
we impose the background equation of motion
$\alpha_1'(z)=\alpha_2'(z)$.
In the following we use $\alpha_1=\alpha_2$
because the constant shift of $\alpha_1$ does not change the action for $\pi$.
Then, we obtain the action,
\begin{align}
\label{scalar_final_full}
S=-\frac{1}{2}\int d^4x
\left[\alpha_1(z+\pi)\partial_m\pi\partial^m\pi
+\alpha_3(z+\pi)\Big(2\partial_z\pi(x)+\partial_m\pi\partial^m\pi\Big)^2\right]
-\frac{1}{2}\int d^4x\frac{d}{dz}A_1(z+\pi)\,,
\end{align}
where
$A_1(z)=2\int^{z} d{z'}\alpha_1(z')$ and the second term is a total derivative.
Note that the derivative with respect to $z$ in the last term of Eq.~\eqref{scalar_final_full} acts
not only explicitly on $z$ but also on $\pi(z)$, i.e., $dA_1(z+\pi)/dz = A_1'(z+\pi) (1+\partial_z \pi)$.
Up to the second order in $\pi$,
the bulk action (the first term) can be expanded as
\begin{align}
\label{scalar_final_bulk}
S_{\rm bulk}=-\frac{1}{2}\int d^4x
\,\alpha_1(z)
\left[\partial_{\widehat{m}}\pi\partial^{\widehat{m}}\pi+c_z^2(z)(\partial_z\pi)^2\right]
\quad
{\rm with}
\quad
c_z^2(z)=1+4\frac{\alpha_3(z)}{\alpha_1(z)}
\,,
\end{align}
where $\widehat{m}=t,x,y$
and $c_z$ can be interpreted as the propagating speed in the $z$-direction.
On the other hand,
the total derivative term can be expanded as
\begin{align}
\label{scalar_t.d.}
S_{\rm t.d.}=-\frac{1}{2}\int d^4x
\frac{d}{dz}
\left[A_1(z)+2\alpha_1(z)\pi+\alpha_1'(z)\pi^2+\mathcal{O}(\pi^3)\right]\,,
\end{align}
where note that there can arise linear order terms in $\pi$
from the total derivative term.
We will revisit its physical meaning in the next section.

\subsection{Physical spectra for single scalar domain-wall}
\label{subsec:spectra_scalar}

We next take a close look at the effective action \eqref{scalar_final_full}
and discuss physical spectra for a single scalar brane case.
For simplicity, let us consider the case $\alpha_3(z)=0$ in this subsection.\footnote{
The assumption here is just for simplicity
and our result should hold for more general setups qualitatively.}
Then, the brane
profile is characterized by the free function $\alpha_1(z)$ in the effective action.
Generically,
$\alpha_1(z)$ is related to the order parameter $\bar{\phi}'(z)$ as
\begin{align}
\alpha_1(z)\sim \bar{\phi}'(z)^2
\end{align}
because the $g^{zz}$ operator in the unitary gauge action
typically arises from
\begin{align}
g^{\mu\nu}\partial_\mu\bar{\phi}\partial_\nu\bar{\phi}=\bar{\phi}'^2g^{zz}\,.
\end{align}
To illustrate the physical spectra,
it is convenient to take the well-studied domain-wall profile obtained in Sec.~\ref{subsec:real_scalar},
\begin{align}
\bar{\phi}(z)=v\tanh \beta z\,,
\end{align}
and the corresponding function $\alpha_1(z)$ of the form
\begin{align}
\alpha_1(z)=
\bar{\phi}'(z)^2=\frac{\beta^2 v^2}{\cosh^4 \beta z}\,.
\end{align}
Here, the constants $v$ and $\beta$
characterize the domain-wall profile and, in particular,
$\beta$ specifies the thickness of the brane.
We then determine the physical spectra for the NG field $\pi$.
From the bulk action \eqref{scalar_final_bulk},
the linear order equation of motion follows as\footnote{
Notice that the linear term in the total derivative term \eqref{scalar_t.d.} vanishes
because $\alpha_1(z)=0$ outside the domain-wall $|z|\gg 1/\beta$,
where the $z$-diffeomorphism invariance is unbroken.
We will revisit this point in Sec.~\ref{section_multi}.}
\begin{align}
\pi''-4\beta\tanh\beta z \,\pi'+\partial_\perp^2\pi=0
\end{align}
in the coordinate space.
Here, the prime denotes the derivative with respect to $z$
and $\partial_\perp^2\equiv\partial_{\widehat{m}}\partial^{\widehat{m}}=-\partial_t^2+\partial_x^2+\partial_y^2$.
By the Fourier transformation in the $x^{\widehat{m}}$ directions
along the brane,
we rewrite it as
\begin{align}
\pi_{k_\perp}''-4\beta\tanh\beta z \,\pi_{k_\perp}'-k_\perp^2\pi_{k_\perp}=0
\quad
{\rm with}
\quad
\pi_{ k_\perp}(z)=\int d^3x_\perp
\pi(x^{\widehat{m}},z)e^{-ix^{\widehat{m}}k_{\perp\widehat{m}}}\,,
\end{align}
whose linearly independent solutions are given by
\begin{align}
\label{spectra_1}
\pi_{ k_\perp}=1\,,\,\,12\beta z+8\sinh 2\beta z+\sinh 4\beta z
\quad
{\rm for}
\quad
k_\perp^2=0 \,,
\end{align}
and
\begin{align}
\label{spectra_2}
\pi_{ k_\perp}=\exp\left(\pm 2\beta z\sqrt{1+\frac{k_\perp^2}{4\beta^2}}\right)\left[\left(1+\frac{k_\perp^2}{6\beta^2}\right)\cosh 2\beta z\mp\sqrt{1+\frac{k_\perp^2}{4\beta^2}}\sinh 2\beta z+\frac{k_\perp^2}{6\beta^2}\right]
\quad
{\rm for}
\quad
k_\perp^2\neq0\,.
\end{align}
We notice that
only the constant mode, $\pi_{k_\perp}=1$, has a finite value throughout the space,
whereas the other modes diverge outside the brane.
Since the $\pi$ field corresponds to the translational transformation parameter,
the constant mode, $\pi_{k_\perp}=1$, generates a shift of brane position
without changing the brane profile
and can be interpreted as the standard gapless NG mode
propagating along the brane.
It is also convenient to express the solutions
in terms of the canonically normalized field $\pi_{k_\perp}^c=\alpha_{1}^{1/2}\pi_{ k_\perp}$:
\begin{align}
\label{spectra_canonical_1}
\pi_{ k_\perp}^c=\frac{\beta v}{(1+\cosh 2\beta z)/2}\,,\,\,
\beta v\frac{12\beta z+8\sinh 2\beta z+\sinh 4\beta z}{(1+\cosh 2\beta z)/2}
\quad
{\rm for}
\quad
k_\perp^2=0 \,,
\end{align}
and
\begin{align}
\label{spectra_canonical_2}
\pi_{ k_\perp}^c=\beta v\exp\left(\pm 2\beta z\sqrt{1+\frac{k_\perp^2}{4\beta^2}}\right)\frac{\left(1+\frac{k_\perp^2}{6\beta^2}\right)\cosh 2\beta z\mp\sqrt{1+\frac{k_\perp^2}{4\beta^2}}\sinh 2\beta z+\frac{k_\perp^2}{6\beta^2}}{(1+\cosh 2\beta z)/2}
\quad
{\rm for}
\quad
k_\perp^2\neq0\,.
\end{align}
This normalization provides how the energy of each mode distributes in the $z$-direction.
For example, it is clear that the energy of the gapless NG mode localizes on the brane.
We also notice that the solutions in Eq.~\eqref{spectra_canonical_2}
have a finite energy density for $-k_\perp^2>4\beta^2$
as well as the first solution in Eq.~\eqref{spectra_canonical_1}.
More concretely,
the two modes in Eq.~\eqref{spectra_canonical_2}
behave like massive modes with the mass $2\beta$
outside the brane
$|z|\gg 1/\beta$,
\begin{align}
\pi_{ k_\perp}^c\sim
\exp(\pm ik_zz)
\quad
{\rm with}
\quad
k_\perp^2+k_z^2=-4\beta^2\,.
\end{align}
Also,
gauge transformation parameters
corresponding to the two modes
diverge outside the brane
$|z|\gg1/\beta$,
as is suggested by Eq.~\eqref{spectra_2}.
We therefore interpret them as bulk propagations
of the original scalar field $\phi(z)$,
rather than standard NG modes.
In  Appendix~\ref{App:Nambu-Goto}
we show that
the low-energy effective action
after integrating out those gapped modes
is nothing but the Nambu-Goto action.

\medskip
To summarize,
there exist two types of physical modes around the single scalar brane background:
the standard massless NG mode localizing on the brane
and the massive modes propagating in the bulk direction.
In particular,
only the standard localized NG mode is relevant
in the low-energy scale $E\ll\beta$
and the standard coset construction takes into account  this degrees of freedom only.
Conversely, if $\beta$ is much smaller than a typical scale of excitation energy,
the massive modes are not negligible
and 
they should be included in the low-energy effective theory.

\subsection{Inclusion of local Lorentz symmetry breaking}
\label{subsec:inclusion}
We then discuss the case when a nonzero spin field has a space-dependent condensation.
To illustrate the degrees of freedom and residual symmetries
in the unitary gauge,
let us consider a (space-like) vector $A_m$ on the Minkowski space as a concrete example.
Suppose that a vector field $A_m$ has a space-dependent condensation of the form
\begin{align}
\label{order_A_m}
\left\langle A_m(x)\right\rangle=\delta_m^3v(z)\,.
\end{align}
Here and in what follows,
we use integers $0,1,2,3$ to denote
the $t,x,y,z$-directions of the local Lorentz index.
Since $A_m(x)$ has a Lorentz charge,
the local Lorentz symmetry is broken
as well as  
the $z$-diffeomorphism invariance.
Following the EFT recipe,
we then introduce the vierbein $e_\mu^m$
to gauge the Lorentz symmetry.
Schematically, we write the degrees of freedom and
symmetries after introducing the vierbein as
\begin{align}
A_m(x)\,,\,\,e_\mu^m(x)
\quad
+\quad
\text{(3+1)-dim diffs}\,,\,\,
\text{(3+1)-dim local Lorentz}.
\end{align}
In order to take the unitary gauge,
it is convenient to note the decomposition
\begin{align}
A_m(x)=\Lambda_m\,^3(x)v\big(z+\pi(x)\big)
\quad
{\rm with}
\quad
\Lambda_m\,^n(x)\in SO(3,1)\,,
\end{align}
where
$\Lambda_m\,^3(x)$
specifies the direction of $A_m$
and corresponds to the NG field for the local Lorentz symmetry.
On the other hand,
$\pi(x)$ specifies the amplitude of $A_m$
and corresponds to the NG field for the $z$-diffeomorphism.
Using the local Lorentz and diffeomorphism invariance,
we can remove those NG fields
to set $\Lambda_m\,^3=\delta_m^3$ and $\pi=0$,
at least as long as the fluctuations are small.
In such a unitary gauge,
the only dynamical degrees of freedom are the vierbein $e_\mu^m$
and the residual symmetries
are the $(2+1)$-dimensional local Lorentz and diffeomorphism invariance along the $t,x,y$-directions.
Schematically, we write this setup as
\begin{align}
e_\mu^m(x)
\quad
+\quad
\text{(2+1)-dim diffs}\,,\,\,
\text{(2+1)-dim local Lorentz}.
\end{align}

\medskip
We then construct the effective action
based on these degrees of freedom
and residual symmetries.
Schematically,
let us decompose the effective action
into the three types of contributions as
\begin{align}
S=S_{P}+S_{L}+S_{PL}\,.
\end{align}
Here, $S_P$ is the effective action \eqref{scalar_simple}
and breaks the diffeomorphism invariance only:
\begin{align}
\label{S_P_unitary}
S_{P}&=-\frac{1}{2}\int d^4x\sqrt{-g}
\left[
\alpha_1(z)+\alpha_2(z) g^{zz}+\alpha_3(z)\left(g^{zz}-1\right)^2
\right]\,.
\end{align}
On the other hand,
$S_L$ and $S_{PL}$ break the local Lorentz invariance
and, in particular, $S_L$ represents terms that may exist even if the diffeomorphism is unbroken
(the following $\beta_i$'s are constants when the diffeomorphism is unbroken).
At the lowest order with respect to fluctuations and derivatives,
they are given by\footnote{
Note that there are some ambiguities
in the expression of the effective action.
For example,
we can use $e_3^z-1$ instead of $e_3^\mu n_\mu-1$
to define $S_{PL}$.
However, the difference between
the two descriptions can be absorbed by the redefinition of
$\alpha_i$'s in \eqref{S_P_unitary}
and it turns out that the generic effective action at the lowest order
is given by \eqref{S_P_unitary}, \eqref{S_L_unitary}, and \eqref{S_PL_unitary}.}
\begin{align}
\label{S_L_unitary}
S_{L}&=\int d^4x\sqrt{-g}
\left[
-\frac{\beta_1(z)}{4}\left(\nabla_\mu e_\nu^3-\nabla_\nu e_\mu^3\right)^2
-\frac{\beta_2(z)}{2} \left(\nabla^\mu e_\mu^3\right)^2
-\frac{\beta_3(z)}{2}\left(e^\nu_3\nabla_\nu e_\mu^3\right)^2
\right]\,,\\
\label{S_PL_unitary}
S_{PL}&=\int d^4x\sqrt{-g}\,\gamma_1(z)\left(e^\mu_3n_\mu-1\right)\,,
\end{align}
where
$\beta_i(z)$ and $\gamma_1(z)$ are scalar functions depending on $z$
and the unit vector $\displaystyle n_\mu=\delta^z_\mu/\sqrt{g^{zz}}$ breaks the $z$-diffeomorphism invariance explicitly.
We next introduce the NG fields by the St\"{u}ckelberg method and decouple the gauge degrees of freedom.
As in Sec.~\ref{subsec:scalar_domain},
we first introduce the NG fields $\pi$ for the $z$-diffeomorphism by the field-dependent gauge transformation \eqref{stuckelberg_diff}.
Similarly, we introduce NG fields $\xi_{\widehat{m}}$'s
for the local Lorentz transformation
in the $3$-$\widehat{m}$ plane ($\widehat{m}=0,1,2$)~as
\begin{align}
e_m^\mu(x)\to \widetilde{e}_m^\mu(x)=\Lambda_m{}^n(x)e_n^\mu(x)
\quad
{\rm with}
\quad
\Lambda_m{}^n(x)=\left(\exp\left[\xi^{\widehat{\ell}}(x)\Sigma_{\widehat{\ell}\,3}\right]\right)_m{}^n\in SO(3,1)\,,
\end{align}
where $\Sigma_{mn}$'s are generators of $SO(3,1)$.
In particular,
$e_3^\mu(x)$ is transformed as
\begin{align}
e_3^\mu\to \widetilde{e}_3^\mu=\Lambda_3{}^me_m^\mu
=\left[\delta_3^m\left(1-\frac{1}{2}\xi_{\widehat{m}}\xi^{\widehat{m}}\right)+\delta^m_{\widehat{m}}\xi^{\widehat{m}}+\mathcal{O}(\xi^3)\right]e_m^\mu\,.
\end{align}
Since the full diffeomorphism and local Lorentz invariance
can be restored
by assigning nonlinear transformation rules
on the NG fields $\pi$ and $\xi_{\widehat{m}}$,
we can set $e_\mu^m=\delta_\mu^m$
using the full gauge degrees of freedom.
After these procedures,
$S_L$ and $S_{PL}$ take the form
\begin{align}
\nonumber
S_{L}&=\int d^4x
\left[
-\frac{\beta_1(z+\pi)}{4}\left(\partial_m \Lambda_{3n}-\partial_n \Lambda_{3m}\right)^2
-\frac{\beta_2(z+\pi)}{2} \left(\partial_m \Lambda_3{}^m\right)^2
-\frac{\beta_3(z+\pi)}{2}\left(\Lambda_{3}{}^n\partial_n \Lambda_{m}^{3}\right)^2
\right]
\\*
\label{S_L_NG}
&=\int d^4x
\left[
-\frac{\beta_1(z)}{4}\left(\partial_{\widehat{m}} \xi_{\widehat{n}}-\partial_{\widehat{n}} \xi_{\widehat{m}}\right)^2
-\frac{\beta_2(z)}{2} \left(\partial^{\widehat{m}} \xi_{\widehat{m}}\right)^2
-\frac{\beta_1(z)+\beta_3(z)}{2}\left(\partial_z\xi_{\widehat{m}}\right)^2
+\ldots
\right]\,,
\\[1mm]
\label{S_PL_NG}
S_{PL}&=\int d^4x\,\gamma_1(z+\pi)\left[
\Lambda_3{}^m\frac{\delta_m^z+\partial_m\pi}{\sqrt{1+2\partial_z\pi+\partial_n\pi\partial^n\pi}}-1\right]
=\int d^4x\left[-\frac{\gamma_{1}(z)}{2}\left(\xi_{\widehat{m}}-\partial_{\widehat{m}}\pi\right)^2+\ldots\right]\,,
\end{align}
where the dots stand for the cubic and higher order in perturbations.
Here, it should be noticed that
$S_{L}$ contains the kinetic terms for $\xi_{\widehat{m}}$,
whereas $S_{PL}$ contains the mass term for $\xi_{\widehat{m}}$
and mixing interactions between $\xi_{\widehat{m}}$ and $\pi$.
Also note that Eqs.~\eqref{S_L_NG} and \eqref{S_PL_NG} do not contain linear order terms.
The background equation of motion therefore requires
$\alpha_1'(z)=\alpha'_2(z)$ so that the tadpole terms in $S_P$ vanishes.
The effective action for NG modes
are then given by Eqs.~\eqref{scalar_final_full}-\eqref{scalar_t.d.},
\eqref{S_L_NG}, and \eqref{S_PL_NG}.

\subsection{Qualitative features of nonzero spin branes}
\label{subsec:qualitative}

\begin{figure}[t]
\begin{center}
\includegraphics[width=80mm, bb=0 0 506 291]{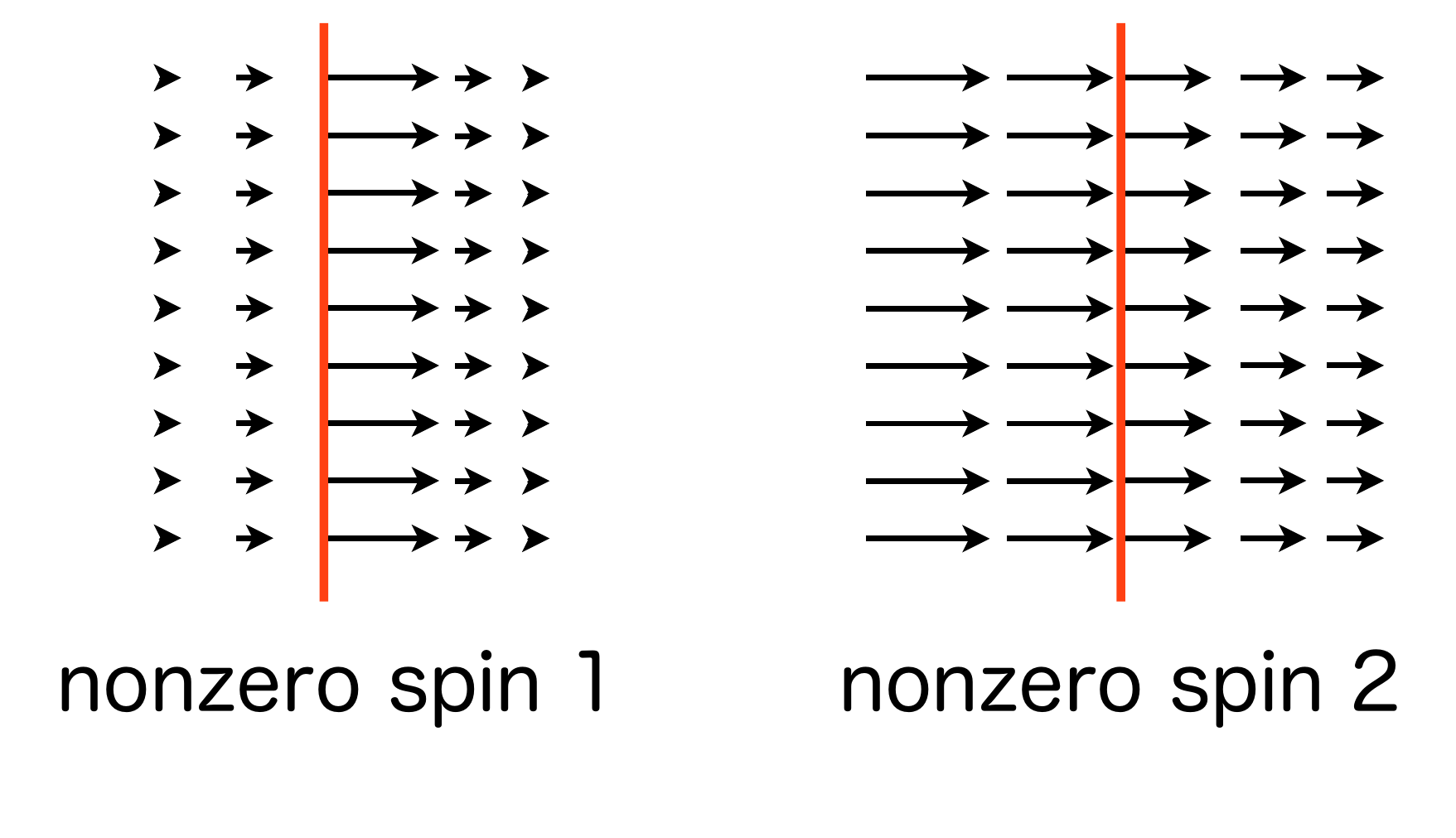}
\end{center}
\vspace{-5mm}
\caption{Two examples for nonzero spin branes.
In the first example (left figure),
both of the diffeomorphism and local Lorentz symmetries
are broken only on the brane.
On the other hand,
in the second example (right figure),
the local Lorentz symmetry is broken in the whole spacetime.}
\label{fig:nonzero}
\end{figure}

We then apply the obtained effective action
to single brane backgrounds.
For simplicity,
let us concentrate on the case $\alpha_3=\beta_2=\beta_3=0$
and consider the following second order action:\footnote{
Although we set several parameters to be zero for simplicity,
our results should hold for more general setups qualitatively. 
Also note that the $\beta_2$ coupling induces kinetic terms with a wrong sign and such terms are prohibited by the stability of the background,
though it is not prohibited only from the symmetry point of view.}
\begin{align}
\label{qualitarive_action}
S=\int d^4x
\left[
-\frac{\alpha_1(z)}{2}\partial_m\pi\partial^m\pi
-\frac{\beta_1(z)}{4}
\left[\left(\partial_{\widehat{m}} \xi_{\widehat{n}}-\partial_{\widehat{n}} \xi_{\widehat{m}}\right)^2
+2\left(\partial_z\xi_{\widehat{m}}\right)^2
\right]
-\frac{\gamma_1(z)}{2}\left(\xi_{\widehat{m}}-\partial_{\widehat{m}}\pi\right)^2
\right]\,.
\end{align}
Among the three functions $\alpha_1(z)$, $\beta_1(z)$, and $\gamma_1(z)$
characterizing the brane profile,
$\alpha_1(z)$ and $\gamma_1(z)$
are associated with the breaking of $z$-diffeomorphism invariance.
These two functions therefore have support
on the brane
and vanish outside,
just as $\alpha_1(z)$ for a single scalar brane.
On the other hand,
$\beta_1(z)$ does not necessarily vanish outside the brane
and has a nonzero value as long as the local Lorentz symmetry is broken.
More concretely,
it is convenient to introduce
a function $v(z)$ parametrizing the local Lorentz symmetry breaking,
just as we did in Eq.~\eqref{order_A_m} for vector condensation.
In terms of $v(z)$,
the three functions are typically given by
\begin{align}
\alpha_1(z)\sim \gamma_1(z)\sim v'(z)^2\,,
\quad
\beta_1(z)\sim v(z)^2\,,
\end{align}
where, for simplicity, we assumed that $v(z)$ is the only field breaking the $z$-diffeomorphism invariance.
One important point is that
there exist several types of $\beta_1$ profiles
even for single brane backgrounds.
In the rest of this subsection,
we discuss how the low-energy dynamics
depends on the $\beta_1$ profile
using two typical examples depicted in Fig.~\ref{fig:nonzero}.

\medskip
In the first case (the left figure),
local Lorentz symmetry is broken only on the brane.
A typical $v(z)$ profile is given by
\begin{align}
v(z)=\frac{v_0}{\cosh\beta z}
\end{align}
and the functions $\alpha_1(z)$, $\beta_1(z)$, and $\gamma_1(z)$ vanish outside 
the brane
$|z|\gg1/\beta$.
Similarly to the discussion in Sec.~\ref{subsec:spectra_scalar},
the NG modes $\pi$ and $\xi_{\widehat{m}}$ cannot have a $z$-dependence
at the energy scale $E\ll\beta$,
because their kinetic terms vanish outside the brane.
The action \eqref{qualitarive_action}
can then be reduced effectively to the three-dimensional one
\begin{align}
S=\int d^3x
\left[
-\frac{A}{2}\partial_{\widehat{m}}\pi\partial^{\widehat{m}}\pi
-\frac{B}{4}
\left(\partial_{\widehat{m}} \xi_{\widehat{n}}-\partial_{\widehat{n}} \xi_{\widehat{m}}\right)^2
-\frac{C}{2}\left(\xi_{\widehat{m}}-\partial_{\widehat{m}}\pi\right)^2
\right]\,,
\end{align}
where $A$, $B$, and $C$ are constants defined by
\begin{align}
 A=\int_{-\infty}^\infty dz\,\alpha_1(z)\,,
 \quad
B=\int_{-\infty}^\infty dz\,\beta_1(z)\,,
\quad
C=\int_{-\infty}^\infty dz\,\gamma_1(z)\,.
\end{align}
In terms of the normalization factor $v_0$ and
the thickness $1/\beta$,
these parameters can be estimated as
\begin{align}
A\sim v_0^2\beta\,,
\quad
B\sim\frac{v_0^2}{\beta}\,,
\quad
C\sim v_0^2\beta\,.
\end{align}
Since $\xi_{\widehat{m}}$ acquires
a mass $m\sim\beta$,
the
dynamics at the energy scale $E\ll\beta$
is governed by the NG mode $\pi$ for the $z$-diffeomorphism.
In particular,
we can integrate out the $\xi_{\widehat{m}}$ field as
\begin{align}
\label{integrate_out_xi}
\xi_{\widehat{m}}=\partial_{\widehat{m}}\pi+\mathcal{O}(E^2/\beta^2)\,.
\end{align}
The low-energy dynamics is then reduced to the same one as the scalar brane.
As we revisit in Sec.~\ref{subsub:single_revisit},
Eq.~\eqref{integrate_out_xi} corresponds to the inverse-Higgs constraint
in the standard coset construction.

\medskip
We next consider the second case (the right figure),
where local Lorentz symmetry is broken also outside the brane
and a typical $v(z)$ profile is given by
\begin{align}
v(z)=\overline{v}(1+\delta \tanh \beta z)\,.
\end{align}
Since the functions $\alpha_1(z)$ and $\gamma_1(z)$
localize on the brane,
these two contributions can be reduced to
the three-dimensional ones
at the energy-scale $E\ll\beta$:
\begin{align}
\label{three_1}
S=
\int d^4x
\left[
-\frac{\beta_1(z)}{4}
\left[\left(\partial_{\widehat{m}} \xi_{\widehat{n}}-\partial_{\widehat{n}} \xi_{\widehat{m}}\right)^2
+2\left(\partial_z\xi_{\widehat{m}}\right)^2
\right]
\right]
+\int d^3x
\left[
-\frac{A}{2}\partial_{\widehat{m}}\pi\partial^{\widehat{m}}\pi
-\frac{C}{2}\left(\xi_{\widehat{m}}-\partial_{\widehat{m}}\pi\right)^2
\right]\,,
\end{align}
where $A$ and $C$ can be estimated as
\begin{align}
A=\int_{-\infty}^\infty dz\,\alpha_1(z)
\sim\overline{v}^2\delta^2\beta\,,
\quad
C=\int_{-\infty}^\infty dz\,\gamma_1(z)
\sim\overline{v}^2\delta^2\beta\,.
\end{align}
Let us then consider the parameter region $\delta\ll1$ in particular:
\begin{align}
\label{three_2}
S\sim
-\overline{v}^2
\left\{\int d^4x
\left[\left(\partial_{\widehat{m}} \xi_{\widehat{n}}-\partial_{\widehat{n}} \xi_{\widehat{m}}\right)^2
+2\left(\partial_z\xi_{\widehat{m}}\right)^2
\right]
+\delta^2\beta\int d^3x
\left[
\partial_{\widehat{m}}\pi\partial^{\widehat{m}}\pi
+(\xi_{\widehat{m}}-\partial_{\widehat{m}}\pi)^2
\right]\right\}\,,
\end{align}
where we dropped some numerical coefficients for simplicity.
An interesting point is that
there exists a hierarchy in the energy scale.
First, at the low-energy limit $E\ll \delta\beta(\ll \beta)$,
the equations of motion inside the brane
are of the form
\begin{align}
\xi_{\widehat{m}}=\partial_{\widehat{m}}\pi+\mathcal{O}\Big(E^2/(\delta^2\beta^2)\cdot\xi\Big)\,,
\quad
\partial_{\widehat{m}}^2\pi=2\partial^{\widehat{m}}(\xi_{\widehat{m}}-\partial_{\widehat{m}}\pi)
=\mathcal{O}\Big(E^3/(\delta^2\beta^2)\cdot\xi\Big)\,,
\end{align}
which result in the equation of motion $\partial_{\widehat{m}}^2\pi=0$ for a massless field on the brane.
Just as the first case,
the equation of motion for $\xi_{\widehat{m}}$ (inside the brane)
corresponds to the inverse-Higgs constraint
and
the dynamics of $\pi$
is reduced to the same one as the scalar
brane
case.
On the other hand,
at the energy scale $\delta \beta \ll E\ll \beta$,
the action \eqref{three_2} can be further approximated as
\begin{align}
S\sim
-\overline{v}^2
\left\{\int d^4x
\left[\left(\partial_{\widehat{m}} \xi_{\widehat{n}}-\partial_{\widehat{n}} \xi_{\widehat{m}}\right)^2
+2\left(\partial_z\xi_{\widehat{m}}\right)^2
\right]
+\delta^2\beta\int d^3x
\left(
\partial_{\widehat{m}}\pi\partial^{\widehat{m}}\pi
-\xi^{\widehat{m}}\partial_{\widehat{m}}\pi
\right)\right\}\,,
\end{align}
where $\xi_{\widehat{m}}$ and $\pi$
can be thought of as massless fields propagating
in the four-dimension and localizing on the brane,
respectively.
The bulk field $\xi_{\widehat{m}}$
and the localized field $\pi$
then interact to each other by the $\xi^{\widehat{m}}\partial_{\widehat{m}}\pi$ interaction.
The dynamics at this energy scale is different from
both of the single scalar brane
case and the first example for single nonzero spin branes.

\begin{figure}[t]
\begin{center}
\hspace{-10mm}\includegraphics[width=140mm, bb=0 0 580 305]{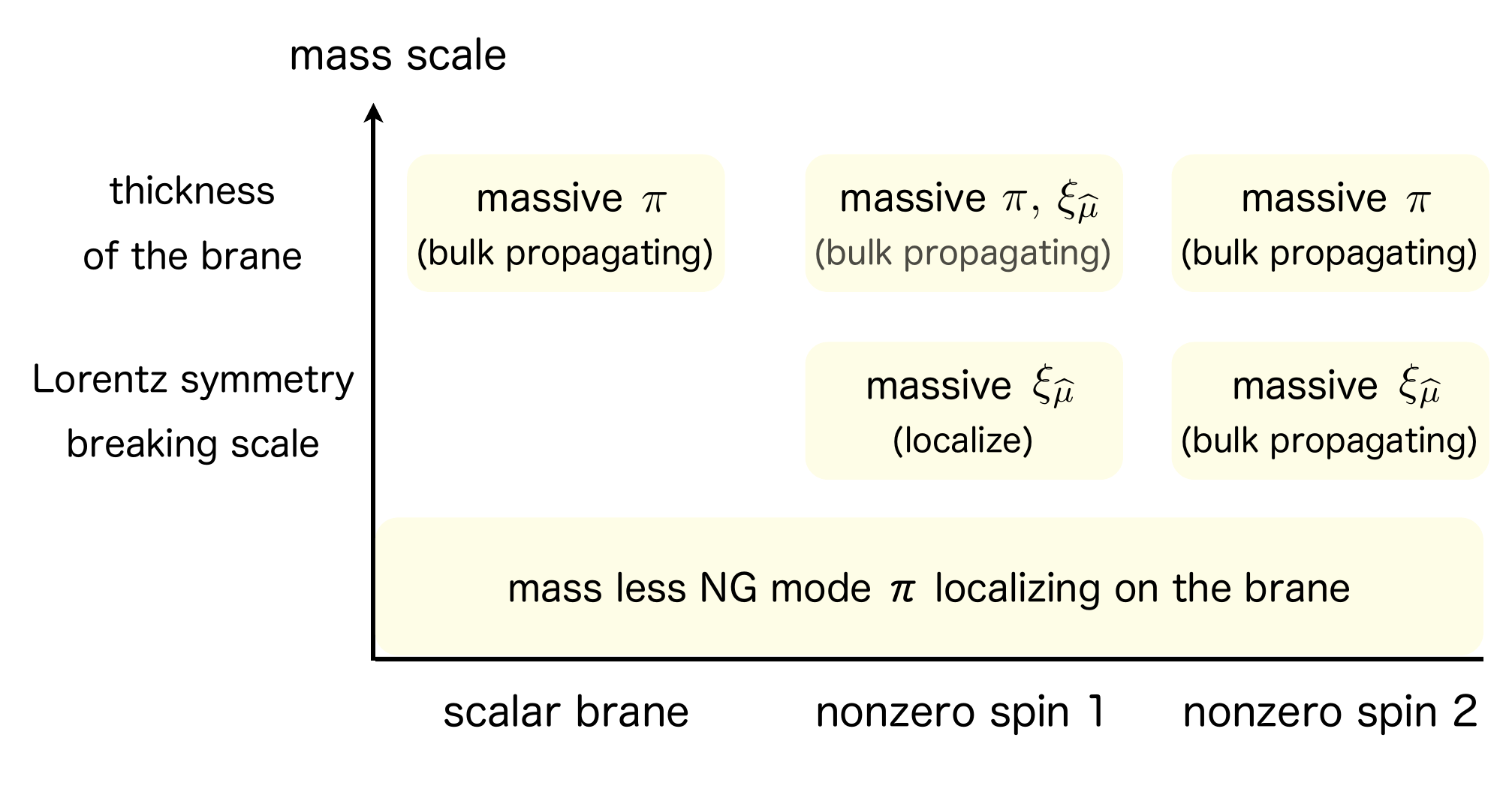}
\end{center}
\vspace{-5mm}
\caption{Physical spectra for three types of branes.
While the massless spectrum is the same between the three,
the massive spectrum depends on symmetry breaking patterns in the local picture.}
\label{fig:spectrum}
\end{figure}

\medskip
In Fig.~\ref{fig:spectrum},
we summarize the qualitative features of three types of single
brane backgrounds
discussed in this section:
one for a single scalar brane
and
two for a single nonzero spin brane.
In the low-energy limit,
the dynamics of $\pi$ (after integrating out $\xi_{\widehat{m}}$) in each setup
results in the same one,
which can be captured by the standard coset construction
based on the global symmetry picture.
However,
this degeneracy is resolved beyond the low-energy limit
and the resolving energy scale is not necessarily high
compared with the scale $\beta$ of the
brane
thickness.
Such a scale can be in our interests
and
we need to specify the symmetry breaking pattern
based on the local symmetry picture
to investigate such an intermediate scale.
This is one point of our paper.

\section{One-dimensional periodic modulation}
\begin{figure}[t]
\begin{center}
\includegraphics[width=160mm, bb=0 0 580 171]{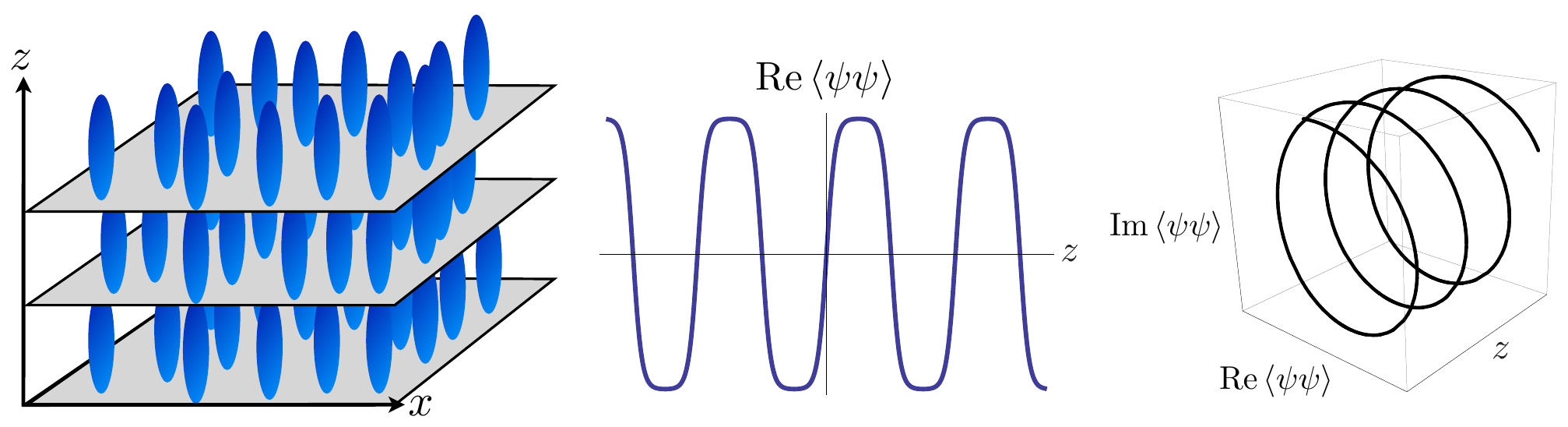}
\end{center}
\vspace{-5mm}
\caption{Examples of one-dimensional periodic modulation.
In the smectic-A phase (the left figure),
the layer structure breaks the diffeomorphism symmetry
on the whole spacetime.
In the FFLO phase of superconductor (the center and right figures),
the chiral condensation arises in an inhomogeneous~way.}
\label{fig:modulation}
\end{figure}

\setcounter{equation}{0}
\label{section_multi}
As we discussed in the previous section,
the effective action for diffeomorphism symmetry breaking
contains free functions of coordinates
and their profiles are directly related to the breaking pattern. 
For example,
in the single brane
case,
the $z$-diffeomorphism invariance is broken
only on the brane
and the functions $\alpha_i$'s in Eq.~\eqref{scalar_simple_pi}
are localized on it.
In this section
we discuss one-dimensional periodic modulation,
i.e., a system in which the condensation is periodic in one direction,
by changing the profile of those functions.
As depicted in Fig.~\ref{fig:modulation},
such symmetry breaking patterns
are realized in condensed matter systems
such as the smectic-A phase of liquid crystals~\cite{deGennesText}
and the Fulde-Ferrell-Larkin-Ovchinnikov (FFLO) phase of superconductor~\cite{Fulde:1964zz,larkin:1964zz}.
With these types of applications in mind,
we extend our analysis to nonrelativistic systems in Minkowski space
and discuss generic features in the dispersion relations of NG modes
in the presence of one-dimensional periodic modulation.

\medskip
Let us first extend the previous effective action
to nonrelativistic systems.
There are several possibilities of generalization to nonrelativistic systems.
Here we assume that the system has spacetime translations and spatial rotational symmetries.
In particular
we do not consider Galilean symmetry for simplicity.\footnote{
When the system originally enjoys Galilean symmetry,
some modifications may be required.
See also Appendix~\ref{Sec:nonrela}.
Under this simplification,
the only difference from the relativistic case
is that we 
consider spatial diffeomorphism, instead of the full diffeomorphism,
to construct the effective action in the unitary gauge.}
The rest  of the discussions is completely parallel to the relativistic case
and it is straightforward to perform the construction of the effective action.
The effective action for the $z$-diffeomorphism symmetry breaking
is then given~by
\begin{align}\label{eq:SP}
S_P&=\frac{1}{2}\int d^4x
\left[
\widetilde{\alpha}_1(z+\pi) \dot{\pi}^2
-\alpha_1(z+\pi)(\partial_i\pi)^2
-\alpha_3(z+\pi)\Big(2\partial_z\pi(x)+(\partial_i\pi)^2\Big)^2
-\frac{d}{dz}A_1(z+\pi)\right]\,,
\end{align}
where $\pi$ is the NG field for the $z$-diffeomorphism
and $\displaystyle A_1(z)=2\int^{z} dz'\alpha_1(z')$.
In contrast to the relativistic case,
the functions $\widetilde{\alpha}_1$ and $\alpha_1$
in front of the temporal and spatial kinetic terms
are independent.
Up to second order in $\pi$,
the bulk contribution (the first three terms)
and the total derivative contribution (the last term)
can be expanded as
\begin{align}
\label{nonrela_S_P_bulk}
S_{P,\,{\rm bulk}}&=\frac{1}{2}\int d^4x
\Big[\,\widetilde{\alpha}_1(z)\dot{\pi}^2
-\alpha_1(z)(\partial_{i}\pi)^2
-4\alpha_3(z)(\partial_z\pi)^2\,\Big]
\,,\\
\label{nonrela_S_P_td}
S_{P,\,{\rm t.d.}}&=-\frac{1}{2}\int d^4x
\frac{d}{dz}
\left[A_1(z)+2\alpha_1(z)\pi+\alpha_1'(z)\pi^2\right]\,,
\end{align}
where note that $S_{P,\,{\rm t.d.}}$
contains a linear order term if $\alpha_1\neq0$ at the boundary.
This feature will be important in the following discussions.

\medskip
We then discuss systems with one-dimensional periodic modulation
based on this effective action.
Suppose that the condensation is periodic in the $z$-direction
and is characterized by a discrete symmetry $z\to z'=z+a$
with $a$ being the periodicity.
Since the functions $\widetilde{\alpha}_1$ and $\alpha_i$'s
are periodic because of the residual discrete symmetry,
they generically have support on the whole spacetime.
In particular,
$\alpha_1(z)$, if it exists, does not vanish at the boundary
and leads to a linear order term
in Eq.~\eqref{nonrela_S_P_td},
which is in sharp contrast to the single brane
case.
Let us take a closer look at this linear order term
and discuss its implications for the dispersion relations of NG modes.
First,
this linear order term is not relevant as long as the NG field $\pi$ decays at spatial infinity
$\displaystyle\lim_{z\to\pm\infty}\pi(x)=0$
since it is a total derivative.
It is essentially because we impose the background (bulk) equations of motion in the effective action construction,
which guarantee the stability of backgrounds
under small perturbations with a fixed boundary condition.
However,
the boundary linear term becomes relevant
if we consider a configuration with $\displaystyle\lim_{z\to\pm\infty}\pi(x)\neq0$
as this implies the existence of configurations with a lower energy,
just as tadpoles in the bulk action.
For example,
let us consider a configuration of the form
\begin{align}
\pi(x)=\epsilon z
\quad(\epsilon:{\rm constant})\,,
\end{align}
which corresponds to a rescaling
$z\to z'=(1-\epsilon)z$.
The energy contributions $E_{\rm t.d.}$ from Eq.~\eqref{nonrela_S_P_bulk} and $E_{\rm bulk}$ from Eq.~\eqref{nonrela_S_P_td}
for this configuration are given by
\begin{align}
E_{\rm t.d.}\sim\alpha_1 \epsilon V\,,
\quad
E_{\rm bulk}\sim(\alpha_1+4\alpha_3) \epsilon^2 V\,,
\end{align}
where $V$ is the spatial volume.
It then turns out that
there exists a low-energy direction along a small negative $\epsilon$.
We therefore conclude that the $\alpha_1$ term is prohibited
when the background energy is a local minimum
in the configuration space
and the effective action at the lowest order derivative
is
\begin{align}
\nonumber
S_P&=\frac{1}{2}\int d^4x
\left[
\widetilde{\alpha}_1(z+\pi) \dot{\pi}^2
-\alpha_3(z+\pi)\Big(2\partial_z\pi(x)+(\partial_i\pi)^2\Big)^2
\right]
\\
&=\frac{1}{2}\int d^4x
\Big[\,\widetilde{\alpha}_1(z)\dot{\pi}^2
-4\alpha_3(z)(\partial_z\pi)^2+\mathcal{O}(\pi^3)\,\Big]\,.
\end{align}
The corresponding dispersion relation for the NG mode $\pi$ is
\begin{align}
\label{dispersion_2nd}
\omega^2 
= c_1 k_z^2\,,
\end{align}
where $\omega$ is the energy, $k_z$ is the momentum in the first Brillouin zone,
$|k_z|\leq \pi/a$,
and $c_1$ is the coefficient depending on $\widetilde{\alpha}_1$ and $\alpha_3$.
It should be emphasized
that the momentum $k_\bot^2=k_x^2+k_y^2$ in the $x,y$-directions
does not appear in the dispersion relation Eq.~\eqref{dispersion_2nd}
up to this order~\cite{Landau:StatisticalPhysics}.
By including higher order terms in the effective action,
we can explicitly show that
higher order derivative corrections to the dispersion relations
are schematically
\begin{align}
\label{dispersion_modulation1}
\omega^2\sim c_{1} k_z^2
+c_{2} k_zk_\bot^2+c_{3} k_\bot^4
\end{align}
up to the second order in $k_z$ and $k_\bot^2$, where $c_i$ are constants.\footnote{
Note that
counting of the scaling dimension changes
when the dispersion relation is anisotropic.
For Eq.~\eqref{dispersion_modulation1},
$k_z$ and $k_\bot^2$ have the same scaling dimension,
so that the terms displayed there
are the lowest-order in the derivative expansion.}

\medskip
It would be also interesting to illustrate
that such higher order correction terms can arise
after integrating out massive NG fields for local rotations.
The effective action for such symmetry breaking pattern
can be easily obtained by extending the construction in the previous section
to nonrelativistic systems.
Here, we again consider that the system has space-time translations and spatial rotational symmetries.
The second order action for NG fields is then given by
\begin{align}
\label{S_L_nonrela2}
S_{L}
&=\int d^4x
\left[
\frac{\widetilde{\beta}_1+\beta_1}{2}(\partial_t\xi_{\hat{i}})^2
-\frac{\beta_1}{4}\left(\partial_{\hat{i}} \xi_{\hat{j}}-\partial_{\hat{j}} \xi_{\hat{i}}\right)^2
-\frac{\beta_2}{2} \left(\partial_{\hat{i}} \xi_{\hat{i}}\right)^2
-\frac{\beta_1+\beta_3}{2}\left(\partial_z\xi_{\hat{i}}\right)^2
\right]\,,
\\*[1mm]
S_{PL}&
=\int d^4x\left[-\frac{\gamma_1}{2}\left(\xi_{\hat{i}}-\partial_{\hat{i}}\pi\right)^2+\frac{\gamma_1}{2}\dot{\pi}^2+\ldots\right]\,,
\end{align}
where $\hat{i}=x,y$ and $\xi_{\hat{i}}$'s are NG fields for rotations in the $\hat{i}$-$z$ plane.
Also we assume that $\widetilde{\beta}_1$, $\beta_i$'s, and $\gamma_1$ are constants for simplicity,
though they
have $z$-dependence in general.
Just as in the relativistic case,
$S_{PL}$ contains a mass term of $\xi_{\hat{i}}$
and mixing interactions between $\pi$ and $\xi_{\hat{i}}$.
In the low energy limit,
the equation of motion for $\xi_{\hat{i}}$ is reduced to
\begin{align}
\xi_{\hat{i}}=\partial_{\hat{i}}\pi\,.
\end{align}
Substituting it to Eq.~\eqref{S_L_nonrela2},
we obtain the effective interaction of the form
\begin{align}
S_{\rm eff}&=\int d^4x
\left[
\frac{\widetilde{\beta}_1+\beta_1}{2}(\partial_t\partial_{\hat{i}}\pi)^2
-\frac{\beta_2}{2} \left(\partial_{\hat{i}}^2\pi\right)^2
-\frac{\beta_1+\beta_3}{2}\left(\partial_z\partial_{\hat{i}}\pi\right)^2
+\frac{\gamma}{2}\dot{\pi}^2
\right] \,,
\end{align}
which gives the corrections to the coefficients of $k_z^2$ and $k_\bot^4$ in the dispersion relation. 

\medskip
To summarize,
in the systems with one-dimensional periodic modulation,
the (locally) minimum energy condition
constrains the dispersion relations of NG modes for the broken diffeomorphism as in
Eq.~\eqref{dispersion_modulation1}.
In particular,
the massive NG fields for local rotations 
can for example induce
higher derivative corrections in the dispersion relations.

\medskip
We note that
such discussions on dispersion relations suggest that
the one dimensional order can be realized only at zero temperature.
It is because the finite temperature effect breaks the order parameter in the thermodynamic limit: The contribution of thermal fluctuation of the NG mode to the order parameter is proportional to 
\begin{equation}
\begin{split}
\langle \pi^2(x)\rangle = T\int \frac{d^2k_\perp dk_z}{(2\pi)^{3}}\frac{1}{k_z^2 +c^2 k_\perp^4} = \frac{T}{4\pi c}  \ln\frac{\Lambda}{\mu},
\end{split}
\end{equation}
where $c=\beta_2/(4\alpha_3)$ and we introduced UV and IR cutoffs, $\Lambda$ and $\mu$. At $\mu\to0$, this correction is logarithmically divergent; it leads to the vanishing order parameter~\cite{Landau:StatisticalPhysics}. 
A typical example is a smectic-A phase of liquid crystals, in which the order parameter vanishes, and the quasi-long range order appears~\cite{chaikin}.
Also note that
if there exist an external field that explicitly breaks rotation symmetry such as a magnetic field,
the term $k_\perp^2$ appears in the dispersion relation of $\pi$. As the result, the fluctuation of $\pi$ to the order parameter is suppressed, and the order parameter remains finite~\cite{Landau:StatisticalPhysics}.

\section{Mixture of internal and spacetime symmetries}
\label{Sec:mix}
\setcounter{equation}{0}
Our approach to construct the effective action 
is applicable not only to spacetime symmetry breaking
but also to the breaking of a mixture of spacetime and internal symmetries.
In this section,
as a simplest example,
we discuss
the case when a global internal $U(1)$ symmetry and a translation symmetry
are broken to the diagonal Abelian symmetry.

\subsection{Complex scalar field model}
\label{subsec:complex}
Let us begin with a complex scalar field model
and illustrate concretely the degrees of freedom and residual symmetries
in the unitary gauge.
Suppose that
a complex scalar field
follows the symmetry transformation rule
\begin{align}
\label{mixture_scalar}
U(1):\phi(x)\to\phi'(x)=e^{i\lambda}\phi(x)\,,
\quad
{\rm translation}:
\phi(x)\to\phi'(x)=\phi(x+\epsilon)\,,
\end{align}
where the transformation parameters
$\lambda$ and $\epsilon^\mu$
are constants.
When it has a background condensation
\begin{align}
\label{mixture_background}
\langle\phi(x)\rangle=r_0\,e^{iut}
\quad
(r_0 \text{ and } u: \text{real constants})\,,
\end{align}
the internal $U(1)$ and time-translational symmetries
are broken to the diagonal one
\begin{align}
\label{mixture_unbrokn_global}
\phi(x)\to\phi'(x)=e^{i\lambda}\phi(x+\epsilon)
\quad
{\rm with}
\quad
\lambda=-u\epsilon^{0}\,.
\quad
\epsilon^i=0\,.
\end{align}
We then reinterpret this symmetry breaking
from the local symmetry point of view.
For this purpose,
let us gauge the internal $U(1)$
and translation symmetry
by introducing a gauge field $A_\mu$ for the internal $U(1)$ symmetry and the metric field $g_{\mu\nu}$.
Their $U(1)$ gauge transformations are
\begin{align}
A_\mu(x)\to A'_\mu(x)=A_\mu(x)-\partial_\mu\lambda(x)\,,
\quad
g_{\mu\nu}(x)\to g'_{\mu\nu}(x)=g_{\mu\nu}(x)
\end{align}
and their diffeomorphism transformations are
\begin{align}
A_\mu(x)\to A'_\mu(x)=
\frac{\partial (x^\nu+\epsilon^\nu)}{\partial x^\mu}
A_\nu(x+\epsilon)\,,
\quad
g_{\mu\nu}(x)\to g'_{\mu\nu}(x)=\frac{\partial (x^\rho+\epsilon^\rho)}{\partial x^\mu}\frac{\partial (x^\sigma+\epsilon^\sigma)}{\partial x^\nu}g_{\rho\sigma}(x+\epsilon)\,.
\end{align}
Also the transformation rule of the scalar field $\phi$ is
given by replacing the transformation parameters $\lambda$ and $\epsilon^\mu$ in Eq.~\eqref{mixture_scalar}
by local ones $\lambda(x)$ and $\epsilon^\mu(x)$.
Correspondingly,
the unbroken diagonal symmetry \eqref{mixture_unbrokn_global}
is gauged as
\begin{align}
\label{diagonal_gauge}
\phi(x)\to\phi'(x)=e^{i\lambda\left(x+\epsilon(x)\right)}\phi\left(x+\epsilon(x)\right)
\quad
{\rm with}
\quad
\lambda(x)=-u\epsilon^{0}(x)\,,
\quad
\epsilon^i=0\,,
\end{align}
which is realized by performing
the time-diffeomorphism $\epsilon^0(x)$
after the internal $U(1)$ gauge transformation $\lambda(x)$.
The background \eqref{mixture_background} is then
characterized by the symmetry breaking
of the internal $U(1)$ gauge and diffeomorphism symmetries
to the diagonal gauge symmetry \eqref{diagonal_gauge}
and spatial diffeomorphism symmetries.

\medskip
We next discuss the degrees of freedom
and the residual symmetries in the unitary gauge.
First, the setup before taking the unitary gauge
can be schematically written as
\begin{align}
\phi(x)\,,
\,\,
A_\mu(x)\,,
\,\,
g_{\mu\nu}(x)
\quad
+
\quad
\text{internal $U(1)$ gauge, diffs}.
\end{align}
To take the unitary gauge,
it is convenient to note the decomposition
\begin{align}
\phi(x)=\big(r_0+\sigma(x)\big)e^{i\left(ut+\pi(x)\right)}\,,
\end{align}
where $\pi(x)$ and $\sigma(x)$ are real scalar fields.
Since internal $U(1)$ gauge transformations and time diffeomorphisms generate local shifts of $\pi(x)$,
it can be identified with the NG field 
and we can impose the unitary gauge condition $\pi(x)=0$.
Just as the background \eqref{mixture_background},
the gauge condition $\pi(x)=0$
is invariant under the diagonal gauge transformations \eqref{diagonal_gauge}
and spatial diffeomorphisms.
Schematically, the dynamical degrees of freedom
and the residual symmetries in the unitary gauge
are given by
\begin{align}
\label{mixture_comp_setup}
\sigma(x)\,,
\,\,
A_\mu(x)\,,
\,\,
g_{\mu\nu}(x)
\quad
+
\quad
\text{diagonal gauge, spatial diffs}.
\end{align}
Note that $\sigma(x)$ is interpreted as a matter field,
which cannot be absorbed by gauge transformations
and is generically massive.

\subsection{Construction of effective action}
\label{subsec:mixture_coonstruction}
In the previous subsection
we illustrated the unitary gauge setup
using a complex scalar field model.
More generally,
the
dynamical degrees of freedom and the residual symmetries
in the unitary gauge for this type of symmetry breaking
are given by
the minimal setup
\begin{align}
\label{mixture_minimal}
A_\mu(x)\,,
\,\,
g_{\mu\nu}(x)
\quad
+
\quad
\text{diagonal gauge, spatial diffs}
\end{align}
and possibly with additional matter fields
such as $\sigma(x)$ in Eq.~\eqref{mixture_comp_setup}.
In this subsection,
we construct the effective action for the minimal setup \eqref{mixture_minimal} concretely.

\medskip
First,
just as the case of diffeomorphism symmetry breaking,
the
ingredients of the effective action covariant under spatial diffeomorphisms
are given by
\begin{align}
\text{scalar functions of $t$},\,\, A_\mu(x)\,,\,\,g_{\mu\nu}\,,\,\,R_{\mu\nu\rho\sigma}\,,
\,\,\text{and their covariant derivatives}.
\end{align}
The general unitary gauge action
is then constructed from these ingredients
in an invariant way under the diagonal gauge transformation.
To write down such an effective action,
it is convenient to note the diagonal gauge transformation of
the gauge field
\begin{align}
A_\mu(x)\to A'_\mu(x)
=\frac{\partial \tilde{x}^\nu}{\partial x^\mu}A_\nu(\tilde{x})
+u\partial_\mu\epsilon(x)
\quad
{\rm with}
\quad
\tilde{x}^\mu=x^\mu+\delta_0^\mu\epsilon(x)\,.
\end{align}
Since the time-coordinate $t$ is invariant under the diagonal transformation
\begin{align}
t\to t=\tilde{t}-\epsilon(x)\,,
\end{align}
we can find the following combination
covariant under the diagonal transformation:
\begin{align}
u\partial_\mu t+A_\mu(x) \to
u\left[\frac{\partial\,\tilde{t}}{\partial x^\mu}-\partial_\mu\epsilon(x)\right]+\frac{\partial \tilde{x}^\nu}{\partial x^\mu}A_\nu(\tilde{x})
+u\partial_\mu\epsilon(x)
=\frac{\partial \tilde{x}^\nu}{\partial x^\mu}\left[u\frac{\partial \tilde{t}}{\partial \tilde{x}^\nu}+A_\nu(\tilde{x})\right]\,.
\end{align}
Note that other ingredients such as the metric field $g_{\mu\nu}$
are also covariant.
We therefore conclude that
the time-coordinate $t$ and the gauge field $A_\mu$ can appear
only in the form $u\partial_\mu t +A_\mu=u\delta_\mu^0 +A_\mu$
and therefore ingredients of the unitary gauge effective action are
\begin{align}
\label{mixture_unitary_ingredients}
u\delta_\mu^0 +A_\mu\,,
\,\,
g_{\mu\nu}\,,
\,\,
R_{\mu\nu\rho\sigma}\,,
\,\,
\text{and their covariant derivatives.}
\end{align}
The lowest few terms in fluctuations and derivatives
are then given by
\begin{align}
\label{mixture_simple}
S=-\frac{1}{2}\int d^4x\sqrt{-g}
\left[\alpha\left(u^2g^{00}+2uA^0+A_\mu A^\mu\right)+\beta\Big(u^2\left(1+g^{00}\right)+2uA^0+A_\mu A^\mu\Big)^2\right]\,,
\end{align}
where
$\alpha$ and $\beta$ are constants independent of $t$.
Also note that we used $g^{\mu\nu}(u\delta_\mu^0 +A_\mu)(u\delta_\nu^0 +A_\nu)=u^2g^{00}+2uA^0+A_\mu A^\mu$.

\medskip
We next introduce the effective action for NG field $\pi(x)$ by performing the St\"{u}ckelberg method.
For example,
we can introduce the NG field by a field-dependent $U(1)$ gauge transformation\footnote{
In the present symmetry breaking pattern,
there are several ways to introduce the NG field
because both of the internal $U(1)$ gauge
and time diffeomorphism symmetries are broken.
For example,
we can perform a field-dependent time diffeomorphism
instead of the $U(1)$ gauge transformation.
However,
the resulting effective action is equivalent up to field redefinition of the NG field.}
\begin{align}
A_\mu(x)\to A'_\mu(x)=A_\mu(x)+\partial_\mu\pi(x)\,.
\end{align}
After this transformation,
the ingredients \eqref{mixture_unitary_ingredients} are given by
\begin{align}
\label{mixture_Stuckelberg_ingredients}
u\delta_\mu^0 +\partial_\mu\pi+A_\mu\,,
\,\,
g_{\mu\nu}\,,
\,\,
R_{\mu\nu\rho\sigma}\,,
\,\,
\text{and their covariant derivatives.}
\end{align}
Note that the internal $U(1)$ gauge and diffeomorphism invariance
can be recovered
by assigning the following nonlinear transformation rule
to the NG field
\begin{align}
U(1):\pi(x)\to \pi'(x)=\pi(x)+\lambda(x)\,,
\quad
{\rm diffs}:\pi(x)\to \pi'(x)=\pi(x+\epsilon(x))+u\epsilon^0(x)\,.
\end{align}
We can now set that $A_\mu=0$ and $g_{\mu\nu}=\eta_{\mu\nu}$
using the nonlinearly realized full gauge symmetries.
The ingredients of the action for the NG field are then
\begin{align}
\label{mixture_NG_ingredients}
u\delta_\mu^0 +\partial_\mu\pi\,,
\,\,
\eta_{\mu\nu}\,,
\,\,
\text{and their derivatives.}
\end{align}
Also the effective action corresponding to the unitary gauge action \eqref{mixture_simple} is given by
\begin{align}
\nonumber
S&=-\frac{1}{2}\int d^4x
\left[\alpha\left(-u^2-2u\partial_0\pi+\partial_\mu\pi\partial^\mu\pi\right)+\beta\left(-2u\partial_0\pi+\partial_\mu\pi\partial^\mu\pi\right)^2\right]
\\
\label{mixture_simple_NG}
&=\frac{\alpha}{2}\int d^4x
\left[
 c_s^{-2}\Big(\dot{\pi}^2-c_s^2(\partial_i\pi)^2\Big)
+(1-c_s^{-2})\left(\frac{1}{u}\dot{\pi}\partial_\mu\pi\partial^\mu\pi-\frac{1}{4u^2}(\partial_\mu\pi\partial^\mu\pi)^2\right)\right]
\,,
\end{align}
where we dropped temporal total derivatives and the constant term.
The propagating speed $c_s$ of the NG field $\pi(x)$
is given by $ c_s^{-2}=1-\frac{4\beta u^2}{\alpha }$.
Note that the obtained effective action takes a similar form to that for time-diffeomorphism symmetry breaking.
The only difference is that
the coefficients are constant instead of functions of time. In particular, 
because of this,
there are no linear order terms in Eq.~\eqref{mixture_simple_NG}
and the background equation of motion is satisfied
from the beginning.
Also note that
linear order terms in temporal total derivatives,
dropped in the second line of Eq.~\eqref{mixture_simple_NG},
are not problematic in contrast to those in spatial total derivatives.
It is because temporal total derivatives
affect only initial conditions
and they are not relevant once we specify the initial conditions.

\subsection{Comments on breaking of spatial translation}
Before closing this section,
we would like to make a comment on the case
when spatial translation, rather than time translation,
is broken in a mixed way with a global internal $U(1)$ symmetry.
One typical condensation pattern for such symmetry breaking
is given by
\begin{align}
\langle\phi(x)\rangle=r_0\,e^{iuz}
\quad
(r_0 \text{ and } u: \text{real constants})
\end{align}
in the complex scalar model of Sec.~\ref{subsec:complex}.
This kind of symmetry breaking are discussed in the context of dense QCD matter~\cite{Deryagin:1992rw,Shuster:1999tn,Park:1999bz,Rapp:2000zd,
Kojo:2009ha,Kojo:2010fe,Kojo:2011cn,Nickel:2009ke,Nickel:2009wj,Carignano:2010ac,Abuki:2011pf,Buballa:2014tba
}  for example.
The effective action for this symmetry breaking
can be constructed in a parallel way to the construction in Sec.~\ref{subsec:mixture_coonstruction}.
If we work in the nonrelativistic system and assume that Galilean symmetry does not exist from the beginning, the effective action is
\begin{align}
\label{nonreal_mixture}
S&=\frac{1}{2}\int d^4x
\left[\widetilde{\alpha}\dot{\pi}^2-\alpha\left(u^2+2u\partial_z\pi+(\partial_i\pi)^2\right)-\beta\left(2u\partial_z\pi+(\partial_i\pi)^2\right)^2\right]\,.
\end{align}
The important point is that the spatial total derivative term
in the action \eqref{nonreal_mixture}
contains a linear order term.
Just as we discussed in Sec.~\ref{section_multi},
such linear order terms in spatial total derivatives
are prohibited by the requirement that
the background energy is at a local minimum
in the configuration space.
We then have $\alpha=0$
and the dispersion relation of the NG mode $\pi(x)$
is schematically given by
\begin{align}
\label{dispersion_mixture}
\omega^2\sim c_1'k_z^2
+c_2'k_zk_\bot^2+c_3'k_\bot^4
\end{align}
up to the second order in $k_z$ and $k_\bot^2$,
where $c_i'$ are coefficients, and
the last two terms on the right-hand side
come from higher derivative terms in the effective action.
For example,
the NG mode in a model of the FFLO phase indeed accommodates
this type of dispersion relations~\cite{Radzihovsky, Radzihovsky:2011rf, Samokhin}.
We will revisit this issue via symmetry arguments
of the present paper elsewhere~\cite{spiralproject}.

\section{Coset construction revisited}\label{sec:CosetConstructionRevisited}
\setcounter{equation}{0}

In this section
we revisit the coset construction for spacetime symmetry breaking,
based on our discussion.
In the first two subsections,
we introduce nonlinear realization for broken spacetime symmetries~\cite{Volkov:1973vd, Ogievetsky},
and show that the parameterization of NG fields in nonlinear realization is closely related to the local symmetry picture.
In Sec.~\ref{subsec:ingredients_coset},
we summarize the general ingredients of the effective action
for spacetime symmetry breaking.
In particular,
we discuss the relation between the Maurer-Cartan one form
and connections for spacetime symmetries.
We also comment on the difference from the internal symmetry case.
In Sec.~\ref{subsec:inverse_coset},
we revisit the role of the inverse Higgs constraints,
focusing on the relation to our approach based on the local picture.
We classify the physical meaning of the inverse Higgs constrains
based on the coordinate dimension of broken symmetries.

\medskip
Throughout this section,
for simplicity,
we concentrate on symmetry breaking in the case of  the Minkowski space
and assume that the system originally enjoys
translation symmetries in all directions.

\subsection{Local decomposition of spacetime symmetries}
\label{subsec:local_coset}

In the first two subsections
we introduce nonlinear realization for broken spacetime symmetries
and discuss its properties.
For this purpose,
it is convenient to consider a local field $\Phi(x)$ that belongs to 
a linear irreducible representation of spacetime and internal symmetries,
which
can be related to a field $\Phi(0)$ at the origin by a translation:
\begin{align}
\label{Phi_x_to_0}
\Phi(x)=\Omega_P(x)\Phi(0)
\quad
{\rm with}
\quad
\Omega_P(x)=e^{x^mP_m}
\,,
\end{align}
where $P_m$ is the translation generator.
Correspondingly,
we can relate symmetry transformations of
$\Phi(x)$
to those of $\Phi(0)$.
For example,
we can rewrite the special conformal transformations of $\Phi(x)$ as\footnote{
We define the symmetry generators such that
\begin{align}
\nonumber
&[D,P_m]=-P_m\,,
\quad
[D,K_m]=K_m\,,
\quad
[K_m,P_n]=2(\eta_{mn}D+L_{mn})\,,
\\
&[L_{mn},P_\ell]=-\eta_{m\ell }P_n+\eta_{n\ell}P_m\,,
\quad
[L_{mn},K_\ell]=-\eta_{m\ell }K_n+\eta_{n\ell }K_m\,,
\quad
[L_{mn},L_{rs}]=-\eta_{mr}L_{ns}+\text{$3$ terms}\,,
\end{align}
and other commutators vanish.}
\begin{align}
b^m K_m\Phi(x)&=
\Omega_P(x)\left[(-2x^nx^\ell b_\ell +x^2b^n) P_n+2b^mx^nL_{mn}+2b_mx^m D+b^mK_m\right]\Phi(0)
\,,
\end{align}
where $L_{mn}$, $D$, and $K_m$
are generators of Lorentz transformations,
dilatations, and special conformal transformations, respectively.
Since the origin $x=0$ is invariant under Lorentz transformations,
dilatations, and special conformal transformations,
the last three terms in the brackets act linearly on $\Phi(0)$.
Moreover,
when $\Phi$ is a primary field,
the special conformal generator $K_m$ acts trivially on $\Phi(0)$,
\begin{align}
\label{K_decomposition_primary}
b^m K_m\Phi(x)&=
\Omega_P(x)\left[(-2x^nx^\ell b_\ell +x^2b^n) P_n+2b^mx^nL_{mn}+2b_mx^m D\right]\Phi(0)
\,.
\end{align}
It is then natural to identify
the last two terms in the brackets
as local Lorentz and local Weyl transformations
at the point $x$.
More explicitly,
one may rewrite~\eqref{K_decomposition_primary} as
\begin{align}
\label{K_decomposition_primary2}
b^m K_m\Phi(x)&=
(-2x^nx^\ell b_\ell +x^2b^n) P_n\Phi(x)+\Omega_P(x)\left[2b^mx^nL_{mn}+2b_mx^m D\right]\Omega_P^{-1}(x)\Phi(x)\notag\\
&=\Big[1+\Omega_P(\tilde{x})\big[2b^m\tilde{x}^nL_{mn}+2b_m\tilde{x}^m D\big]\Omega_P^{-1}(\tilde{x})\Big]\Phi(\tilde{x}) -\Phi(x)+\mathcal{O}(b^2)
\,,
\end{align}
where $\tilde{x}^n=x^n-2x^nx^\ell b_\ell +x^2b^n+\mathcal{O}(b^2)$.
The expression~\eqref{K_decomposition_primary2} corresponds to the local decomposition
of spacetime symmetries discussed in Sec.~\ref{sec:strategy}:
the first term is the Lorentz transformation
and dilatation around the point $x^m$.
Note that
the transformation $x\to \tilde{x}$ is identified with an inverse transformation of $x\to x'^n=x^n+2x^nx^\ell b_\ell -x^2b^n+\mathcal{O}(b^2)$
if we use the notation in Sec.~\ref{sec:strategy}.

\medskip
More generally,
the decomposition~\eqref{Phi_x_to_0}
allows us to express arbitrary spacetime symmetry transformations
in terms of diffeomorphisms, local Lorentz transformations,
and local (an)isotropic Weyl transformations,
just as we did in Sec.~\ref{sec:strategy}.
Suppose that the spacetime and internal symmetry algebra
contains symmetry generators with the coordinate dimension $n\geq0$
as well as the translation symmetry generators $\mathfrak{g}_P$.
Let us also introduce the Lie algebra $\mathfrak{g}_{n}$
of symmetry generators with the coordinate dimension $n$,
which satisfies the commutation relations
\begin{align}
\label{commtation_spacetime}
[\mathfrak{g}_{m}\,,\mathfrak{g}_{n}]
=\mathfrak{g}_{m+n}\,.
\end{align}
When the space and time coordinates
have the same scaling dimension,
$\mathfrak{g}_{n}$'s are schematically represented in the coordinate space as
\begin{align} 
\sim x^{m_1}x^{m_2}\ldots x^{m_{n+1}}\partial_m \,\,\quad\text{and}\qquad \sim x^{m_1}x^{m_2}\ldots x^{m_{n}}T_a \,,
\end{align}
and $\mathfrak{g}_P=\mathfrak{g}_{-1}$ in particular.
Here $T_a$ is a generator of Lie algebra.
Note that when the internal symmetry belongs to an abelian group, the spacetime and internal symmetry may mix.\footnote{
A typical example is the Galilean boost generator, which is expressed as $t\partial_m-x_m mT_0$. Here $m$ is the mass of the particle and $T_0$ is the abelian generator corresponding to the particle number.
See Appendix~\ref{Sec:nonrela} for details.}
Just as we did for infinitesimal transformations above,
we can rewrite any spacetime symmetry transformation $g$
in the form, 
\begin{align}
\label{local_decomposition_Omega_full}
g\Phi(x)
=\Omega_0({x'}^{-1}(x);g)\Omega_1({x'}^{-1}(x);g)\ldots\Phi({x'}^{-1}(x))\,,
\end{align}
where ${x'}^{-1}(x)$ is the inverse function of $x'(x)$ associated with the coordinate transformation, $x\to x'=x'(x)$.
$\Omega_n(x;g)$ is the element of  the Lie group $G$ with $\mathfrak{g}_n$ around the point $x$.
In general, $\Omega_n(x;g)$ depends on both $x$ and $g$.
When $\Phi$ is a primary field,
$\Omega_n(x;g)$ with $n\geq1$ acts on $\Phi(x)$ trivially,
so that we have
\begin{align} \label{gPhi}
g\Phi(x')
=\Omega_0(x;g)\Phi(x)\,.
\end{align}
We can then identify $\Omega_0(x;g)$ 
as the local Lorentz, local (an)isotropic Weyl, and  internal transformations.
In this way
the expression~\eqref{Phi_x_to_0}
provides the local decomposition of spacetime symmetries.

\subsection{Nonlinear realization}
\label{subsec:nonlinear}
We then introduce nonlinear realization for broken spacetime  and internal symmetries~\cite{Volkov:1973vd, Ogievetsky},
and discuss its relation to the local decomposition in the previous subsection.
Suppose that an original global symmetry group $G$
is broken to a subgroup $H$,
where $G$ and $H$
include both of internal and spacetime symmetries.
To construct the effective action,
it is convenient to
decompose the symmetry generators
as
\begin{align}
\mathfrak{g}
=\mathfrak{g}_P\oplus\widehat{\mathfrak{g}}
=\mathfrak{g}_P
\oplus\mathfrak{g}_{0}
\oplus\mathfrak{g}_{1}+\ldots\,,
\end{align}
where $\mathfrak{g}_P$ and $\widehat{\mathfrak{g}}$ are for translation and non-translational symmetry generators.
The non-translational part $\widehat{\mathfrak{g}}$
is made from subalgebras of
$\mathfrak{g}_{n}$ of spacetime and internal symmetry generators
with the coordinate dimension $n$.
We further decompose them into the residual symmetry parts $\mathfrak{h}$'s
and the broken symmetry parts $\mathfrak{m}$'s
as
\begin{align}
\mathfrak{g}=\mathfrak{h}\oplus \mathfrak{m}\,,
\quad
\mathfrak{g}_P=\mathfrak{h}_P\oplus \mathfrak{m}_P\,,
\quad
\widehat{\mathfrak{g}}=\widehat{\mathfrak{h}}\oplus \widehat{\mathfrak{m}}\,,
\quad
\mathfrak{g}_n=\mathfrak{h}_n\oplus \mathfrak{m}_n\,\,(n\geq0)\,.
\end{align}
In contrast to the internal symmetry case, Eq.~\eqref{Lie_algebra_internal},
we assume that
\begin{align}
[\,\widehat{\mathfrak{h}}\,,\mathfrak{g}_P\oplus  \widehat{\mathfrak{m}}\,]=\mathfrak{g}_P\oplus  \widehat{\mathfrak{m}}
\end{align}
rather than $[\,\mathfrak{h}\,,\mathfrak{m}\,]=\mathfrak{m}$ in the following.\footnote{
In most of the symmetry breaking patterns in our interests,
$[\,\widehat{\mathfrak{h}}\,,\widehat{\mathfrak{m}}\oplus \mathfrak{g}_P\,]=\widehat{\mathfrak{m}}\oplus \mathfrak{g}_P$, but $[\,\mathfrak{h}\,,\mathfrak{m}\,]\neq\mathfrak{m}$.
For example,
when rotation symmetries are broken,
we have $[\,\mathfrak{h}_{P}\,,\mathfrak{m}\,]=\mathfrak{h}_P$.}
In this case, we need to employ all the translation generators in addition to broken ones
for parametrizing the coordinate of the coset space.
Correspondingly,
we use representatives of the coset $G/\widehat{H}$
rather than $G/H$
to realize the original symmetry group $G$~\cite{Volkov:1973vd, Ogievetsky}:
\begin{align}
\label{Omega_spacetime_coset}
\Omega=\Omega_P\Omega_{0}\Omega_{1}\ldots
\quad
{\rm with}
\quad
\Omega_P=e^{Y^m(\Mc{x}) P_m}\,,
\quad
\Omega_n=e^{\pi_n(\Mc{x})}
\,\,(n\geq0)\,,
\end{align}
where $\pi_n(\Mc{x})\in\mathfrak{m}_n$
are NG fields
for broken non-translational symmetries
and $x^m(\Mc{x})$'s are the Minkowski coordinates.
We note that $\Mc{x}^\mu$'s are not the Minkowski coordinates,
but rather the unitary gauge coordinates, as we will see.\footnote{In this section, we employ the bar symbol for the unitary gauge coordinates
to distinguish those with Minkowski's.}
One useful choice of the unitary coordinate is\footnote{
If we are not interested in dynamics in $\Mc{x}^a$-directions,
we set $Y^m(\Mc{x})=(\Mc{x}^{\widehat{m}},\pi^a)$
and define NG fields as $\Mc{x}^{a}$-independent fields.
Though such a simplification is often performed in the literature,
we keep $\Mc{x}^a$-dependence for generality.}
\begin{align}
\label{def_of_Y^m}
Y^{\widehat{m}}(\Mc{x})=\Mc{x}^{\widehat{m}}\,,
\quad
Y^a(\Mc{x})=\Mc{x}^a+\pi^a(\Mc{x})\,,
\end{align}
where
the indices $\widehat{m}$ and $a$
denote directions with and without translation invariance, respectively,
and $\pi^a$'s are NG fields for broken translation symmetries.
Under a global left $G$ transformation, the representative transforms as 
\begin{equation}
\begin{split}\label{transformation_coset}
\Omega(Y,\pi)\to \Omega(Y',\pi')=g \Omega(Y,\pi) h^{-1}(\pi,g).
\end{split}
\end{equation}
For the translation $x\to x'+a$,  the NG fields transform as $Y'^m(\Mc{x})=Y^m(\Mc{x})-a^m$, and $\pi_n'(\Mc{x})=\pi_n(\Mc{x})$.
In general, the NG fields $\pi_n'$ transform nonlinearly.
Here we notice that the expression~\eqref{Omega_spacetime_coset}
takes a similar form as the local decomposition~\eqref{local_decomposition_Omega_full} of spacetime symmetries.
Indeed,
from the global left $G$ transformation property of $\Omega$,
it turns out that
NG fields $\pi_n$'s are identified with
transformation parameters for
$\mathfrak{g}_n$ transformations around the point $Y^m(\Mc{x})$.
In particular,
$\pi_0(\Mc{x})$ should be understood as
NG fields for local Lorentz and local (an)isotropic Weyl transformations,
rather than those for global ones.
Also,
$\pi_n$'s with $n\geq1$ correspond to redundant NG fields
because primary fields at a point $Y^m(\bar{x})$
is invariant under the $\mathfrak{g}_n$ transformations
around the same point $Y^m(\bar{x})$ for $n\geq1$.

\medskip
Such identification can be also 
understood from the Maurer-Cartan one form
\begin{align}
\label{MC_spacetime}
J_\mu=\Omega^{-1}\partial_\mu \Omega
=\widehat{\Omega}^{-1}\left(\partial_\mu Y^mP_m\right)\widehat{\Omega}
+\widehat{\Omega}^{-1}\partial_\mu\widehat{\Omega}
\quad
{\rm with}
\quad
\widehat{\Omega}=\Omega_{0}\Omega_{1}\ldots\,,
\end{align}
where we used $[\,\widehat{\mathfrak{g}}\,,\widehat{\mathfrak{g}}\,]=\widehat{\mathfrak{g}}$.
In the coset construction,
its $\mathfrak{g}_p$-component is used as the vierbein:
\begin{align}
\label{vierbein_local}
e_\mu^m=\left[J_\mu\right]_{P_m}
=\big[\,\widehat{\Omega}^{-1}\left(\partial_\mu Y^nP_n\right)\widehat{\Omega}\,\big]_{P_m}
=\left[\,\Omega_0^{-1}\left(\partial_\mu Y^nP_n\right)\Omega_0\,\right]_{P_m}\,,
\end{align}
where $[\,A\,]_{P_m}$ denotes the $P_m$-component of $A$
and we used $[\,\mathfrak{g}_m\,,\mathfrak{g}_n\,]=\mathfrak{g}_{m+n}$ at the last equality.
We notice that
the vierbein~\eqref{vierbein_local} depends on $Y^m$ and $\Omega_0$ only,
and it is independent of $\Omega_n$'s with $n\geq1$.
Also $\Omega_0$ just transforms the vierbein
without changing the Minkowski coordinate $Y^m$.
Such properties are again
consistent with the interpretation
that $\Omega_0$ represents NG fields for local symmetry transformations
and
$\Omega_n$'s with $n\geq1$ do not generate physical degrees of freedom.

\medskip
To summarize,
the representative of the coset space~\eqref{Omega_spacetime_coset}
is closely related to the local symmetry picture, rather than the global one.
In particular,
all the NG fields can be described by $Y^m$, $\Omega_{0}$ only,
and their identification should be based on the local picture.

\subsection{Ingredients of the effective action}
\label{subsec:ingredients_coset}
We next take a closer look at the ingredients of the effective action
based on the identification of NG fields in the previous subsection.
For simplicity,
we concentrate on the relativistic case
in the following.
Suppose that local fields $\Phi^A$'s have background condensations,
\begin{align}
\langle\Phi^A(x)\rangle=\bar{\Phi}^A(x)\,,
\end{align}
where $A$ denotes both the internal and local Lorentz indices
and $\bar{\Phi}^A$'s are spacetime-dependent in general.
Just as we usually do for internal symmetry breaking (see, e.g., Ref.~\cite{WeinbergText}),
let us decompose $\Phi^A$
into the NG field part and the matter part $\widetilde{\Phi}^A$ as
\begin{align}
\label{phi_parametrization}
\Phi^A(x) 
= \Omega_P(\pi)\left[{\Omega_0(x,\pi_0)}\right]^A{}_B\tilde{\Phi}^{B}(x) \,,
\end{align}
where 
$\Omega_0(x,\pi_0)=\Omega_{\rm int}\Omega_L\Omega_D$ with
$\Omega_{\rm int}$, $\Omega_L$, and $\Omega_D$
being representatives for broken internal, local Lorentz, and local Weyl symmetries,
respectively.\footnote{
Though we keep $\Omega_{\rm int}$, $\Omega_L$, and $\Omega_D$
in our discussions,
we turn on NG fields only for broken symmetries.
For example, if the Lorentz symmetry is not broken,
we set $\Omega_L=1$.}
Note that $\Omega_n$'s with $n\geq1$ do not appear here
because they are redundant NG fields and do not transform $\Phi^A$'s.
It is also useful to introduce the unitary gauge coordinate $\Mc{x}^\mu(x)=x^\mu+\pi^\mu(x)$ and rewrite Eq.~\eqref{phi_parametrization} to
\begin{align}
\Phi^A(x) 
= \left[{\Omega_0(\bar{x},\pi_0)}\right]^A{}_B\tilde{\Phi}^{B}(\bar{x}) \,.
\end{align}
From Eq.~\eqref{gPhi}, $\Phi'^A(x)$ transforms under $G$ as 
\begin{align}
\Phi'^A(x') = g\Phi^A(x')&=
 \left[\Omega_0(\Mc{x}',g) {\Omega_0(\Mc{x}',\pi_0)}\right]^{A}_{~~B} \tilde{\Phi}^B(\Mc{x}') \notag\\
&= \left[ {\Omega_0(\Mc{x}',\pi'_0(\pi,g)})\right]^{A}_{~~B} \tilde{\Phi}'^B(\Mc{x}') 
\end{align}
with
$\Mc{x}'(x')=\Mc{x}(x)$, $\tilde{\Phi}'^A(\Mc{x}')=[h(\pi,g)]^A_{~~B}\tilde{\Phi}^B(\Mc{x})$,
which follow the same transformation rule as those of the coset construction \eqref{transformation_coset}.
In this unitary gauge coordinate, the Minkowski coordinate is expressed by the inverse function $Y^m(\Mc{x})$ such that $Y^m(\Mc{x}(x))=x^m$.
Then, the vierbein $e_\mu^m$
is given by
\begin{align}
\label{vierbein_unitary_1}
e_\mu^m=\partial_\mu Y^m(\Mc{x})\,.
\end{align}
Just as the internal case,
the Maurer-Cartan type one form arises from the derivative of $\Phi^A$
as
\begin{align}
\label{MC_no_redundant}
\partial_\mu\Phi^A=\left[\Omega_{\rm int}\Omega_L\Omega_D\right]^A{}_B\left[\left(\Omega_{\rm int}^{-1}\partial_\mu\Omega_{\rm int}+\Omega_L^{-1}\partial_\mu\Omega_L+\Omega_D^{-1}\partial_\mu\Omega_D\right)^B{}_C\,\widetilde{\Phi}^C+\partial_\mu\widetilde{\Phi}^B\right]\,,
\end{align}
which suggests that the three terms in the parentheses
play the role of connections.
Indeed,
if we introduce the internal gauge field $A_\mu$ and
the Weyl gauge field $W_\mu$,
the inverse local transformation $(\Omega_{\rm int}\Omega_L\Omega_D)^{-1}$
maps the configuration $A_\mu=W_\mu=0$
to the configuration
\begin{align}
\label{MC_to_connections1}
A_\mu=\Omega_{\rm int}^{-1}\partial_\mu \Omega_{\rm int}\,,
\quad
W_\mu=\Omega_D^{-1}\partial_\mu \Omega_D\,.
\end{align}
Similarly,
the vierbein~\eqref{vierbein_unitary_1} and the corresponding
spin connection $S_\mu=\frac{1}{2}S_\mu^{mn}L_{mn}$
are mapped to
\begin{align}
\label{MC_to_connections2}
e_\mu^m=\left[\Omega_D^{-1}\Omega_L^{-1}(\partial_\mu Y^mP_m)
\Omega_L\Omega_D\right]_{P_m}\,,
\quad
S_\mu=\Omega_L^{-1}\partial_\mu\Omega_L
+\frac{1}{2}\left(e^m_\mu e^n_\nu-e^n_\mu e^m_\nu\right)
W^\nu L_{mn}\,.
\end{align}
Note that,
via those identifications,
Eq.~\eqref{MC_no_redundant} can be reduced to
the Weyl covariant derivative~\eqref{Weyl_covariant_der}~as
\begin{align}
\partial_\mu\Phi^A=\left[\Omega_{\rm int}\Omega_L\Omega_D\right]^A{}_BD_\mu\widetilde{\Phi}^B\,.
\end{align}
Also note that the vierbein coincides with
the $P_m$-component~\eqref{vierbein_local}
of the Maurer-Cartan one form.
In this way,
the Maurer-Cartan one form~\eqref{MC_spacetime}
with $\Omega_n=1$ ($n\geq1$)
can be identified with connections and the vierbein.

\medskip
Just as the internal symmetry case,
the above decomposition provides ingredients of the effective action.
To illustrate the difference from the internal symmetry case,
let us consider constructing the effective action
for NG fields without matter fields,
whose ingredients can be obtained by
setting that $\widetilde{\Phi}^A=\bar{\Phi}^A$.
The original fields $\Phi^A$'s and their derivatives
are then given by
\begin{align}
\Phi^A&=[\Omega_{\rm int}\Omega_L\Omega_D]^A{}_B\,\bar{\Phi}^B\,,
\\*
\partial_\mu\Phi^A&=[\Omega_{\rm int}\Omega_L\Omega_D]^A{}_B\left[
(\Omega_{\rm int}^{-1}\partial_\mu\Omega_{\rm int}+\Omega_L^{-1}\partial_\mu\Omega_L+\Omega_D^{-1}\partial_\mu\Omega_D)^B{}_C\,\bar{\Phi}^C
+\partial_\mu\bar{\Phi}^B\right]\,.
\end{align}
One difference from the internal symmetry case is that
local Lorentz indices can be coupled to the translation generator $P_m$
at the same time as they are representations of local Lorentz symmetry.
For example,
when the condensation has a local Lorentz index, $\bar{\Phi}^n$,
there can be a coupling of the form
\begin{align}
{\rm tr}\left[(e_\mu^mP_m) (\Phi^nP_n)\right]\,,
\end{align}
whose expression
after the inverse local transformation $(\Omega_{\rm int}\Omega_L\Omega_D)^{-1}$
is given by
\begin{align}
{\rm tr}\left[(e_\mu^mP_m) (\bar{\Phi}^nP_n)\right]=e_\mu^m\bar{\Phi}_m
\quad
{\rm with}
\quad
e_\mu^m=[\Omega_D^{-1}\Omega_L^{-1}(\partial_\mu Y^mP_m)
\Omega_L\Omega_D]_{P_m}\,,
\end{align}
where the trace for $P_m$ is defined as ${\rm tr}\left[P_m P_n\right]=\eta_{mn}$.
Another difference is that $\bar{\Phi}^A$
can be spacetime-dependent.
For example,
when the condensation is inhomogeneous in the $z$-direction,
$\bar{\Phi}^A(\Mc{z})$,
we obtain functions of $\Mc{z}$ from terms without derivatives like
\begin{align}
\Phi^A\Phi_A\to \bar{\Phi}^A(\Mc{z})\bar{\Phi}_A(\Mc{z})\,.
\end{align}
Similarly,
the derivative $\partial_\mu\Phi^A$ leads to
functions of $\Mc{z}$ and their derivatives
as well as the Maurer-Cartan type one form
\begin{align}
\partial_\mu\Phi^A&=[\Omega_{\rm int}\Omega_L\Omega_D]^A{}_B\left[
(\Omega_{\rm int}^{-1}\partial_\mu\Omega_{\rm int}+\Omega_L^{-1}\partial_\mu\Omega_L+\Omega_D^{-1}\partial_\mu\Omega_D)^B{}_C\,\bar{\Phi}^C
+\delta_\mu^{\Mc{z}}\partial_\Mc{z}\bar{\Phi}^B\right]\,.
\end{align}
With those modifications to the internal symmetry case,
the general effective action can be constructed
from the one forms $\Omega_{\rm int}^{-1}\partial_\mu\Omega_{\rm int}$,
$\Omega_L^{-1}\partial_\mu\Omega_L$,
and $\Omega_D^{-1}\partial_\mu\Omega_D$,
the vierbein $e_\mu^m$, and
functions of coordinates in the inhomogeneous directions.
Note that the volume element also contains NG field through the determinant of vierbein.
Since those one forms are related to the connections
$S_\mu$, $W_\mu$, and $A_\mu$,
it is obvious that
those ingredients are the same as the ones
in the approach based on the local picture.

\subsection{Inverse Higgs constraint}
\label{subsec:inverse_coset}

In the coset construction for spacetime symmetry breaking,
one imposes the so-called inverse Higgs constraints
to remove the redundant NG fields~\cite{Ogievetsky,Ivanov:1975zq}
and the massive degrees of freedom~\cite{Nicolis:2013sga,Endlich:2013vfa,Brauner:2014aha}.
For a broken (global) symmetry generator $A\in {\mathfrak{m}}$,
we compute its commutator with the translation generator $P_m$,
which contains both the broken and unbroken symmetry generators in general:
\begin{align}
[P_m,A]\sim B+C
\quad
{\rm with}
\quad
B\in {\mathfrak{m}}\,,
\quad
C\in {\mathfrak{h}}\,.
\end{align}
When the commutator contains broken symmetry generators,
$B\neq0$,
we remove the NG field for $A$
by imposing a certain constraint
in a consistent way with the symmetry structure.
Typically, we require that the $B$-component of the Maurer-Cartan one form is zero:
\begin{align}
\label{inverse_Higgs}
[\Omega^{-1}\partial_\mu\Omega]_B=0\,,
\end{align}
which generically relates the NG field for $A$
to a derivative of the NG field for $B$.
The effective action is then constructed from the Maurer-Cartan one form
with the condition~\eqref{inverse_Higgs} imposed.
At the end of this section,
we revisit the role of such inverse Higgs constraints
and redundant NG fields,
focusing on their counterparts in the approach based on the local symmetry viewpoint.
In particular,
we show that its physical meaning
is different between the case $A\in \mathfrak{g}_{(n)}$ with $n\geq1$
and the case $A\in \mathfrak{g}_{(0)}$.

\subsubsection{Redundant NG fields for special conformal symmetry}
\label{subsub:redundant}
An illustrative example for the first case
is the redundant NG fields for special conformal symmetry~\cite{Ogievetsky,Ivanov:1975zq,Low:2001bw,Isham:1970gz,Salam:1970qk}.
Suppose that the conformal symmetry group is broken to its subgroup.
To perform the coset construction,
let us first classify the symmetry generators by the coordinate dimension as
\begin{align}
\mathfrak{g}_{-1}=\{ P_m\}\,,
\quad
\mathfrak{g}_{0}=\{ L_{mn}\,,D\}\,,
\quad
\mathfrak{g}_{1}=\{ K_m\}\,.
\end{align}
Based on this classification,
we introduce the representative of the coset space $\Omega$ as
\begin{align}
\Omega=e^{Y^mP_m}\Omega_0\Omega_1
\quad
{\rm with}
\quad
\Omega_0=\Omega_L\Omega_D\,,
\quad
\Omega_1=\Omega_K\,.
\end{align}
Here $\Omega_K=e^{\chi^m K_m}$ describes NG fields for special conformal transformations,
which should be interpreted as redundant ones
as we have discussed.
We then calculate the corresponding Maurer-Cartan one form
$J_\mu=\Omega^{-1}\partial_\mu\Omega$.
First, its $\mathfrak{g}_P$-component
is the vierbein
\begin{align}
\label{vierbein_conformal}
e_\mu^mP_m=\Omega_0^{-1}\partial_\mu Y^mP_m\Omega_0\,.
\end{align}
On the other hand,
the $\mathfrak{g}_{0}$-component is given by
\begin{align}
\Omega_0^{-1}\partial_\mu\Omega_0
-[\chi^mK_m,e^n_\mu P_n]
=
\Omega_L^{-1}\partial_\mu \Omega_L
+\Omega_D^{-1}\partial_\mu\Omega_D
-2\chi_me^m_\mu D-2(\chi^me^n_\mu)L_{mn}\,,
\end{align}
which is reduced to the connections in~\eqref{MC_no_redundant}
if we set $\chi^m=0$.
The $\mathfrak{g}_{1}$-component is given by
\begin{align}
\Big[\partial_\mu\chi^m+\chi^2 e_\mu^m
+\left[\Omega_L^{-1}\partial_\mu\Omega_L\right]_{L_{mn}} \chi_n
+\left(\left[\Omega_D^{-1}\partial_\mu\Omega_D\right]_D-2\chi_ne^n_\mu\right)\chi^m\Big]K_m
\,,
\end{align}
which vanishes when $\chi^m=0$.
Using the relations~\eqref{MC_to_connections1} and~\eqref{MC_to_connections2},
we can rearrange the $\mathfrak{g}_{0}$- and $\mathfrak{g}_{1}$-components
in terms of the spin connection $S_\mu^{mn}$
and the Weyl gauge field $W_\mu$ as
\begin{align}
\label{MC_L}
[J_\mu]_{L_{mn}}&=S_\mu^{mn}-(e_\mu^mW^n-e_\mu^nW^m)
+2(e_\mu^m\chi^n-e_\mu^n\chi^m)\,,
\\
\label{MC_D}
[J_\mu]_D&=W_\mu-2\chi_\mu\,,
\\
\label{MC_K}
[J_\mu]_{K_m}&=
\nabla_\mu\chi^m
+W^m\chi_\mu
+(\chi^2-W^\nu\chi_\nu) e_\mu^m
+\left(W_\mu-2\chi_\mu\right)\chi^m\,,
\end{align}
where $\nabla_\mu\chi^m=\partial_\mu\chi^m+S_\mu^{mn}\chi_n$,
and the local Minkowski indices and the global coordinate indices
are converted to each other by the vierbein~\eqref{vierbein_conformal}
as $\chi_\mu=e^m_{\mu}\chi_m$ and $W^m=e^m_{\mu}W^\mu$.

\medskip
We now discuss the role of inverse Higgs constraints.
Suppose that the special conformal symmetry is broken
and, for simplicity,
let us assume that the translation symmetry is unbroken.
The commutator relevant to inverse Higgs constraints is then
\begin{align}
[P,K]\sim D+L\,.
\end{align}
Since at least one of the dilatation symmetry and the Lorentz symmetry
is broken if the special conformal symmetry is broken,
we remove the NG field $\chi^m$ for the special conformal transformation
by imposing the inverse Higgs constraint.
This statement corresponds to the fact
that $\chi^m$ is a redundant NG field
and does not generate physical degrees of freedom~\cite{Ogievetsky,Ivanov:1975zq,Low:2001bw}.
We then take a closer look at the inverse Higgs constraints
in the following two cases.

\begin{enumerate}
\item Broken dilatation and unbroken Lorentz.

Let us first consider the case
when the dilatation symmetry is broken,
but the Lorentz symmetry is unbroken.
In this case,
the inverse Higgs constraints~\cite{Ivanov:1975zq} can be stated as
\begin{align}
[J_\mu]_D=0
\quad
\leftrightarrow
\quad
\chi_\mu=\frac{1}{2}W_\mu\,.
\end{align}
Using this constraint,
the $L_{mn}$-component~\eqref{MC_L}
is reduced to the spin connection
\begin{align}
[J_\mu]_{L_{mn}}&=S_\mu^{mn}
\,.
\end{align}
On the other hand,
the $K_m$-component~\eqref{MC_K} becomes
\begin{align}
\label{K_m_simplified}
[J_\mu]_{K_m}&=
e^{m\nu}\frac{1}{2}\left(\nabla_\mu W_\nu
+W_\mu W_\nu
-\frac{1}{2}g_{\mu\nu}
W^2\right)\,.
\end{align}
As discussed in Ref.~\cite{Iorio:1996ad},
the Weyl transformations of the combination in the parentheses
can be related to those of the Ricci tensor $R_{\mu\nu}$ as
\begin{align}
\Delta\left[\nabla_{\mu}W_\nu+W_\mu W_\nu-\frac{1}{2}g_{\mu\nu}W_\rho W^\rho\right]
=\Delta\left[ \frac{1}{2-d}\left(R_{\mu\nu}-\frac{1}{2(d-1)}g_{\mu\nu}R\right)\right]\,,
\end{align}
where $\Delta$ denotes Weyl transformations
and $d$ is the spacetime dimension. 
Since the metric constructed from
the vierbein~\eqref{vierbein_conformal} is conformally flat,
we can further rewrite~\eqref{K_m_simplified}
in terms of the Ricci tensor as
\begin{align}
[J_\mu]_{K_m}
=\frac{e^{m\nu}}{2(2-d)}\left(R_{\mu\nu}-\frac{1}{2(d-1)}g_{\mu\nu}R\right)\,.
\end{align}
In this way,
the $K_m$-component
reproduces the Ricci tensor in the unitary gauge~\cite{Hinterbichler:2012mv}.
To summarize,
the $L_{mn}$- and $K_m$- components
of the Maurer-Cartan one form
reproduce the spin connection and the Ricci tensor
and we have a vanishing $D$-component.
In particular,
the Weyl gauge field $W_\mu$ does not appear explicitly.

\medskip
This is indeed consistent with the local symmetry picture.
As we mentioned earlier,
any conformal system on the Minkowski space
can be reformulated in a local Weyl invariant way
by introducing an appropriate curved spacetime action (the Ricci gauging).
The unitary gauge effective action
should then be written without using Weyl gauge fields
explicitly.

\medskip
\item Broken Lorentz and broken dilatation.

We next
consider the case when
both the dilatation and the Lorentz symmetry are broken.
For concreteness,
let us assume that the Lorentz symmetry associated with the $3$-direction,
i.e., $L_{3\widehat{n}}=-L_{\widehat{n}3}$ with $\widehat{n}\neq3$,
is broken.
We now have two types of
inverse Higgs constraints:
\begin{align}
[J_\mu]_{L_{mn}}&=S_\mu^{mn}-(e_\mu^mW^n-e_\mu^nW^m)
+2(e_\mu^m\chi^n-e_\mu^n\chi^m)=0\,,
\quad
[J_\mu]_D&=W_\mu-2\chi_\mu=0\,.
\end{align}
Since the role of the inverse Higgs constraints here
is to remove redundant NG fields consistently,
we do not have to impose both conditions.
Indeed,
the global transformation does not
mix the two constraints,
so that we can impose one of them alone.
By imposing the second condition
\begin{align}
[J_\mu]_D=0
\quad
\leftrightarrow
\quad
\chi_\mu=\frac{1}{2}W_\mu\,,
\end{align}
the other components of the Maurer-Cartan one form
can be reduced to
\begin{align}
[J_\mu]_{P_m}=e_\mu^m\,,
\quad
[J_\mu]_{L_{mn}}=S_\mu^{mn}\,,
\quad
[J_\mu]_{K_m}=\frac{e^{m\nu}}{2(2-D)}\left(R_{\mu\nu}-\frac{1}{2(D-1)}g_{\mu\nu}R\right)\,.
\end{align}
Just as the first example,
the inverse Higgs constraint guarantees that
the Weyl gauge field does not appear explicitly
in the unitary gauge effective action.
\end{enumerate}

To summarize,
the role of inverse Higgs constraints of this type
is to remove redundant NG fields.
In particular,
in the relativistic case,
they are closely related to
whether the original system
permits the Ricci gauging or not.
Correspondingly,
the inverse Higgs constraints
convert the Weyl gauge fields
into the Ricci tensors,
so that the obtained action does not contain Weyl gauge fields explicitly.

\subsubsection{Single brane revisited}
\label{subsub:single_revisit}
An illustrative example
for the second case is the single brane,
which we discussed in Sec.~\ref{sec:single}.
From the global symmetry point of view,
all the examples there are characterized by
the symmetry breaking from the $(1+3)$-dimensional Poincar\'e symmetry
to the $(1+2)$-dimensional one.
In the coset construction
we then introduce
NG fields for both the broken translation and broken (global) Lorentz symmetries.
The representative of the coset space and the nonzero components of the Maurer-Cartan one form are
\begin{align}
\Omega=\Omega_P\Omega_L\,,
\quad
[J_\mu]_{P_m}=e_\mu^m\,,
\quad
[J_\mu]_{L_{mn}}=S_\mu^{mn}\,.
\end{align}
Notice that, accompanied by functions of $z$ and matters,
general ingredients of the effective action for nonzero spin branes
in Sec.~\ref{subsec:inclusion}
can be obtained.
As we have discussed,
the NG fields in the nonlinear realization
are identified with the local symmetry transformation parameters
and they generate physical degrees of freedom only when the corresponding local symmetries are broken.
Therefore, NG fields for the Lorentz symmetries
are physical ones for the nonzero spin branes,
but redundant ones for the scalar branes
in this construction.
It is in a sharp contrast to the first case discussed in Sec.~\ref{subsub:redundant},
where NG fields for higher dimensional generators are always redundant ones.

\medskip
We next discuss the role of inverse Higgs constraints.
The relevant commutators here are those of broken Lorentz symmetry generators 
and translation generators given by
\begin{align}
[P_m,L_{3\widehat{n}}]=\eta_{3m}P_{\widehat{n}}-\eta_{\widehat{n}m}P_{3}\,,
\end{align}
which contains the broken generator, $P_3$, on the right hand side.
This commutator suggests that
the mass term of NG fields for Lorentz symmetries
and their mixing interaction with NG fields for diffeomorphisms
can be constructed from the $P_3$ component of the Maurer-Cartan one form~\cite{Nicolis:2013sga,Endlich:2013vfa,Brauner:2014aha}.
In our construction,
there are several options for the inverse Higgs constraints:\footnote{In particular the conditions are different from Eq.~\eqref{inverse_Higgs}.
It is because we chose the unitary gauge coordinate as Eq.~\eqref{def_of_Y^m}
and the Maurer-Cartan one form does not vanish even if NG fields vanish, $\pi=0$.}
\begin{align}
\label{two_options}
[J_\mu]_{P_3}=e_\mu^3=n_\mu\,,
\quad
[J_\mu]_{P_3}=e_\mu^3=\delta_\mu^3\,,
\end{align}
where $\displaystyle n_\mu=\frac{\delta^z_\mu}{\sqrt{g^{zz}}}$ is a unit vector perpendicular to the brane.
Both of the conditions are satisfied by the background configuration,
$\pi=0$,
and are also consistent with the original symmetry.
Also,
the second condition is equivalent to a combination of the first one, $e_\mu^3=n_\mu$,  and $g^{zz}=1$.
Finally, let us illustrate their physical interpretations:

\begin{enumerate}
\item The condition $e_\mu^3=n_\mu$.

This condition provides three constraints
that make three NG fields for local Lorentz symmetries freeze out.
Indeed,
it exactly coincides with the procedure
in Sec.~\ref{subsec:qualitative}
to integrate out the massive Lorentz NG fields,
because the interaction~\eqref{S_PL_unitary}
leads to the constraint $n^\mu e_\mu^3=1\leftrightarrow e_\mu^3=n_\mu$
in the low energy limit.
It should be noticed that
the removed NG fields are physical massive ones
for the nonzero spin brane case,
but redundant ones for the scalar brane case.

\item The condition $g^{zz}=1$.

It is nothing but the condition~\eqref{g^zz=1}
to remove the gapped modes in the diffeomorphism NG field.
The resulting effective action then turns out to be
the Nambu-Goto action for the gapless NG mode localizing on the brane.
Note that
the ambiguity in~\eqref{two_options}
corresponds to the choice whether we integrate out
the gapped modes in the diffeomorphism NG field or not.

\end{enumerate}
In this way
the inverse Higgs constraints for nonzero spin/scalar branes
remove massive/redundant NG fields for Lorentz symmetries
and gapped modes in the diffeomorphism NG field.

\medskip
It might be useful to note that
the general effective action for
diffeomorphism symmetry breaking
in Sec.~\ref{subsec:scalar_domain}
can be constructed
without introducing redundant NG fields
for Lorentz symmetries.
Consider the following representative of the coset space and the corresponding Maurer-Cartan one form:
\begin{align}
\Omega=\Omega_P\,,
\quad
[J_\mu]_{P_m}=e_\mu^m\,.
\end{align}
Ingredients of the effective action
are then this Maurer-Cartan one form,
functions of the coordinate $z$,
matter fields,
and their covariant derivatives.
It is obvious that
those ingredients reproduce the general ingredients
discussed in Sec.~\ref{subsec:scalar_domain}.
We can then construct the effective action for scalar branes
before integrating out gapped modes for example.

\bigskip
To summarize,
the conventional inverse Higgs constraints can be classified
into the following three types
by their physical meanings:
\begin{enumerate}
\item When spacetime symmetries of the coordinate dimension $n\geq1$,
$\mathfrak{g}_n$ with $n\geq1$,
are broken,
the role of inverse Higgs constraints is to remove redundant NG fields.
In particular,
in the relativistic systems, it is closely related to whether the original system permits the Ricci gauging~or~not.

\item When global spacetime symmetries of the coordinate dimension $0$, $\mathfrak{g}_0$, are broken as well as translation symmetries,
we introduce NG fields for $\mathfrak{g}_0$
in the coset construction.
However,
if the broken local symmetries are only diffeomorphisms,
NG fields for $\mathfrak{g}_0$ are redundant ones
and the inverse Higgs constraints remove them.
Also
we do not necessarily have to introduce NG fields for $\mathfrak{g}_0$
in our construction,
as long as we include gapped modes
in the effective action.

\item On the other hand,
when local Lorentz or local (an)isotropic Weyl symmetries are broken
(as well as diffeomorphism symmetries),
the corresponding physical NG fields acquire a mass.
Under certain conditions,
the inverse Higgs constraint can be identified with the procedure
to take the low-energy limit and integrate out massive NG fields.
\end{enumerate}

\section{Application to gravitational systems}
\setcounter{equation}{0}
\label{EFT_gravity}
Before closing this paper,
we would like to make a brief comment on the applications of our formulation
to gravitational systems.
As we mentioned in the introduction,
the EFT approach for inflation~\cite{Cheung:2007st}
is based on the symmetry argument in the unitary gauge.
In the unitary gauge, the
relevant degrees of freedom in single field inflation
are the metric $g_{\mu\nu}$ only
and the residual symmetries
are the time-dependent spacial diffeomorphisms.
This setup is essentially the same as the scalar branes
and indeed our discussion in Sec.~\ref{subsec:scalar_domain}
is parallel to that of
 reference~\cite{Cheung:2007st}:
the only differences between the two cases
are the background spacetime and whether we decouple the gravity sector or not.
By keeping the gravity sector without decoupling,
we can apply our strategy in Sec.~\ref{sec:strategy}
to gravitational systems.
We will apply our approach to gravitational systems
such as inflationary models with different symmetry breaking patterns elsewhere.

\section{Summary}
\setcounter{equation}{0}

In this paper
we discussed the EFT approach
for spacetime symmetry breaking
from the local symmetry point of view.
The identification of NG fields
and the construction of the effective action
are based on the local picture of symmetry breaking, 
i.e., the breaking of diffeomorphism, local Lorentz, and (an)isotropic Weyl symmetries
as well as the internal symmetries including possible central extensions
in nonrelativistic systems. This picture distinguishes, e.g., whether the condensations have Lorentz charges (spins),
while the standard coset construction based on the global symmetry breaking picture with the inverse Higgs constraints does not.
The distinction enable us to provide a correct identification of the physical NG fields because
they are generated by local transformations of condensations.

\medskip
In order to illustrate the difference between the global and local pictures
of spacetime symmetry breaking, in Sec.~\ref{sec:single}, we discussed codimension one branes,
in which global translation and rotation symmetries are broken. In the global picture, the low energy degrees of freedom is the NG field for the broken translation.
In the local picture, these degrees of freedom correspond to the NG fields for the broken diffeomorphism.
For scalar branes, both picture gives the same EFT. However, the situation is different for nonzero spin branes.
In this case the condensation has a spin, so that, in addition to NG field for diffeomorphism breaking, there appear
{\it massive} NG fields associated with local Lorentz symmetry breaking as the physical degrees of freedom, which nonlinearly transform under global broken symmetry.
One might think such massive modes are irrelevant in the low energy EFT.
This is true for the EFT at the low energy scale compared with the mass, and the EFT will be the same as that in the global picture.
However, when the scale of order parameters for translation and rotation breaking have different scales,
the mass could be smaller than other typical mass scales of the system, and thus, the massive modes may become relevant as the low-energy degrees of freedom.
For example in cosmology,
massive fields with the Hubble scale mass
affect the cosmological perturbations
(see, e.g., Refs.~\cite{Chen:2009zp,Baumann:2011nk,Noumi:2012vr} for recent discussions),
so that
massive modes associated with
symmetry breaking can be relevant
when they have a mass less than or comparable to
the Hubble scale.

\medskip
In Secs.~\ref{section_multi} and~\ref{Sec:mix},
we also discussed a system in which the condensation is periodic in one direction.
We found that the dispersion relations of NG modes for the broken diffeomorphism are constrained by the minimum energy condition, in contrast to the codimension one brane case.
Such a property would be important, e.g., in the inhomogeneous chiral condensation phase~\cite{spiralproject}.

\medskip
In Sec.~\ref{sec:CosetConstructionRevisited}
we revisited the coset construction
from the local symmetry point of view.
It was pointed out that the inverse Higgs constraints have two physical meanings\cite{Nicolis:2013sga,Endlich:2013vfa,Brauner:2014aha}:
removing redundant NG fields and massive fields. 
The standard coset construction does not distinguish these two.
Based on the relation between the Maurer-Cartan one form and connections for spacetime symmetries,
we classified these meanings of inverse Higgs constraints
by the coordinate dimension of broken symmetries.
Inverse Higgs constraints for spacetime symmetries with a higher dimension
remove redundant NG fields and,
in particular,
those for the special conformal symmetry are closely related to the fact that
the original system admits Ricci gauging.
Those for dimensionless symmetries
can be further classified by the local symmetry breaking pattern,
just as the codimension one brane case in Sec.~\ref{sec:single}.

\medskip
Although we mainly focused on the relativistic case,
it would be interesting to extend the discussion to the nonrelativistic case.
It would be also interesting to include supersymmetry in our discussion.
We defer such studies to future work.

\bigskip

\section*{Acknowledgments}
We would like to thank Luca Delacr\'etaz, Tomoya Hayata, Katsumi Ito, Kazuhiko Kamikado, Takuya Kanazawa and Yu Nakayama for valuable discussions.
Y.H. and T.N. also thank Institute for Advanced Study at the Hong Kong University of Science and Technology,
where a part of this work was done.
Y.H. is partially supported by JSPS KAKENHI Grants Numbers 24740184, and by the RIKEN iTHES Project.
T.N. is supported in part
by Special Postdoctoral Researchers
Program at RIKEN
and the RIKEN iTHES Project.
G.S. is supported in part by the DOE grant DE-FG-02-95ER40896 and the HKGRC grants HKUST4/CRF/13G, 604213, 16304414.
\\

\appendix
\section{Spacetime symmetry in nonrelativistic systems}
\label{Sec:nonrela}
\setcounter{equation}{0}

In this appendix
we extend the discussion in Sec.~\ref{sec:strategy}
to nonrelativistic systems.
After some geometrical preliminaries,
we discuss local properties of nonrelativistic spacetime symmetries.
We then summarize how
they can be gauged and embedded into 
local symmetries.

\subsection{Geometrical preliminaries}
\label{subsec:preliminaries}

\paragraph{$3+1$ decomposition}

In nonrelativistic systems,
there exists a particular time direction,
and constant-time slices
specify a spatial foliation structure.
Correspondingly,
spacetime symmetries in nonrelativistic systems
should preserve the foliation structure.
To discuss
such systems and symmetries,
it is convenient to introduce
a time-like vector field $n_\mu$
perpendicular to the spatial slices
normalized as
\begin{align}
g^{\mu\nu}n_\mu n_\nu=-1\,.
\end{align}
The induced metric $h_{\mu\nu}$
on the slices is then given by
\begin{align}
h_{\mu\nu}=g_{\mu\nu}+n_\mu n_\nu\,.
\end{align}
We also introduce
the projectors onto the temporal and spatial directions as
\begin{align}
\label{projectors}
\text{temporal projector: }-n^\mu n_\nu\,,
\quad
\text{spatial projector: }h^\mu_\nu\,.
\end{align}
In the following,
we often write the temporal component and spatial projection
of a vector $v^\mu$ as
\begin{align}
\tem{v}=n_\mu v^\mu\,,
\quad
\spa{v}^\mu=h^\mu_\nu v^\nu\,.
\end{align}

\paragraph{Decomposition of local Lorentz indices}
It is also convenient to decompose local Lorentz indices into the temporal and spatial directions in a similar way.
Using the projectors,
\begin{align}
\label{local_Lorentz_projectors}
\text{temporal projector: }\delta^m_0 \delta_n^0\,,
\quad
\text{spatial projector: }\delta^m_n-\delta^m_0 \delta_n^0\,,
\end{align}
we decompose the vierbein $e^\mu_m$ as
\begin{align}
e^\mu_m=h^\mu_\nu e^\nu_n \left(\delta^n_m-\delta^n_0\delta^0_m\right)
+h^\mu_\nu e^\nu_0 \delta^0_m
-n^\mu n_\nu e^\nu_0\delta_m^0
-n^\mu n_\nu e^\nu_n\left(\delta^n_m-\delta^n_0\delta^0_m\right)\,,
\end{align}
where the second and the fourth terms
mix the temporal/spatial coordinate indices
and the spatial/temporal local Lorentz indices.
Such terms can be eliminated by performing local Lorentz boost transformations such that the temporal directions
of the global coordinate and the local Lorentz frame coincide with each other.
Indeed,
we can always impose the gauge condition,
\begin{align}
\label{temporal_gauge}
e_0^\mu=n^\mu
\end{align}
to obtain the vierbein of the form,
\begin{align}
\label{spatial_vierbein}
e^\mu_m=\tilde{e}^\mu_m+n^\mu\delta_m^0
\quad
{\rm with}
\quad
\tilde{e}^\mu_m=h^\mu_\nu e^\nu_n \left(\delta^n_m-\delta^n_0\delta^0_m\right)\,.
\end{align}
Note that
the gauge condition~\eqref{temporal_gauge}
is invariant under diffeomorphisms and local rotations. 
In the rest of this appendix,
we impose the gauge condition~\eqref{temporal_gauge}
and use the expression~\eqref{spatial_vierbein} of the vierbein.

\medskip
The spatial projection $\tilde{e}^\mu_m$ can then be identified with
the spatial dreibein.
First, its square reproduces the spatial induced metric:
\begin{align}
\tilde{e}_m^\mu \tilde{e}_n^{\nu}\eta^{mn}=h^{\mu\nu}\,,
\quad
\tilde{e}_m^\mu \tilde{e}_n^\nu h_{\mu\nu}
=\tilde{e}_m^\mu \tilde{e}_n^\nu g_{\mu\nu} 
=\eta_{mn}+\delta_m^0\delta^0_n\,.
\end{align}
Also the relations,
\begin{align}
\tilde{e}^m_\mu
=\left(\delta^m_n-\delta^m_0\delta^0_n\right)e^n_\mu
=e^m_\nu h^\nu_\mu\,,
\quad
n^\mu \delta_m^0
=-n^\mu n_\nu e^\nu_m = e^\mu_n \delta^n_0\delta_m^0\,,
\end{align}
guarantee that the decomposition of coordinate indices and that of local Lorentz indices
are consistent.
More concretely,
we can use the notation $\spa{v}^m$ with the local Lorentz index consistently:
\begin{align}
\spa{v}^m=\left(\delta^m_n-\delta^m_0\delta^0_n\right)v^n
=e^m_\nu \spa{v}^\nu
=\tilde{e}^m_\nu \spa{v}^\nu\,.
\end{align}
The temporal projection is also consistent between the two:
\begin{align}
\tem{v}=n_\mu v^\mu=n_\mu e^\mu_m v^m=v_0=-v^0\,.
\end{align}

\subsection{Local properties of nonrelativistic spacetime symmetries}

We now discuss local properties of nonrelativistic spacetime symmetries
under some plausible assumptions on the foliation structure
and symmetry transformations.

\paragraph{Nonrelativistic ansatz}

When we take a nonrelativistic limit of relativistic systems,
the time direction is typically identified with that in a rest frame
of massive free particles.
It would then be natural to assume that the time-like vector $n_\mu$ generates time-like geodesics
and satisfies
\begin{align}
\label{geodesics}
n^\nu\nabla_\nu n^\mu=0\,.
\end{align}
As we mentioned earlier,
spacetime symmetries in nonrelativistic systems should preserve
the foliation structure.
Coordinate transformations preserving
the foliation structure
(foliation preserving diffeomorphism transformations)
can be defined as\footnote{
If we choose the coordinate system
such that $x^t$ coincides with the time direction,
a concrete form of $n_\mu$ is given by
\begin{align}
n_\mu=-\frac{\delta_\mu^t}{\sqrt{-g^{tt}}}\,.
\end{align}
Correspondingly, the geodesic condition~\eqref{geodesics}
and the foliation preserving condition~\eqref{foliation_preserving_condition}
can be stated as
\begin{align}
\partial_ig^{tt}=\partial_i\epsilon^t=0\,.
\end{align}}
\begin{align}
\label{foliation_preserving_condition}
x^\mu\to x'^\mu = x^\mu-\epsilon^\mu(x)
\quad
{\rm with}
\quad
h_\mu^\lambda \mathcal{L}_\epsilon n_\lambda\equiv h_\mu^\lambda (\epsilon^\nu\nabla_\nu n_\lambda+ n_\nu\nabla_\lambda \epsilon^\nu)=0\,,
\end{align}
where $\mathcal{L}_\epsilon$ is the Lie derivative along $\epsilon^\mu$. From Eq.~\eqref{geodesics},
we obtain $h_\mu^\nu\partial_\nu\tem{\epsilon}=0$,
which guarantees that the time-component of the transformation parameter
is constant on each slice.
In the rest of this appendix,
we assume
that the time-like vector $n_\mu$
satisfies the geodesic assumption~\eqref{geodesics}
and the nonrelativistic spacetime symmetries
satisfy the condition~\eqref{foliation_preserving_condition}. 

\paragraph{Local decomposition}

As we discussed in Sec.~\ref{subsec:local},
local properties of spacetime symmetry are determined by the covariant derivative of the corresponding coordinate transformation parameter $\epsilon^\mu$.
In nonrelativistic systems,
it is convenient to decompose $\nabla_\mu\epsilon_\nu$
using the projectors~\eqref{projectors}
as
\begin{align}
\nonumber
\nabla_\mu\epsilon_\nu
&=-\nabla_\mu(n_\nu\tem{\epsilon})+\nabla_\mu{\spa{\epsilon}}_\nu
\\
\nonumber
&=-K_{\mu\nu}\tem{\epsilon}-n_\nu \partial_\mu\tem{\epsilon}
+\nabla_\mu{\spa{\epsilon}}_\nu
\\
&=-K_{\mu\nu}\tem{\epsilon}
+n_\mu n_\nu \left(n^\rho\partial_\rho\tem{\epsilon}\right)
+\nabla_\mu{\spa{\epsilon}}_\nu\,,
\end{align}
where $K_{\mu\nu}=h_\mu^\rho\nabla_\rho n_\nu$ is the extrinsic curvature
on the spatial slices,
and
we used the geodesic condition~\eqref{geodesics}
and the foliation preserving condition, $h_\mu^\nu\partial_\nu\tem{\epsilon}=0$,
at the second and the third equalities, respectively.
The last term can be further decomposed as
\begin{align}
\nonumber
\nabla_\mu{\spa{\epsilon}}_\nu
&=h_\mu^\alpha h_\nu^\beta\nabla_\alpha{\spa{\epsilon}}_\beta
-h_\mu^\alpha n_\nu n^\beta\nabla_\alpha{\spa{\epsilon}}_\beta
-n_\mu n^\alpha \nabla_\alpha{\spa{\epsilon}}_\nu
\\*
&=h_\mu^\alpha h_\nu^\beta\nabla_\alpha{\spa{\epsilon}}_\beta
+K_{\mu\rho}\spa{\epsilon}^\rho n^\nu
-n_\mu  h_\nu^\rho n^\alpha\nabla_\alpha{\spa{\epsilon}}_\rho\,,
\end{align}
where we used $n_\alpha \spa{\epsilon}^\alpha=0$
and the geodesic condition~\eqref{geodesics}
at the second equality.
We then have
\begin{align}
\label{deformation_nonrela_components}
\nabla_\mu\epsilon_\nu
=
n_\mu n_\nu \left(n^\rho\partial_\rho\tem{\epsilon}\right)
+\left(h_\mu^\alpha h_\nu^\beta\nabla_\alpha{\spa{\epsilon}}_\beta
-K_{\mu\nu}\tem{\epsilon}
\right)
+\left(K_{\mu\rho}\spa{\epsilon}^\rho n_\nu
-n_\mu  h_{\nu}^\rho n^\alpha\nabla_\alpha{\spa{\epsilon}}_\rho\right)\,,
\end{align}
where the first term
represents local rescalings in the temporal direction.
For later use,
we define
\begin{align}
\tem{\lambda}=-n^\mu\partial_\mu \tem{\epsilon}\,.
\end{align}
The second term in Eq.~\eqref{deformation_nonrela_components}
describes deformations of spatial coordinates
and it can be decomposed~as
\begin{align}
h_\mu^\alpha h_\nu^\beta\nabla_\alpha{\spa{\epsilon}}_\beta
-K_{\mu\nu}\tem{\epsilon}
={\spa{\omega}}_{\mu\nu}+{\spa{s}}_{\mu\nu}+\spa{\lambda}h_{\mu\nu}\,,
\end{align}
where the antisymmetric part ${\spa{\omega}}_{\mu\nu}$,
the symmetric traceless part ${\spa{s}}_{\mu\nu}$,
and the trace part $\spa{\lambda}$
generate
local rotations,
anisotropic spatial rescalings,
and isotropic spatial rescalings,
respectively.
Spatial isotropy
requires that ${\spa{s}}_{\mu\nu}=0$
to obtain
\begin{align}
h_\mu^\alpha h_\nu^\beta\nabla_\alpha{\spa{\epsilon}}_\beta
-K_{\mu\nu}\tem{\epsilon}
={\spa{\omega}}_{\mu\nu}+\spa{\lambda}h_{\mu\nu}\,.
\end{align}
The last term in Eq.~\eqref{deformation_nonrela_components}
mixes the temporal and spatial directions.
If we introduce parameters $b^\pm_\mu$'s as
\begin{align}
b^\pm_\mu=-h_{\mu\nu} n^\rho\nabla_\rho \spa{\epsilon}^\nu\pm K_{\mu\nu}\spa{\epsilon}^\nu\,,
\end{align}
we can rewrite the last term in Eq.~\eqref{deformation_nonrela_components} as
\begin{align}
K_{\mu\rho}\spa{\epsilon}^\rho n_\nu
-n_\mu  h_{\nu\rho} n_\alpha\nabla^\alpha\spa{\epsilon}^\rho
=\frac{1}{2}n_\mu(b^+_\nu+b^-_\nu)
+\frac{1}{2}n_\nu(b^+_\mu-b^-_\mu)\,.
\end{align}
Here note that $n^\mu b^\pm_\mu=0$.
It should be also noted that when the extrinsic curvature is zero,
$b^+_\mu=b^-_\mu$ is the temporal derivative of $\spa{\epsilon}^\mu$.
As it suggests,
$b^\pm_\mu$ can be thought of as local Galilei boosts.

\medskip
To summarize,
using the quantities introduced above,
we can decompose $\nabla_\mu\epsilon_\nu$
for nonrelativistic spacetime symmetries
as
\begin{align}
\label{nonrela_decompsition}
\nabla_\mu\epsilon_\nu
={\spa{\omega}}_{\mu\nu}
+\frac{1}{2}n_\mu(b^+_\nu+b^-_\nu)
+\frac{1}{2}n_\nu(b^+_\mu-b^-_\mu)
-\tem{\lambda}n_\mu n_\nu 
+\spa{\lambda}h_{\mu\nu}
\,,
\end{align}
where ${\spa{\omega}}_{\mu\nu}$, $\tem{\lambda}$, and $\spa{\lambda}$
describe local rotations, temporal rescalings,
and spatial rescaling, respectively.
The parameters $b_\mu^\pm$'s are associated with
local Galilei boosts.

\paragraph{Transformation rule of $n^\mu$, $h^{\mu\nu}$,
and $\tilde{e}^\mu_m$}
To understand the physical interpretation
of the above decomposition,
it would be useful to note the transformation rule
of the unit vector $n^\mu$, the induced metric $h^{\mu\nu}$,
and the spatial dreibein $\tilde{e}^\mu_m$
under infinitesimal foliation preserving diffeomorphisms.
First,
their general coordinate transformations
are given by
\begin{align}
\label{diffs_n_mu}
\delta n^\mu&=-n^\rho\nabla_\rho \epsilon^\mu
+\epsilon^\rho\nabla_\rho n^\mu
\,,
\\*
\label{diffs_h_munu}
\delta h^{\mu\nu}
&=-\big(h^{\mu\rho} \nabla_\rho\epsilon^\nu+h^{\nu\rho} \nabla_\rho\epsilon^\mu\big)
+\epsilon^\rho \nabla_\rho h^{\mu\nu}
\,,
\\*
\label{diffs_dreibein}
\delta \tilde{e}^\mu_m
&=-\tilde{e}_m^\nu\nabla_\nu \epsilon^\mu
+\epsilon^\rho\partial_\rho \tilde{e}_m^\mu
+\epsilon^\rho\,\Gamma^\mu_{\rho\nu}\tilde{e}_m^\nu
\,.
\end{align}
By using
the geodesic condition~\eqref{geodesics}
and the foliation preserving condition~\eqref{foliation_preserving_condition},
they can be reduced to the form,\footnote{
For notational simplicity,
we use $b_\pm^\mu$
to denote $g^{\mu\nu}b^\pm_\nu$.}
\begin{align}
\delta n^\mu
=-\tem{\lambda} n^\mu
+b_+^{\mu}
\,,
\quad
\delta h^{\mu\nu}
=-2\spa{\lambda}h^{\mu\nu}\,,
\quad
\delta \tilde{e}^\mu_m
=\spa{\omega}^\mu{}_\nu\tilde{e}^\nu_m-\spa{\lambda}\tilde{e}^\mu_m-\epsilon^\rho \widetilde{S}_{\rho\,m}{}^n\,\tilde{e}^\mu_n\,,
\end{align}
where we defined
\begin{align}
\label{Stilde}
\widetilde{S}_\mu^{mn}=
\left(\delta^m_r-\delta^m_0\delta_r^0\right)
S_\mu^{rs}
\left(\delta^n_s-\delta^n_0\delta_s^0\right)
=\tilde{e}^m_\nu\partial_\mu \tilde{e}^{\nu n}+\tilde{e}^m_\lambda\Gamma^\lambda_{\mu\nu}\tilde{e}^{\nu n}\,.
\end{align}
Note that
the transformations of
spatial quantities $h^{\mu\nu}$ and $\tilde{e}^\mu_m$
(with upper indices)
depend only on the spatial components
${\spa{\omega}}_{\mu\nu}$ and $\spa{\lambda}$.
In particular,
the spatial metric $h^{\mu\nu}$ (and $h_{ij}$ also)
is invariant under transformations with $\spa{\lambda}=0$.
Such properties
are consistent with the interpretation that
${\spa{\omega}}_{\mu\nu}$ and $\spa{\lambda}$
generate local rotations and spatial rescalings.

\subsection{Examples: Galilean, Schr\"odinger and Galilean conformal symmetries}
\label{subsec:examples}

Before discussing embedding of nonrelativistic spacetime symmetries into local ones,
let us perform the local decomposition
for concrete nonrelativistic spacetime symmetries
in this subsection.
As illustrative examples,
we consider Galilean, Schr\"odinger, and Galilean conformal symmetries on the Minkowski space.

\paragraph{Galilean symmetry}
Galilean symmetry is generated by
translations $P_\mu$, rotations $J_{ij}$, and Galilei boosts $B_i$.
Their algebras can be obtained by taking the nonrelativistic limit of the Poincar\'e algebra,
except for a possible central extension in the commutator
of spatial translations and Galilei boosts 
\begin{align}
\label{central_extension}
[P_i,B_j]= -\delta_{ij}M\,,
\end{align}
where the central charge $M$ is associated with the mass energy
and it can be identified with the internal $U(1)$ charge associated with the particle number conservation.
As is suggested by the commutator~\eqref{central_extension},
the Galilei boost generates
internal $U(1)$ transformations
as well as the spacetime transformation.
Using the notation in Sec.~\ref{subsec:local_coset},
we can express the Galilei boost as
\begin{align}
v^iB_i=tv^i\partial_i-(v\cdot x) M\,,
\end{align}
where $v^i$ is the transformation parameter.
Since its spacetime transformation part takes the form
\begin{align}
\epsilon^t=0\,,
\quad
\epsilon^i=v^it\,,
\end{align}
nonzero components in the decomposition~\eqref{nonrela_decompsition}
are given by
\begin{align}
b^i_+=b^i_-=-v^i\,,
\end{align}
which is consistent with the
observation that $b_\pm^\mu$'s
are associated with local Galilei boosts.
Note that local decompositions of other generators
are the same as the relativistic case.

\paragraph{Schr\"odinger symmetry}
We next consider the Schr\"odinger symmetry~\cite{Hagen:1972pd,Niederer:1972zz},
which is generated by
\begin{align}
\widetilde{D}=2 t\partial_t+ x^i\partial_i\,,
\quad
\widetilde{K}=t^2\partial_t+ tx^i\partial_i-\frac{1}{2} x^2M\,,
\end{align}
and Galilean symmetry generators.
Nonzero components in the decomposition~\eqref{nonrela_decompsition}
for $\lambda\widetilde{D}$ are
\begin{align}
\frac{1}{2}\tem{\lambda}=\spa{\lambda}=\lambda\,.
\end{align}
On the other hand,
those for $\mu\widetilde{K}$ are
\begin{align}
\frac{1}{2}\tem{\lambda}=\spa{\lambda}=\mu t\,,
\quad
b_+^i=b_-^i=-\mu x^i\,.
\end{align}
Here $\lambda$ and $\mu$ are transformation parameters.
We notice that both of $\widetilde{D}$ and $\widetilde{K}$
have the rescaling components
satisfying $\tem{\lambda}=2\spa{\lambda}$.
In other words,
the dynamical exponent is $z=2$.

\paragraph{Galilean conformal symmetry}
Finally,
let us consider the Galilean conformal symmetry~(see, e.g., Ref.~\cite{Bagchi:2009my} for references).
For this purpose,
it is convenient to introduce the extended Galilean conformal algebra generated by
\begin{align}
L^{(n)}=(n+1)t^nx^i\partial_i+t^{n+1}\partial_t\,,
\quad
M_i^{(n)}=t^n\partial_i\,,
\quad
J_{ij}^{(n)}=t^n(x^i\partial_j-x^j\partial_i)\,,
\end{align}
where $n$ is an arbitrary integer.
In terms of these operators,
the Galilean conformal symmetry generators
are given by $L^{(n)}$ and $M_i^{(n)}$
with $n=0,\pm1$,
and $J_{ij}^{(0)}$.
Using a function $\Lambda(t)$ of time,
the coordinate transformation associated with $L^{(n)}$'s can be recast as
\begin{align}
\epsilon^t=\Lambda(t)\,,
\quad
\epsilon^i=\Lambda'(t)x^i\,,
\end{align}
and nonzero components in the decomposition~\eqref{nonrela_decompsition} are
\begin{align}
\tem{\lambda}=\spa{\lambda}=\Lambda'(t)\,,
\quad
b_+^i=b_-^i=-\Lambda''(t)x^i\,.
\end{align}
Note that the dynamical exponent is $z=1$.
On the other hand,
coordinate transformations associated with $M_i^{(n)}$'s
take the form,
\begin{align}
\epsilon^t=0\,,
\quad
\epsilon^i=B^i(t)\,,
\end{align}
and nonzero components are
\begin{align}
b_+^i=b_-^i=-B'^i(t)\,,
\end{align}
which can be thought of as
a time-dependent generalization of Galilei boosts.
Similarly,
$J_{ij}^{(n)}$'s can be regarded as
a time-dependent generalization of spatial rotations.

\subsection{Embedding into local symmetries}
As we have seen in the previous subsection,
nonrelativistic spacetime symmetries
generically have a particular dynamical exponent $z$
and the decomposition~\eqref{nonrela_decompsition}
takes the form,
\begin{align}
\label{nonrela_decompsition_generic}
\nabla^\mu\epsilon^\nu
=\spa{\omega}^{\mu\nu}
+\frac{1}{2}n^\mu(b_+^\nu+b_-^\nu)
+\frac{1}{2}n^\nu(b_+^\mu-b_-^\mu)
+\lambda\left(-z\,n^\mu n^\nu 
+h^{\mu\nu}\right)
\,.
\end{align}
Let us concentrate on such symmetries
in the following.
They also admit central extensions.
In this subsection
we first discuss how nonrelativistic spacetime symmetries
without central extensions can be embedded into local symmetries.
We then extend discussions to the case with central extensions.

\paragraph{Without central extensions}
Let us begin with the case without central extensions.
In this case,
the transformation rules of local fields
are determined
by their local rotation charge and scaling dimension.
Suppose that a local field $\Phi(x)$ follows a representation $\widetilde{\Sigma}_{mn}$ and has a scaling dimension $\Delta_\Phi$,
where $\widetilde{\Sigma}_{mn}$ is projected on to the spatial direction:
$\widetilde{\Sigma}_{0n}=\widetilde{\Sigma}_{m0}=0$.
It is then transformed as\footnote{
Note that fields with coordinate indices can be decomposed
into local fields following some representations
of local rotations,
by using the vierbein.
For example,
a gauge field $A_\mu$ can be decomposed as
\begin{align}
A^\mu=-n^\mu \tem{A}+\spa{A}^\mu
=n^\mu A^{0}+\tilde{e}^\mu_m\spa{A}^m\,.
\end{align}
Here $\tem{A}=-A^{0}$
and $\spa{A}^m$
are a scalar and a spatial vector, respectively,
and their transformation rules follow from Eq.~\eqref{nonrela_Phi_1}.
In this way,
any local field can be expressed
in terms of local fields with the transformation rule~\eqref{nonrela_Phi_1},
the time-like vector $n_\mu$,
the spatial induced metric $h_{\mu\nu}$,
and the spatial dreibein $\tilde{e}^\mu_m$.}
\begin{align}
\label{nonrela_Phi_1}
\delta\Phi&=\Delta_\Phi\lambda(x)\Phi+\frac{1}{2}\spa{\omega}^{mn}(x)\widetilde{\Sigma}_{mn}\Phi+\epsilon^\mu(x)\nabla_\mu\Phi
\,,
\end{align}
where $\spa{\omega}^{mn}=\tilde{e}^m_\mu\tilde{e}^n_\nu\spa{\omega}^{\mu\nu}$.
The covariant derivative is defined by
\begin{align}
\label{nonrela_cov_der}
\nabla_\mu\Phi
=\partial_\mu\Phi+\frac{1}{2}S_\mu^{mn}\widetilde{\Sigma}_{mn}\Phi
=\partial_\mu\Phi+\frac{1}{2}\widetilde{S}_\mu^{mn}\widetilde{\Sigma}_{mn}\Phi\,,
\end{align}
where $\widetilde{S}_\mu^{mn}$ is given by Eq.~\eqref{Stilde}.
Rewriting Eq.~\eqref{nonrela_Phi_1} as
\begin{align}
\label{nonrela_Phi_2}
\delta\Phi
&=\Delta_\Phi\lambda(x)\Phi+\frac{1}{2}\Big(\spa{\omega}^{mn}(x)+\epsilon^\mu(x)\widetilde{S}_\mu^{mn}(x)\Big)\widetilde{\Sigma}_{mn}\Phi+\epsilon^\mu(x)\partial_\mu\Phi
\,,
\end{align}
we notice that
the three terms can be thought of as
anisotropic Weyl transformations,
local rotations, and diffeomorphisms.
Since the transformation rule of $\Phi$
under anisotropic Weyl transformations,
local rotations, and diffeomorphisms is given by
\begin{align}
\text{anisotropic Weyl}:\delta \Phi=\Delta_\Phi\tilde{\lambda}\Phi\,,
\,\,\,\,
\text{local rotation}:\delta \Phi=\frac{1}{2}\tilde{\omega}_\bot^{mn}\widetilde{\Sigma}_{mn}\Phi\,,
\,\,\,\,
\text{diffs}:
\delta\Phi=\tilde{\epsilon}^\mu\partial_\mu\Phi\,,
\end{align}
the transformation~\eqref{nonrela_Phi_2}
can be reproduced by the parameter choice:
\begin{align}
\label{embed_nonrelativistic}
\tilde{\lambda}=\lambda\,,
\quad
\tilde{\omega}_\bot^{mn}=\spa{\omega}^{mn}+\epsilon^\mu \widetilde{S}_\mu^{mn}\,,
\quad
\tilde{\epsilon}^\mu=\epsilon^\mu\,,
\end{align}
where $\tilde{\lambda}$, $\tilde{\omega}_\bot^{mn}$, and $\tilde{\epsilon}$
are transformation parameters of
anisotropic Weyl transformations,
local rotations, and diffeomorphisms, respectively.
Similarly,
the transformation rule
of $h^{\mu\nu}$, $\tilde{e}^\mu_m$, and $n^\mu$
under local symmetries are given by
\begin{align}
\label{anisotropic_Weyl}
\text{anisotropic Weyl}:&\quad\,\,
\delta h^{\mu\nu}=2\tilde{\lambda} h^{\mu\nu}\,,
\quad\,
\delta \tilde{e}_m^\mu=\tilde{\lambda} \tilde{e}_m^\mu\,,
\quad\hspace{7.2mm}\,\,
\delta n^\mu=z\tilde{\lambda} n^\mu\,,
\\*
\label{local_rotation}
\text{local rotation}:&
\quad\,
\delta h^{\mu\nu}=0\,,
\quad\quad
\quad\,\,\,
\delta \tilde{e}_m^\mu=\tilde{\omega}_{\bot \,m}{}^n \tilde{e}_n^\mu\,,
\quad\,\,
\delta n^\mu=0\,,
\end{align}
and Eqs.~\eqref{diffs_n_mu}-\eqref{diffs_dreibein},
where $z$ is the dynamical exponent.
It then turns out that
the spatial induced metric $h^{\mu\nu}$ and the spatial dreibein $\tilde{e}^\mu_m$,  ($h_{ij}$, $\tilde{e}_i^m$, and $\delta n_\mu$ also)
are invariant under the (global) nonrelativistic spacetime symmetry transformation given by the parameter choice~\eqref{embed_nonrelativistic}.
On the other hand, however,
the time-like vector is not invariant and transforms as
\begin{align}
\label{delta_n_mu}
\delta n^\mu=b^\mu_+\,.
\end{align}

\paragraph{Central extension}
We then consider the case with the central extension.
In this case,
spacetime symmetries can generate internal $U(1)$ transformations as well as spacetime ones,
just as Galilei boosts do.
Using the notation in Sec.~\ref{subsec:local_coset},
let us write such spacetime symmetries as
\begin{align}
\epsilon^\mu(x)\partial_\mu+\alpha(x) M\,,
\end{align}
where $M$ is the internal $U(1)$ generator
and $\alpha(x)$ is the corresponding parameter.
For example,
the Galilei boost $v^iB_i$
can be expressed as $\epsilon^t=0$,
$\epsilon^i=v^it$, and $\alpha=-v_ix^i$,
as we illustrated in Sec.~\ref{subsec:examples}.
When a local field $\Phi$ has an internal $U(1)$ charge $im$,
the transformation rule~\eqref{nonrela_Phi_1}
is extended to
\begin{align}
\label{nonrela_Phi_3}
\delta\Phi
&=\lambda(x)\Delta_\phi\Phi+\frac{1}{2}\spa{\omega}^{mn}(x)\widetilde{\Sigma}_{mn}\Phi+\epsilon^\mu(x)\nabla_\mu\Phi
+i m\alpha(x) \Phi
\,.
\end{align}
Also,
the internal $U(1)$ gauge field is transformed as
\begin{align}
\delta A_\mu = A_\nu\nabla_\mu\epsilon^\nu+\epsilon^\rho\nabla_\rho A_\mu-\partial_\mu\alpha\,.
\end{align}
Note that
the transformation rule
of the temporal component and the spatial projection
of the gauge field is given by
\begin{align}
\delta \tem{A}&=z\lambda \tem{A}+\epsilon^\mu\partial_\mu\tem{A}
-n^\mu\partial_\mu\alpha\,,
\\
\delta A_{\bot m}&=\lambda A_{\bot m}+\omega_{\bot m}{}^nA_{\bot n}+\epsilon^\rho\nabla_\rho A_{\bot m}
-\tilde{e}_m^\mu\partial_\mu\alpha\,,
\end{align}
whose dependence on $\lambda$, $\omega_{\bot m}{}^n$, and $\epsilon^\mu$ is consistent with
Eq.~\eqref{nonrela_Phi_1}.
Since the internal $U(1)$ gauge transformations
of $\Phi$ and $A_\mu$ are given by
\begin{align}
\delta \Phi=im\tilde{\alpha}\Phi\,,
\quad
\delta A_\mu=-\partial_\mu\tilde{\alpha}\,,
\end{align}
the (global) nonrelativistic spacetime symmetry transformation
can be reproduced by the parameter set
$\tilde{\alpha}=\alpha$ and Eq.~\eqref{embed_nonrelativistic},
where $\tilde{\alpha}$ is the internal $U(1)$ gauge transformation parameter.
Note that the transformation rule
of $h^{\mu\nu}$, $\tilde{e}^\mu_m$, and $n^\mu$
is the same as the case without central extensions:
under the (global) nonrelativistic spacetime symmetry transformation,
the spatial metric and dreibein
are invariant,
but the time-like vector transforms as~\eqref{delta_n_mu}.

\bigskip
To summarize,
the transformation rule~\eqref{nonrela_Phi_3}
of standard matter fields
can be naturally reproduced
by embedding global nonrelativistic spacetime symmetries
into diffeomorphisms, local rotations, anisotropic Weyl symmetries, and internal $U(1)$ gauge symmetries
associated with the central extension.
Identification of symmetry breaking patterns and the corresponding NG fields should therefore be based on
those local symmetries:
When the condensation is spacetime dependent,
diffeomorphism invariance is broken.
The local rotation symmetry, the anisotropic Weyl symmetry,
and the internal $U(1)$ symmetry
are broken
when the condensation has a rotation charge, scaling dimension, and internal $U(1)$ charge, respectively.
The effective action construction can then be performed
in a similar way to the relativistic case,
by gauging those local symmetries.
This is one point of this appendix.

\medskip
It should be also noted that
the time-like vector $n^\mu$ and
the internal $U(1)$ gauge field $A_\mu$
transform nonlinearly under nonrelativistic spacetime symmetries with nonvanishing $b^+_\mu$ and $\partial_\mu\alpha$.
This situation is similar to
the Weyl gauge field $W_\mu$ in conformal systems.
As we mentioned in Sec.~\ref{subsec:gaugespacetime},
when we perform Weyl gauging
in conformal field theories,
the Weyl gauge field $W_\mu$
appears in a particular combination
because it is not special conformal invariant by itself.
In the next subsection,
we will illustrate that a similar situation occurs
for $n^\mu$ and $A_\mu$ in Galilei boost invariant systems.

\subsection{Gauging nonrelativistic spacetime symmetries}

We then summarize how global nonrelativistic spacetime symmetries
can be gauged into local ones.
First,
the diffeomorphism symmetry, local rotation symmetry,
and internal $U(1)$ gauge symmetry
associated with the central extension
can be realized
by introducing covariant quantities $n^\mu$, $h^{\mu\nu}$,
and $\tilde{e}^\mu_m$ introduced in appendix~\ref{subsec:preliminaries},
and the gauge field $A_\mu$.
For example,
the free fermion action,
\begin{align}
\label{free_fermion}
S=\int d^4x\left[i\psi^*\partial_t\psi-\frac{1}{2m}|\partial_i\psi|^2\right]\,,
\end{align}
can be reformulated as
\begin{align}
\label{curved_action_nonrela}
S=\int d^4x
\sqrt{-g}\left[
\frac{1}{2}in^\mu\psi^*(\overset{\longleftrightarrow}{\nabla}_\mu+imA_\mu)\psi
-\frac{1}{2m}h^{\mu\nu}(\nabla_\mu-imA_\mu)\psi^*(\nabla_\nu+imA_\nu)\psi
\right]\,,
\end{align}
where the covariant derivative $\nabla_\mu$ is defined by Eq.~\eqref{nonrela_cov_der}
and $\psi^*\overset{\longleftrightarrow}{\nabla}_\mu\psi\equiv \psi^*{\nabla}_\mu\psi-({\nabla}_\mu\psi^*)\psi$.
This curved space action
enjoys the full diffeomorphism symmetry,
the local rotation symmetry, and the internal $U(1)$ gauge symmetry.
Note that
the original action~\eqref{free_fermion}
can be reproduced by setting that
\begin{align}
\label{flat_nonrela}
h^{\mu\nu}=\eta^{\mu\nu}+\delta^\mu_0\delta^\nu_0\,,
\quad
n^\mu=\delta^\mu_0\,,
\quad
A_\mu=0\,.
\end{align}
As we mentioned in the pervious subsection,
the above conditions are not invariant
under the global symmetries
with $b^+_\mu\neq0$, $\partial_\mu\alpha\neq0$, or both.
Indeed,
under a finite Galilei boost,
\begin{align}
\psi'(x)=e^{im\alpha(x)}\psi(x+\epsilon)
\quad
{\rm with}
\quad
\epsilon^t=0\,,
\quad
\epsilon^i=v^it\,,
\quad
\alpha=-v_ix^i-\frac{1}{2}v^2t\,,
\end{align}
the time-like vector $n^\mu$ and the gauge field $A_\mu$
are transformed as
\begin{align}
n'^\mu(x)=n^\mu(x+\epsilon)-\delta^\mu_iv^in^t(x+\epsilon)\,,
\quad
A'_\mu(x)=A_\mu(x+\epsilon)+\delta_\mu^t\Big(v^iA_i(x+\epsilon)+\frac{1}{2}v^2\Big)+\delta_\mu^iv_i\,,
\end{align}
which breaks the conditions~\eqref{flat_nonrela}.
This situation is similar to the special conformal transformation
of the Weyl gauge field $W_\mu$.
Similarly to the previous case,
by rewriting the action~\eqref{curved_action_nonrela} as
\begin{align}
S=\int d^4x
\sqrt{-g}\left[
\frac{1}{2}i\left(n^\mu+h^{\mu\nu}A_\nu\right)(\psi^*\overset{\longleftrightarrow}{\nabla}_\mu\psi)
-\frac{m}{2}\left(2n^\mu A_\mu+h^{\mu\nu}A_\mu A_\nu\right)|\psi|^2
-\frac{1}{2m}h^{\mu\nu}\nabla_\mu\psi^*\nabla_\nu\psi
\right]\,,
\end{align}
we notice that $n^\mu$ and $A_\mu$
appear in the following combinations:
\begin{align}
n^\mu+h^{\mu\nu}A_\nu\,,
\quad
2n^\mu A_\mu+h^{\mu\nu}A_\mu A_\nu\,,
\end{align}
which are Galilei boost invariant.
Note that
such combinations are known to be Milne boost invariant
in the context of the Newton-Cartan geometry.
See, e.g., Ref.~\cite{Jensen:2014aia,Hartong:2014pma} for details.

\medskip
Finally,
let us consider the anisotropic Weyl symmetry.
Just as the Ricci gauging in relativistic systems,
it is known to be possible to introduce anisotropic Weyl invariant 
curved space actions
for some class of nonrelativistic conformal theories.
If such a procedure cannot be performed,
we need to introduce a gauge field $W_\mu$
just as the Weyl gauging in relativistic systems.
If the curved space action is invariant
under global anisotropic Weyl transformations,
we can always introduce a local anisotropic Weyl invariant action
by replacing the covariant derivative $\nabla_\mu$
with the Weyl covariant derivative,\footnote{
See e.g.~\cite{Bergshoeff:2014uea} for recent discussions
on gauging of the anisotropic Weyl symmetry.}
\begin{align}
\label{Weyl_covariant_der_nonrela}
D_\mu\Phi
=\nabla_\mu\Phi+
\left(\Delta_\Phi \,\delta_\mu^\nu-\widetilde{\Sigma}_\mu{}^\nu\right)W_\nu\Phi\,,
\end{align}
where $\widetilde{\Sigma}_\mu{}^\nu=\tilde{e}_\mu^m\, \Sigma_m{}^n\,\tilde{e}_n^\nu$
and the local anisotropic Weyl transformation rule is given by
\begin{align}
\nonumber
\Phi\to\Phi'=e^{\Delta_\Phi\lambda} \Phi\,,
\quad
n^\mu\to n'^\mu=e^{-z\lambda}n^\mu\,,
\quad
h^{\mu\nu}\to h'^{\mu\nu}=e^{-2\lambda}h_{\mu\nu}\,,
\\
\tilde{e}^\mu_m\to \tilde{e}'^{\mu}_m=e^{-\lambda}\,e^\mu_m\,,
\quad
W_\mu\to W_\mu'=W_\mu-\partial_\mu\lambda\,.
\end{align}
In contrast to the relativistic case,
it seems not well understood
under what conditions Weyl gauging can be converted to Ricci gauging.
It would be interesting to investigate this issue
by extending the discussion~\cite{Iorio:1996ad} in relativistic systems.

\section{Derivation of Nambu-Goto action}
\label{App:Nambu-Goto}
\setcounter{equation}{0}

In Sec.~\ref{subsec:spectra_scalar}
we discussed that our effective action
for a single scalar brane
contains gapped modes
in addition to gapless NG modes
localizing on the brane.
In this appendix
we show that
the low-energy effective action
after integrating out massive modes
is nothing but the Nambu-Goto action.
As we have discussed,
the unitary gauge action for $z$-diffeomorphism symmetry breaking
takes the form,
\begin{align}
S&=-\frac{1}{2}\int d^4x\sqrt{-g}
\left[\alpha_1(z)\left(1+g^{zz}\right)+\alpha_3(z)\left(\delta g^{zz}\right)^2
+\mathcal{O}\left((\delta g^{zz})^3\right)\right]
\quad
{\rm with}
\quad
\delta g^{zz}=g^{zz}-1
\end{align}
at the lowest dimension.
To discuss its relation to the Nabmu-Goto action,
it is convenient to rewrite
\begin{align}
\label{rewrite}
S&=-\frac{1}{2}\int d^4x\sqrt{-h}
\left[2\alpha_1(z)+\widetilde{\alpha}_3(\delta g^{zz})^2+\mathcal{O}\left((\delta g^{zz})^3\right)\right]
\,,
\end{align}
where $h$ is the determinant of the induced metric, $h_{\widehat{\mu}\widehat{\nu}}=g_{\widehat{\mu}\widehat{\nu}}$,
on the constant $z$ surfaces.
Here we follow the convention
in Sec.~\ref{sec:single},
e.g.,
$\widehat{\mu}=t,x,y$.
We also introduced $\widetilde{\alpha}_3=\frac{1}{4}\alpha_1(z)+\alpha_3(z)$.
In the following
we show that the integration of gapped modes
provides a constraint $\delta g^{zz}=0$
and the effective action is reduced to the Nambu-Goto action
in the low-energy regime.

\medskip
For this purpose,
let us first write down the second order action for the NG field.
In the unitary gauge coordinate,
the NG field for the broken $z$-diffeomorphism
are eaten by the metric field.
The induced metric, $h_{\widehat{\mu}\widehat{\nu}}$,
and the $z$-component, $g^{zz}$, are given by\footnote{
We define $\pi$ in the unitary gauge coordinate
such that $z_{\rm flat}=z-\pi(x)$,
where $z_{\rm flat}$ is the flat space coordinate,
and $z$ and $x$ are the unitary gauge coordinates.}
\begin{align}
h_{\widehat{\mu}\widehat{\nu}}(x)=\eta_{\widehat{\mu}\widehat{\nu}}+\partial_{\widehat{\mu}}\pi(x)\partial_{\widehat{\nu}}\pi(x)\,,
\quad
g^{zz}=1+2\partial_z\pi+3(\partial_z\pi)^2+(\partial_{\widehat{\mu}}\pi)^2+\mathcal{O}(\pi^3)\,,
\end{align}
where
note that $\partial_z\pi$ appears only in $g^{zz}$.
The second order action then takes the form,
\begin{align}
\nonumber
S_2&=-\frac{1}{2}\int d^4x\left[
\alpha_1(z)(\partial_{\widehat{\mu}}\pi)^2
+4\widetilde{\alpha}_3(\partial_z\pi)^2
\right]
\\
&=-\frac{1}{2}\int d^4x\,
\alpha_1\left[
(\partial_{\widehat{\mu}}\pi)^2
-\pi\left(\frac{\blue{4}\widetilde{\alpha}_3}{\alpha_1}\partial_z^2+\frac{\blue{4}\widetilde{\alpha}_3'}{\alpha_1}\partial_z\right)\pi
\right]
\,,
\end{align}
where we dropped total derivative terms.
The physical spectrum is now determined by
the eigenvalue problem of the operator,
$\frac{\widetilde{\alpha}_3}{\alpha_1}\partial_z^2+\frac{\widetilde{\alpha}_3'}{\alpha_1}\partial_z$.
Note that our analysis in Sec.~\ref{subsec:spectra_scalar}
was for $\alpha_1=\blue{4}\widetilde{\alpha}_3=\frac{\beta^2 v^2}{\cosh^4 \beta z}$ in particular.
There,
we had two types of physical modes:
gapless modes localizing on the brane
and gapped modes propagating in the bulk.
Let us assume that
such a qualitative feature holds generically
in more general setups for a single domain-wall.
We then expand the NG field, $\pi$,
by those modes as
\begin{align}
\pi(x)=\pi_0(x_\bot)+\sum_{\lambda}\sum_{i=\pm} \pi_{\lambda_i}(x_\bot)u_{\lambda_i}(z)\,,
\end{align}
where
$x_\bot$ stands for coordinates in the transverse directions, $t,x,y$,
and
$\displaystyle\sum_{\lambda}$ denotes
both the sum and integral over $\lambda$.
$u_{\lambda_i}$ ($i=\pm$) stands for
two eigenfunctions
with the eigenvalue $\lambda$
satisfying
\begin{align}
\left(\frac{\widetilde{\alpha}_3}{\alpha_1}\partial_z^2+\frac{\widetilde{\alpha}_3'}{\alpha_1}\partial_z\right)
u_{\lambda_i}+\lambda u_{\lambda_i}=0
\quad
{\rm and}
\quad
\int dz\,\alpha_1u_{\lambda_+}u_{\lambda_-}=0\,.
\end{align}
The second order action is now reduced to the form,
\begin{align}
\nonumber
S&=-\frac{1}{2}
\int dz\,\alpha_1(z)
\int d^3x_\bot
(\partial_{\widehat{\mu}}\pi_0)^2
\\
&\quad\,
-\frac{1}{2}\sum_{\lambda}\sum_{i=\pm}
\int dz\,\alpha_1(z)\left(u_{\lambda_i}(z)\right)^2
\int d^3x_\bot\left[
(\partial_{\widehat{\mu}}\pi_{\lambda_i})^2
+\lambda\pi_{\lambda_i}^2
\right]\,,
\end{align}
where
$\lambda$ can be thought of as
the mass squared in three dimension.
Note that such a mass term originates from
the $\widetilde\alpha_3$ term in \eqref{rewrite}.
Also,
the linear equation of motion
for gapped modes reduces to
$\pi_{\lambda_\pm}=0$
in the low energy limit, $|k_{\widehat{\mu}}|^2\ll \lambda$.

\medskip
Finally,
we extend the previous discussion
to the nonlinear level and derive the Nambu-Goto action.
Just as the linear order discussions,
the $\widetilde\alpha_3$ term plays an important role:
\begin{align}
-\frac{1}{2}\int d^4x\sqrt{-h}\,
\widetilde\alpha_3(\delta g^{zz})^2
=-\frac{1}{2}\int d^4x\sqrt{-h}\,\widetilde\alpha_3
\left(2\partial_z\pi+3(\partial_z\pi)^2+(\partial_{\widehat{\mu}}\pi)^2+\mathcal{O}(\pi^3)\right)^2\,,
\end{align}
where note that the factor, $\sqrt{-h}$, does not contain
$\partial_z\pi$.
As we have discussed,
the mass term for the gapped modes, $\pi_{\lambda_i}$,
arises from this interaction at the second order action level.
If we include higher order terms,
there appear mixing interactions between gapped and gapless modes.
In the low-energy limit,
the equation of motion for the gapped mode 
is then given by
\begin{align}
\label{g^zz=1}
2\partial_z\pi+
3(\partial_z\pi)^2+(\partial_{\widehat{\mu}}\pi)^2+\mathcal{O}(\pi^3)=0
\quad
\leftrightarrow
\quad
g^{zz}=1\,,
\end{align}
so that the effective action for the gapless mode reduces to the Nambu-Goto type one,
\begin{align}
S_{\rm eff}=-\frac{1}{2}\int d^4x\sqrt{-h} \alpha_1(z)
=-T\int d^3x\sqrt{-h}
\quad
{\rm with}
\quad
T=\frac{1}{2}\int dz\,\alpha_1(z)\,,
\end{align}
where $h$ contains gapless modes only
and $T$ can be identified with the brane tension.

\bibliographystyle{h-physrev}
\bibliography{EFT}
\end{document}